\documentclass[pra,twocolumn,showpacs,amsmath,amssymb,superscriptaddress,floatfix,longbibliography,nofootinbib]{revtex4-2}
\usepackage{amsmath,amsthm,amssymb,amsfonts,enumitem,float,graphicx,physics,hyperref}
\usepackage{color}
\usepackage[dvipsnames]{xcolor}
\usepackage[normalem]{ulem}
\usepackage{subcaption}
\usepackage{caption}
\usepackage[normalem]{ulem} 
\usepackage{graphicx}
\usepackage{dcolumn}
\usepackage{bm}
\usepackage{physics}
\usepackage{blochsphere}
\usepackage{mathtools}
\usepackage{multirow}

\usepackage{hyperref}
\hypersetup{
    colorlinks=true,       
    linkcolor=cyan,          
    citecolor=magenta,        
    filecolor=magenta,      
    urlcolor=cyan,           
    runcolor=cyan
}
\usepackage[capitalise]{cleveref} 

\def\a{\alpha}
\def\b{\beta}
\def\g{\gamma}
\def\d{\delta}

\def\r{\rho}     

\def\o{\omega}

\def\D{\Delta}

\def\ox{\otimes}
\def\>{\rangle}
\def\<{\langle}
\def\Tr{\mathrm{Tr}}

\newcommand{\ketb}[2]{|{#1}\>\!\<#2|}

\newcommand{\bes} {\begin{subequations}}
\newcommand{\ees} {\end{subequations}}
\newcommand{\bea} {\begin{align}}
\newcommand{\eea} {\end{align}}
\newcommand{\beq}{\begin{equation}}
\newcommand{\eeq}{\end{equation}}
\newcommand{\ave}[1]{\<{#1}\>}

\DeclareMathOperator{\Lap}{Lap}
\DeclareMathOperator{\sinc}{sinc}

\DeclareMathOperator{\Ei}{Ei}
\DeclareMathOperator{\Ci}{Ci}
\DeclareMathOperator{\Si}{Si}

\begin{document}

\title{Markovian and non-Markovian master equations versus an exactly solvable model of a qubit in a cavity}

\author{Zihan Xia}
\author{Juan Garcia-Nila}%
\affiliation{Center for Quantum Information Science \& Technology}
\affiliation{Department of Electrical \& Computer Engineering}

\author{Daniel A. Lidar}
\affiliation{Center for Quantum Information Science \& Technology}
\affiliation{Department of Electrical \& Computer Engineering}
\affiliation{Department of Physics \& Astronomy}
\affiliation{Department of Chemistry\\
University of Southern California, Los Angeles, CA 90089, USA}

\begin{abstract}
Quantum master equations are commonly used to model the dynamics of open quantum systems, but their accuracy is rarely compared with the analytical solution of exactly solvable models. In this work, we perform such a comparison for the damped Jaynes-Cummings model of a qubit in a leaky cavity, for which an analytical solution is available in the one-excitation subspace. 
We consider the non-Markovian time-convolutionless master equation up to the second (Redfield) and fourth orders as well as three types of Markovian master equations: the coarse-grained, cumulant, and standard rotating-wave approximation (RWA) Lindblad equations. We compare the exact solution to these master equations for three different spectral densities: impulse, Ohmic, and triangular. 
We demonstrate that the coarse-grained master equation outperforms the standard RWA-based Lindblad master equation for weak coupling or high qubit frequency (relative to the spectral density high-frequency cutoff $\o_c$), where the Markovian approximation is valid. In the presence of non-Markovian effects characterized by oscillatory, non-decaying behavior, the TCL approximation closely matches the exact solution for short evolution times (in units of $\o_c^{-1}$) even outside the regime of validity of the Markovian approximations. For long evolution times, all master equations perform poorly, as quantified in terms of the trace-norm distance from the exact solution. The fourth-order time-convolutionless master equation achieves the top performance in all cases. Our results highlight the need for reliable approximation methods to describe open-system quantum dynamics beyond the short-time limit.
\end{abstract}

\maketitle

\section{Introduction}
\label{sec:intro}

The study of open quantum systems presents both conceptual and technical challenges due to the complexity and high dimensionality of the environment, or bath. Exact analytical solutions describing the joint system-bath evolution are rarely attainable, necessitating the development of approximation methods to capture the reduced system dynamics~\cite{alicki_quantum_2007,Breuer:book,rivas_open_2012}. To address this challenge, various approaches have been developed to derive master equations that describe the system's evolution using the reduced density matrix, the best known of which is the Markovian Lindblad [or Gorini-Kossakowski-Lindblad-Sudharshan (GKLS)] equation~\cite{Lindblad:76,Gorini:1976uq}. Numerous other master equations have been derived, some of which include non-Markovian effects. In some cases, rigorous error bounds have been derived that quantify the deviation between the solutions of known master equations and the exact solution~\cite{Mozgunov:2019aa}. However, these tend to be rather loose. Hence, it is desirable to compare the predictions of various master equations to non-trivial examples of exactly solvable open-system problems. This has been done, e.g., for the central spin model~\cite{Krovi2007,Jing:2018}.

This is the goal of the present work, where we
study the damped Jaynes-Cummings model 
of a qubit inside a leaky cavity~\cite{Jaynes:1963aa}.
In this model, the qubit system interacts with the cavity electromagnetic field through the dipole approximation~\cite{Garraway:1997aa}. 
The qubit decay rate can be associated with experimentally measurable parameters such as the dipole moment and the energy gap \cite{Carmichael:93}. Experimental proposals for simulating the spin-boson model with an Ohmic spectral density using superconducting circuits have been previously discussed~\cite{Koch2007,Leppakangas,Stehli2023}.
We solve this model analytically assuming the zero temperature limit and the $1$-excitation subspace of the joint qubit-cavity system, where the cavity is populated by at most a single photon. This is similar to previous studies assuming a Lorentzian spectral density~\cite{Breuer:99,Vacchini:2010fv}, but we do this for three different spectral densities: impulse, Ohmic, and triangular (formally defined below). These choices are motivated by experiments involving condensed matter systems such as superconducting qubits interacting with bosonic modes~\cite{transmon-invention}, rather than the original quantum-optical setting of an atom in a cavity that inspired the damped Jaynes-Cummings model. We then compare the exact solution to a number of different Markovian and non-Markovian master equations. We find that in the weak-coupling limit and for large qubit frequencies relative to the spectral density cutoff, where the Markovian approximation holds, the Lindblad equation derived using the coarse-graining approach~\cite{Majenz:2013qw} is more accurate than the standard rotating wave approximation based Lindblad equation. In the non-Markovian regime, the time-convolutionless master equation proves to be accurate in approximating the exact solution for relatively short evolution times. All the approximation methods we consider struggle at long evolution times.

This paper is structured as follows. In \cref{sec:exact}, we introduce the damped Jaynes-Cummings model and derive the exact solutions for all three spectral densities. In \cref{sec:approximate}, we explore various Markovian approximation methods in the context of the damped Jaynes-Cummings model. We begin with the coarse-grained Lindblad equation (CG-LE) in \cref{sec:chap9}, then the cumulant LE (C-LE) in \cref{sec:CGLch10}, and finally the commonly used rotating-wave approximation (RWA)-based LE (RWA-LE, \cref{sec:RWA-LE}). We apply the time-convolutionless (TCL) approach to second (TCL2, also known as the Redfield equation) and fourth orders (TCL4) in \cref{sec:TCL}. \cref{Analysis} is the heart of this work, where we present comparisons between the exact solutions and the various approximation methods. This includes a comparison of the exact solution with Markovian and TCL approximations for the Ohmic spectral density, a comparison of the exact solution with the Markovian CG-LE and RWA-LE, and finally a comparison with the non-Markovian TCL models for the impulse and triangular spectral densities. We summarize our findings and present our conclusions in \cref{sec:summary}. A variety of technical details that complement the main text are presented in the Appendices.

Readers who are already familiar with the different types of master equations may choose to skip \cref{sec:approximate}. All our key analytical results are conveniently accessible via \cref{tab:summary}, which provides the corresponding equation numbers. Readers who are interested primarily in the results of the comparison between the exact model results and the various master equations may choose to skip ahead to \cref{Analysis} and focus on the graphs presented there.

\section{Exact Dynamics}
\label{sec:exact}

\subsection{General open system setup}

The total Hamiltonian of the system and the bath is given by  
\begin{align}
\label{eq:general_H}
    H = H_0 + H_{SB}\ ,
\end{align}
where $H_0=H_S\ox I_B+I_S\ox H_B$ with $H_S$ and $H_B$ being the pure system and bath Hamiltonians, respectively, $I$ the identity operator.

We move to the interaction picture, where all operators transform according to: 
\begin{align}
    X \mapsto \tilde{X}(t)=e^{iH_0 t} X e^{-iH_0 t}\ .
\end{align}

The dynamics of the total system in the interaction picture are governed by the Liouville-von Neumann equation:
\begin{align}
\label{eq:Liouville}
    \frac{d\tilde{\rho}_{SB}}{dt} = -i [ \Tilde{H}_{SB},\tilde{\rho}_{SB}]\ ,
\end{align}
where $\tilde{\rho}_{SB}$ is the density matrix of the total system acting on the Hilbert space $\mathcal{H}_{SB} = \mathcal{H}_S\otimes\mathcal{H}_B$. The joint system-bath state is thus given by:
\begin{align}
\label{eq:ip_rho_sb}
    \tilde{\rho}_{SB}(t) = \tilde{U}(t)\tilde{\rho}_{SB}(0)\tilde{U}^\dagger(t)\ ,
\end{align}
where the unitary evolution operator is:
\begin{align}
\label{eq:ip_u_sb}
    \tilde{U}(t)=T_+\exp(-i\int_0^t \tilde{H}_{SB}(t')dt')\ .
\end{align}
and $T_+$ denotes Dyson time-ordering.

The solution can equivalently be expressed as a Dyson series by integrating and iterating \cref{eq:Liouville}: 
 \begin{align}
 \label{eq:Liouville-exp}
&\tilde{\rho}_{SB}(t)=\rho_{SB}(0)+\sum_{n=1}^{\infty}(-i)^n\int_0^tdt_1\int_0^{t_1}dt_2\cdots\\
&\ \ \int_0^{t_{n-1}}dt_n[\tilde{H}_{SB}(t_1),[\tilde{H}_{SB}(t_2),...[\tilde{H}_{SB}(t_n),\rho_{SB}(0)]]....] \ . \nonumber
\end{align} 

The state of the system is given by the reduced density operator:
\begin{align}
\label{eq:ip_rho_s}
    \tilde{\rho}(t)=\Tr_B[\tilde{\rho}_{SB}(t)]\ ,
\end{align}
where $\Tr_B$ denotes the partial trace over the bath state.

\subsection{Model of a qubit in a leaky cavity}
\label{sec:model}

We analyze the dynamics of a single qubit inside a leaky cavity, coupled to a bosonic bath at zero temperature. By working in the single-excitation subspace this model becomes analytically solvable, and we closely follow the solution method of Refs.~\cite{Garraway:1997aa,Breuer:99} (see also Ref.~\cite{Breuer:book}). However, we consider different bath spectral densities. 

The system Hamiltonian, $H_S$, can be expressed as
\begin{align}\label{eq:HS}
    H_S=\Omega_0 \ketb{1}{1}=\Omega_0\sigma_+\sigma_- \ .
\end{align}
Here, $\sigma_+=\ketb{1}{0}$ and $\sigma_-=\ketb{0}{1}$ are the raising and lowering operators for the qubit, respectively. The qubit ground state is $\ket{0}$ with energy $0$ and its excited state is $\ket{1}$ with energy $\Omega_0$. The bath Hamiltonian, $H_B$, is given by:
\begin{align}\label{eq:HB}
    H_B=\sum_k \omega_k b_k^\dagger b_k=\sum_{k}\omega_k n_k\ .
\end{align}
Here, $b_k$ and $b_k^\dagger$ represent the annihilation and creation operators for the bosonic modes and $n_k$ is the number operator for mode $k$ with energy $\omega_k$ (we set $\hbar=1$ throughout). 
The qubit-cavity interaction Hamiltonian is 
\begin{align}\label{eq:first_HSB}
    H_{SB}=\sigma_+\otimes B + \sigma_-\otimes B^\dagger\ ,
\end{align}
where $B=\sum_k g_k b_k$. Here the $g_k$'s are coupling constants with dimensions of energy.  Introducing complex phases in the couplings can model chirality ~\cite{roccati2024}.

In the interaction picture, the interaction Hamiltonian $H_{SB}$ becomes: 
\bes
\label{eq:HSBt}
\begin{align}
    \tilde{H}_{SB}(t) &= \sigma_+(t)\otimes B(t) + \sigma_-(t)\otimes B^\dagger(t) \\
    \sigma_\pm(t) & =e^{\pm i\Omega_0 t}\sigma_\pm\ , \quad B(t)=\sum_k e^{-i\omega_k t}g_k b_k\ .
\end{align}
\ees

This model is not analytically solvable in general, but it is when we make the assumption that the cavity supports at most one photon. We thus consider the initial joint system-bath state to be: 
\begin{align}
\label{eq:initial_state}
\ket{\phi(0)}=c_0(0)\ket{\psi_0}+c_1(0)\ket{\psi_1}+\sum_k \mathfrak{c}_k(0)\ket{\varphi_k}\ ,
\end{align}
where
\begin{subequations}
\label{eq:substates}
\begin{align}
    \ket{\psi_0}&=\ket{0}_S\otimes\ket{v}_B \\
    \ket{\psi_1}&=\ket{1}_S\otimes\ket{v}_B \\
    \ket{\varphi_k}&=\ket{0}_S\otimes\ket{k}_B\ .
\end{align}
\end{subequations}
Here $\ket{v}_B$ denotes the vacuum state of the cavity, and $\ket{k}_B=b_k^\dagger\ket{v}_B=\ket{0_1,\cdots,0_{k-1},1_k,0_{k+1},\cdots}$ denotes the state with one photon in mode $k$. The subspace spanned by $\{\ket{\psi_0},\ket{\psi_1},\ket{\varphi_k}\}$ is referred to as the $1$-excitation subspace, and is conserved under the Hamiltonian in \cref{eq:general_H}. I.e., the joint state remains in the following form for all time $t$:
\begin{align}
\label{eq:state_t}    
\ket{\phi(t)}=c_0(t)\ket{\psi_0}+c_1(t)\ket{\psi_1}+\sum_k \mathfrak{c}_k(t)\ket{\varphi_k}\ ,
\end{align}
subject to the normalization condition:
\begin{align}
\label{eq:norm_cond} 
|c_0(t)|^2+|c_1(t)|^2+\sum_k |\mathfrak{c}_k(t)|^2 = 1 \ .
\end{align}
The problem remains solvable with a more general bath state assuming interaction with a continuous-mode laser field~\cite{Ahmadi:2024}.

We assume that initially there are no photons in the cavity \cite{Garraway:1997aa}, hence:
\begin{align}
    \mathfrak{c}_k(0)=0\ .
\end{align}

We introduce a spectral density $J(\omega)$ via: 
\begin{align}
\label{eq:integral_spectral}
    \sum_{ k} |g_k|^2 e^{-i\omega_k t} = \int_0^{\infty} d\omega J(\omega) e^{-i\omega t} \ .
\end{align}
The continuum of bath spectral modes is a necessary condition for irreversibility; a discrete spectrum necessarily results in recurrences.
 
To complete the model specification, we consider three different bath spectral densities: an impulse function centered at the cutoff frequency $\omega_c$, an Ohmic function with the same cutoff frequency, and a triangular spectral density with a sharp cutoff, namely: 
\bes
\label{eq:18}
\begin{align}
    J_1(\omega) &= |g|^2\delta(\omega-\omega_c) \\
    J_2(\omega) &= \eta \omega e^{-\omega/\omega_c} \\
    J_3(\omega) &= \eta \omega \Theta(\omega_c-\omega)\ ,
\end{align}
\ees
where the Heaviside function obeys $\Theta(x)=0$ for $x<0$ and $\Theta(x)=1$ for $x\ge 0$.
In the impulse spectral density $J_1(\omega)$, the bath has a singular response at the cutoff frequency, characterized by a Dirac delta function. The Ohmic spectral density $J_2(\omega)$ is ubiquitous in the study of the spin-boson problem~\cite{RevModPhys.59.1}. The dimensionless parameter $\eta$ in the Ohmic spectral density serves as a measure of the coupling strength between the bath and the system, while the ratio of the qubit frequency to the cutoff frequency measures the ratio of the photonic energy gap, which is the energy between the ground state and the excited state, and the number of frequency modes before reaching the cutoff frequency.
The triangular spectral density $J_3(\omega)$ is a sharp-cutoff approximation to the Ohmic spectral density, which we introduce to simplify analytical calculations. The case of a Lorentzian spectral density was studied in Ref.~\cite{Breuer:book}.

\subsection{Exact Solution}

Considering the entire qubit-cavity system as closed, its evolution is governed by the Schr\"odinger equation, which can be expressed as [see \cref{app:deriv_exact_sol}]: 
\begin{align}\label{eq:closed_evol}
    i\partial_t\ket{\phi(t)} &= \sum_k g_k \mathfrak{c}_k(t) e^{i(\Omega_0-\omega_k)t}\ket{\psi_1}\nonumber\\&+\sum_k g_k^* c_1(t) e^{-i(\Omega_0-\omega_k)t}\ket{\varphi_k}\ .
\end{align}
Multiplying this equation by $\bra{\psi_1}$ or $\bra{\varphi_k}$, we obtain the following set of differential equations for the amplitudes:
\begin{subequations}
\label{eq:der_coeff_ci}
\begin{align}
    \dot{c}_0(t)&=0\\
    \label{eq:der_coeff_ci-b}
    \dot{c}_1(t)&=-i\sum_k g_k \mathfrak{c}_k(t) e^{i(\Omega_0-\omega_k)t}\\
    \dot{\mathfrak{c}}_k(t)&=-i g_k^*c_1(t)e^{-i(\Omega_0-\omega_k)t}
\end{align}
\end{subequations}
Integrating, we arrive at:  
\bes
\label{eq:c1t_derivative}
\begin{align}
    c_0(t) &= c_0(0) \\
\label{eq:c1t_derivative-b}
   \mathfrak{c}_k(t)&=-i\int_0^t dt' g_k^* c_1(t') e^{-i(\Omega_0-\omega_k)t'}
\end{align}
\ees
Let us define the memory kernel $f(t)$ as the Fourier transform of the spectral density, shifted by the qubit frequency $\Omega_0$:
\begin{align}
\label{eq:general_kernel}
    f(t)=\int_0^\infty d\omega J(\omega) e^{i(\Omega_0-\omega)t}\ .
\end{align}
Substituting \cref{eq:c1t_derivative-b} into \cref{eq:der_coeff_ci-b}, we obtain:
\begin{align}
\label{eq:general_c1}
    \dot{c}_1(t)=-\int_0^t dt' f(t-t')c_1(t')\ , 
\end{align}
which can be solved via a Laplace transform since the RHS is a convolution. Denoting the Laplace transform $\Lap$ of a general function $g(t)$ by $\hat{g}(s)$, and recalling that $\Lap[\dot{g}(t)] = s\hat{g}(s) -g(0)$, we have: 
\begin{align}
\label{eq:laplace_c1}
\hat{c}_1(s)=\frac{c_1(0)}{s+ \hat{f}(s)}\ ,
\end{align}
and $c_1(t)$ is then found via the inverse Laplace transform of \cref{eq:laplace_c1}.

Next, in order to determine the interaction picture system state by tracing out the bath state, we can utilize \cref{eq:state_t} and find:
\begin{align}\label{eq:state_system}
    \tilde{\rho}(t) &= \Tr_B\left(\ketb{\phi(t)}{\phi(t)}\right)=\left(\begin{array}{cc}
        1-|c_1|^2 & c_0c_1^*(t) \\
        c_0^*c_1(t) & |c_1|^2
    \end{array}\right)\ , 
\end{align}
so that: 
\begin{align}\label{eq:der_rho}
    \dot{\tilde{\rho}}(t)=\left(\begin{array}{cc}
        -\partial_t |c_1|^2 & c_0\dot{c}_1^*(t) \\
       c_0^*\dot{c}_1(t)  & \partial_t|c_1|^2
    \end{array}\right)\ .
\end{align}
At this point, it is straightforward to verify that the dynamics are time-local,
\begin{align}\label{eq:Ks_gen}
    \dot{\tilde{\rho}} &= \mathcal{K}_S(t)\tilde{\rho} \ ,
\end{align}
with a generator given by: 
\begin{align}
\label{eq:ansatz}
    \mathcal{K}_S(t)\tilde{\rho}(t) &= -\frac{i}{2} \mathcal{S}(t)[\sigma_+\sigma_-,\tilde{\rho}(t)]\nonumber\\
    &+\gamma(t)\left(\sigma_-\tilde{\rho}(t)\sigma_+-\frac{1}{2}\{\sigma_+\sigma_-,\tilde{\rho}(t)\}\right)\ ,
\end{align}
provided we identify:
\bes
\label{eq:sandgamma}
\begin{align}
\label{eq:sandgamma-s}
    \mathcal{S}(t) &= -2\Im\left(\frac{\dot{c}_1(t)}{c_1(t)}\right) \\
\label{eq:sandgamma-g}
    \gamma(t) &= -2\Re\left(\frac{\dot{c}_1(t)}{c_1(t)}\right) 
\end{align}
\ees
The rate $\gamma(t)$ can be negative, corresponding to non-Markovian dynamics according to the CP non-divisibility criterion~\cite{rivas2010entanglement}. 

We focus on the excited state population and the coherence, which evolve in time according to:
\bes
\label{eq:rho-exact}
\begin{align}
\label{eq:rho11}
  \tilde{\rho}_{11}(t)&=|c_1(t)|^2=|c_1(0)|^2 \exp{-\int_0^{t}\gamma(t')dt'}\\
\label{eq:rho01}
  \tilde{\rho}_{01}(t)&= c_0{c}_1^*(t) = c_0 {c}_1^*(0) \exp{\frac{1}{2}\int_0^{t}\left(i\mathcal{S}(t')-\gamma(t')\right)dt'} \ .
\end{align}
\ees

Details of the derivation above can be found in \cref{app:deriv_exact_sol}.

We next discuss the solutions for the three spectral densities of \cref{eq:18}. In each case, we express the solution in terms of $c_1(t)$ or its Laplace transform.

\subsection{Exact solution for three different spectral densities}

\subsubsection{$J_1= |g|^2\delta(\omega-\omega_c)$}

As a toy example, we consider the impulse bath spectral density, which replaces the continuum of bath modes with a single mode. Consequently, we do not expect irreversibility and indeed, the solution is oscillatory. To demonstrate this, we circumvent the Laplace transform and instead use \cref{eq:general_kernel} to write:
\begin{align}\label{eq:kernel_J1}
    f(t) &= |g|^2e^{i(\Omega_0-\omega_c)t}\ .
\end{align}
As a result, \cref{eq:general_c1} becomes: 
\begin{align}
    \dot{c}_1(t)&=-|g|^2\int_0^tdt'e^{i(\Omega_0-\omega_c)(t-t')}c_1(t')\ .
\end{align}
Differentiating both sides yields:
\begin{align}
    &\ddot{c}_1(t) -i(\Omega_0-\omega_c)\dot{c}_1(t)+|g|^2c_1(t)= 0\ ,
\end{align}
whose solution is:
\begin{align}
\label{eq:exactJ1}   
\frac{c_1(t)}{c_1(0) } &= e^{i(\Omega_0-\omega_c)t/2}\left(\cos(\frac{t}{2}\delta)-i\frac{\Omega_0-\omega_c}{\delta} \sin (\frac{t}{2}\delta)\right),
\end{align}
where $\delta$ is a real number:
\begin{align}\label{eq:delta_J1}
\delta=\sqrt{(\Omega_0-\omega_c)^2+4g^2}\ .
\end{align}
Since the solution is perfectly periodic, any master equation approximation with a non-trivial dissipator term will deviate from this exact solution for sufficiently long times. Note that the excited state population $|c_1(t)|^2 \in |c_1(0)|^2[\frac{(\Omega_0-\omega_c)^2}{(\Omega_0-\omega_c)^2+4g^2},1]$. 

The Lamb shift and decay rate are now found from \cref{eq:sandgamma} to be:
\begin{subequations}
\label{eq:exact-S-gamma-J1}
\begin{align}
\label{eq:exact-S-J1}
\mathcal{S}(t) &= \left(1-\frac{(\Omega_0-\omega_c)^2}{\delta^2}\right)\frac{\frac{\Omega_0-\omega_c}{\delta}}{\cot^2\left(\frac{t\delta}{2}\right)+\frac{(\Omega_0-\omega_c)^2}{\delta^2}}\\
\label{eq:exact-gamma-J1}
\g(t) &= \left(1-\frac{(\Omega_0-\omega_c)^2}{\delta^2}\right)\frac{\cot\left(\frac{t\delta}{2}\right)}{\cot^2\left(\frac{t\delta}{2}\right)+\frac{(\Omega_0-\omega_c)^2}{\delta^2}}
\end{align}
\end{subequations}

\subsubsection{$J_2(\omega) = \eta \omega e^{-\omega/\omega_c}$}

Now we turn our attention to the Ohmic spectral density.
The memory kernel, as given in \cref{eq:general_kernel}, takes the form:
\begin{align}
\label{eq:kernel_Ohmic}
f(t)=\frac{\eta \omega_c^2 e^{i\Omega_0 t}}{(1+i\omega_c t)^2}\ .
\end{align}	
To obtain the Laplace transform analytically, we integrate by parts and obtain:
\begin{align}
\label{eq:laplace_Ohmic}
\hat{f}(s) = 
&\eta  \Big[ \left(s-i\Omega _0\right) e^{-\frac{\Omega _0+i s}{\omega _c}}
   \left( i\frac{\pi}{2} - \Ci\left(\frac{s-i \Omega _0}{\omega _c}\right)\right. \notag \\
   &\left. \quad -i \Si\left(\frac{s-i \Omega _0}{\omega _c}\right) \right)-i \omega _c\Big]\ ,
\end{align}	
where the sine and cosine integral functions are defined as:
\begin{subequations}
\begin{align}
\label{eq:Si}
\Si(z) &\equiv \int_0^z \frac{\sin(t)}{t} dt =\pi/2-\int_z^{\infty} \frac{\sin(t)}{t}dt\\
\label{eq:Ci}
\Ci(z) &\equiv -\int_z^{\infty} \frac{\cos(t)}{t} dt\\
\label{eq:Ei}
\Ei(z) &\equiv -\int_{-z}^{\infty} \frac{e^{-t}}{t} dt\ ,
\end{align}
\end{subequations}
where for future reference we have also defined the exponential integral function.

Due to the absence of analytic expressions for the Laplace transforms of the trigonometric integral functions, we use the numerical inverse Laplace transform of \cref{eq:laplace_c1} to obtain $c_1(t)$. The Lamb shift and decay rate are then calculated numerically using \cref{eq:sandgamma}.

\subsubsection{
    $J_3(\omega)=\eta \omega \Theta (\omega_c-\omega)$}

Finally, we compute the memory kernel for the triangular spectral density, for which we obtain:
\begin{align}
\label{eq:kernel_J3}
    f(t)=\int_{0}^{\omega_c}\eta \omega e^{it(\Omega_0-\omega)}=\eta e^{it(\Omega_0-\omega_c)} \frac{1-e^{i\omega_c t}+i\omega_c t}{t^2}\ .
\end{align}
The Laplace transform is:
\begin{align}
\hat{f}(s)=-i\eta\omega_c-\eta(s-i\Omega_0)\ln \left(\frac{s-i\Omega_0}{s-i\Omega_0+i\omega_c}\right)\ ,
\end{align}
and once more the (numerical) inverse Laplace transform of \cref{eq:laplace_c1} yields $c_1(t)$.
Finally, the Lamb shift and decay rate are again computed numerically using \cref{eq:sandgamma}.

\section{Approximation methods}
\label{sec:approximate}

In this section, we compute the excited state population predictions for the damped Jaynes-Cummings model of the previous section using three different Markovian master equations and using the TCL to second and fourth orders. In each case, we first provide a brief summary of the underlying theory of the corresponding master equation, both to assist the reader who may be unfamiliar with this theory and to establish our notation.

\subsection{Coarse-Grained Lindblad Equation (CG-LE)}
\label{sec:chap9}

We follow the derivation of Ref.~\cite{Lidar200135}. Let us choose a dimensionless, fixed and orthogonal system operator basis for $\mathcal{B}(\mathcal{H}_S)$ as $\{S_\alpha\}_{\alpha=0}^{M}$ with $S_0=I$ and $M=d^2-1$, where $d=\dim(\mathcal{H}_S)$ and 
\begin{align}
    \Tr(S_\alpha^\dagger S_\beta) =\frac{1}{N_\alpha}\delta_{\alpha\beta} \ ,
\end{align}
where $N_\alpha$ is a normalization factor. 
We can then always write the system-bath interaction Hamiltonian in the following form:
\begin{align}
\label{eq:formHSB}
    H_{SB} = \sum_\alpha g_\alpha S_\alpha \otimes B_\alpha\ ,
\end{align}
where $\{B_\alpha\}$ are dimensionless bath operators and $\{g_\a\}$ are the coupling coefficients. In the interaction picture, with $U_S(t) = e^{-iH_S t}$ and $U_B(t) = e^{-iH_B t}$, we obtain:
\bes
\label{eq:46}
\begin{align}
    \tilde{H}_{SB} &= \sum_\alpha g_\alpha S_\alpha(t) \otimes B_\alpha(t)\\
    \label{eq:pab}
    S_\alpha(t) &= U_S^\dagger(t) S_\alpha U_S(t)=\sum_{\beta} p_{\alpha\beta}(t) S_\beta \\ 
    \label{eq:qab}
    B_\alpha(t) &= U_B^\dagger(t) B_\alpha U_B(t)=\sum_{\beta} q_{\alpha\beta}(t) B_\beta \ ,
\end{align}
\ees
with initial conditions $p_{\alpha\beta}(0)=q_{\alpha\beta}(0)=\delta_{\alpha\beta}$. The dynamics of the total system+bath are described by \cref{eq:ip_rho_sb,eq:ip_u_sb}. The system state is obtained by tracing over the bath and, assuming the initial state is factorized [$\r_{SB}(0) = \r(0)\ox\r_B(0)$], can be represented as a completely positive quantum dynamical map:
\begin{align}
\label{eq:tilderho}
    \tilde{\rho}(t) = \Tr_B[\tilde{\rho}_{SB}(t)]=\sum_{i=0}^{M} \tilde{K}_i(t)\rho(0) \tilde{K}_i^\dagger(t)
\end{align}
where $\{\tilde{K}_i\}$ are Kraus operators in the interaction picture, which are defined as:
\begin{align}
    \tilde{K}_{i=\{\mu\nu\}}(t) = \sqrt{\lambda_\mu}\bra{\nu}\tilde{U}(t)\ket{\mu}\ ,
\end{align}
where the initial bath state is spectrally decomposed as $\r_B(0) = \sum_\mu \lambda_\mu \ketb{\mu}{\mu}$.
Using a standard Dyson series expansion of $\tilde{U}(t)$ similar to \cref{eq:Liouville-exp}, we can write: 
\bes
\label{eq:osr_expansion}
\begin{align}
 \label{eq:osr_expansion-a}
   \tilde{K}_i(t) &=\sqrt{\lambda_\mu} \delta_{\mu\nu} I + \sum_{n=1}^\infty K_i^{(n)}(t)\\
\label{eq:osr_expansion_basis}
    &=\sum_{\a=0}^M b_{i\a}(t)S_\a  = b_{i0}I + \sum_{\a=1}^M \sum_{n=1}^\infty b^{(n)}_{i\a}(t)S_\a \ ,
\end{align}
\ees
where $K_i^{(n)}(t)$ is the $n$th order term in the Dyson series, and the second line is an expansion in the system operator basis.
In the weak-coupling limit ($\max_\a g_\alpha t\ll 1$), the higher-order terms in the expansion of $K_i^{(n)}(t)$ become negligible, i.e.,  $\|K_i^{(n+1)}\|\sim g_\alpha t \|K_i^{(n)}\|$. Therefore, we can approximate the exact Kraus operators by truncating the expansion to first order. This yields:
\bes
\begin{align}
    \tilde{K}_i^{(1)}(t) &= -i\sqrt{\lambda_\mu} \bra{\nu}\int_0^t dt_1\tilde{H}_{SB}(t_1)\ket{\mu} \\
    &= -i\sqrt{\lambda_\mu}\sum_\a g_\a \int_0^t dt_1 S_\a(t_1)\bra{\nu}B_\a(t_1)\ket{\mu} \\
    &= -it\sqrt{\lambda_\mu} \sum_{\alpha\beta\gamma}g_\a S_\b \bra{\nu} B_\gamma \ket{\mu} \Gamma_\a^{\b\gamma}(t)\ ,
    \label{eq:49c}
\end{align}
\ees
where:
\begin{align}
\label{eq:Gamma_CG}
    \Gamma_\alpha^{\beta\gamma}(t)\equiv \frac{1}{t} \int_0^t dt_1 p_{\alpha\beta}(t_1) q_{\alpha\gamma}(t_1)\ .
\end{align}
Then, to match \cref{eq:49c,eq:osr_expansion_basis}, we have:
\begin{align}\label{eq:cg_b_coefficients}
    b^{(1)}_{i\alpha}(t)=-it\sqrt{\lambda_\mu}\sum_{\alpha'\alpha''} g_{\alpha'}\bra{\nu} B_{\alpha''}\ket{\mu} \Gamma^{\alpha\alpha''}_{\alpha'}(t)\ ,
\end{align}
and \cref{eq:osr_expansion} implies that $b_{i0}=\sqrt{\lambda_\mu} \delta_{\mu\nu}$.
Next, we can construct the process matrix $\chi(t)$ (closely related to the Choi matrix), where 
\beq
\chi_{\alpha\beta}(t)=\sum_{i=\mu\nu}b_{i\alpha}(t)b^*_{i\beta}(t)\ ,
\eeq 
and truncate it to lowest order ($n\le 1$) in the Dyson expansion as follows: 
\begin{align}
    \chi_{00}(t)&=\sum_\mu \lambda_\mu =1 \\ 
    \label{eq:chi_alpha_0}
    \chi^{(1)}_{\alpha0}(t) &= -it\sum_{\alpha'\alpha''} g_{\alpha'} \expval{B_{\alpha''}}_B \Gamma_{\alpha'}^{\alpha\alpha''}(t),~~\alpha\geq 1\\ 
    \label{eq:chi_alpha_beta}
    \chi^{(1)}_{\alpha\beta}(t)&=t^2\sum_{\alpha'\alpha''\beta'\beta''}g_{\alpha'}g_{\beta'}^* \expval{B^\dagger_{\beta''}B_{\alpha''}}_B \times\nonumber\\
    &\Gamma^{\alpha\alpha''}_{\alpha'} (t) \left(\Gamma^{\beta\beta''}_{\beta'}(t)\right)^*,~~\alpha,\beta\geq 1 \ ,
\end{align}
where $\expval{X}_B \equiv \Tr(\r_B X)$. Let us also define 
\beq
\expval{X}_j \equiv \frac{1}{\tau}\int_{j\tau}^{(j+1)\tau} X(t) dt\ , 
\eeq
where we call $\tau$ the \emph{coarse-graining timescale}. Note that $\chi_{\alpha\beta}(0) = \d_{\a0}\d_{\b0}$.
Then: 
\begin{align}
\label{eq:CG9_rate}
    a_{\alpha\beta}\equiv\expval{\dot{\chi}^{(1)}_{\alpha\beta}}_0=\frac{1}{\tau} (\chi^{(1)}_{\alpha\beta}(\tau)-\chi_{\alpha\beta}(0))=\frac{\chi^{(1)}_{\alpha\beta}(\tau)}{\tau}
\end{align}
unless $\alpha=\beta=0$, in which case we have $\expval{\dot{\chi}_{00}}_0=0$. 

It can be shown~\cite{Lidar200135} (see also \cite{PhysRevA.60.1944}) that starting from \cref{eq:tilderho}, substituting the various expansions above, rearranging terms, and assuming that \cref{eq:CG9_rate} can be extended to any interval $[t,t+\tau]$ (essentially an assumption of Markovianity), that one arrives at the Lindblad equation in the interaction picture:
\begin{align}
    \dot{\tilde{\rho}}(t) &= -i[H_{LS},\tilde{\rho}(t)]\nonumber\\
    \label{eq:60}
    &+\sum_{\alpha,\beta=1}^M a_{\alpha\beta} (S_\alpha\tilde{\rho}(t)S_\beta^\dagger-\frac{1}{2}\{S_\beta^\dagger S_\alpha,\tilde{\rho}(t)\}) \ ,
\end{align}
where the Lamb shift is given by:
\bes
\begin{align}
    H_{LS} & = \frac{i}{2}\sum_{\alpha} \expval{\dot{\chi}_{\alpha 0}}S_\alpha - \expval{\dot{\chi}_{\alpha 0}}^* S_\alpha^\dagger  \\
    &= \frac{1}{2} \sum_{\alpha} \phi_\alpha S_\alpha + \phi_\alpha^* S_\alpha^\dagger
\end{align}
\ees
with
\begin{align}\label{eq:phi_alpha}
    \phi_\alpha \equiv \sum_{\alpha'\alpha''} g_{\alpha'}\expval{B_{\alpha''}}_B \Gamma^{\alpha\alpha''}_{\alpha'} (\tau)\ ,
\end{align}
and the decoherence rates are:
\begin{align}
    a_{\alpha\beta} = \tau \sum_{\alpha'\alpha''\beta'\beta''} g_{\alpha'}g_{\beta'}^* \expval{B^\dagger_{\beta''}B_{\alpha''}}_B \Gamma_{\alpha'}^{\alpha\alpha''}(\tau)\Gamma_{\beta'}^{\beta\beta''}(\tau)^*\ .
\end{align}
The choice of the \emph{coarse-graining timescale} $\tau$ is crucial. It can be understood as a free optimization parameter, constrained by the inequality 
\beq\label{eq:CG_cond}
\tau_S \ll \tau \ll 1/\o_c\ ,
\eeq 
where $\tau_S$ is the timescale over which $\tilde{\r}(t)$ changes, which arises from the replacement of 
\beq
\label{eq:65}
\langle\dot{\tilde{\rho}}(t)\rangle_0 \equiv \frac{\tilde{\rho}(\tau)-{\rho}(0)}{\tau}
\eeq
by $\dot{\tilde{\rho}}(t)$ in arriving at \cref{eq:60}.

For the spin-boson model \cref{eq:first_HSB}, we have $S_+=\sigma_+$, $S_-=\sigma_-$, $B_k=b_k$ or $b_k^\dagger$. The system-bath interaction Hamiltonian in the interaction picture is given by \cref{eq:HSBt}. 
From \cref{eq:pab,eq:qab}, we obtain: 
\bes
\label{eq:pabqab2}
\begin{align}
    p^{\pm\pm}(t)&=e^{\pm i\Omega_0 t} \ ,  \quad p^{\pm\mp}(t)=0\\
    q^{\pm\pm}_{kk''}(t) &= \delta_{kk''}e^{\mp i\omega_k t} \ ,  \quad q^{\pm\mp}_{kk''}(t)=0  \ ,
\end{align}
\ees
where the $+$ or $-$ superscripts indicate that the corresponding bath operator is $b_k$ or $b_k^\dagger$, respectively.

Assuming that the initial state of the bath (cavity) is the zero-temperature vacuum state $\rho_B(0)=\ketb{v}{v}$, we obtain the standard bosonic expectation values:
\begin{align}\label{eq:useful_b}
    \langle{b_k^\dagger b_l}\rangle_B&=\langle{b_k^\dagger}\rangle_B=\langle{b_k}\rangle_B=\langle{b_k^\dagger b_l^\dagger}\rangle_B=\langle{b_k b_l}\rangle_B = 0 \nonumber\\
    \langle b_k b_l^\dagger\rangle_B&=\delta_{kl}.
\end{align}
Then, by utilizing \cref{eq:Gamma_CG}, 
we obtain a slightly modified expression for $\Gamma$ up to first order:
\begin{align}
    \Gamma_{\alpha,(\alpha' k)}^{\beta,(\beta' k'')}(t) =\frac{1}{t} \int_0^t dt_1 p^{\alpha\beta}(t_1) q^{\alpha'\beta'}_{kk''}(t_1)\ ,
\end{align}
where $\alpha,\beta,\alpha',\beta'\in\{+,-\}$. 
Using \cref{eq:pabqab2}, we have the following non-zero $\Gamma$'s:
\bes
\label{eq:Gamma_CG9}
\begin{align}
    \Gamma_{+,(+ k)}^{+,(+ k)}(t) &= \frac{e^{i(\Omega_0-\omega_k) t}-1}{i(\Omega_0-\omega_k)t}\\ 
    \Gamma_{-,(- k)}^{-,(- k)}(t) &= \frac{e^{-i(\Omega_0-\omega_k) t}-1}{-i(\Omega_0-\omega_k)t} \ .
\end{align}
\ees
We also have a slightly modified expression for $b$ by using \cref{eq:cg_b_coefficients}:
\begin{align}\label{eq:cg_b_coeff_new}
    b_{\mu\nu,\alpha} &= -it\sqrt{\lambda_\mu}\sum_{(\alpha' k')} g^{\alpha'}_{k'}\bra{\mu}B^{\alpha'}_{k'}\ket{\nu}\Gamma^{\alpha, (\alpha'k')}_{\alpha,(\alpha'k')}\ ,
\end{align}
where $B^\alpha_k$ represents $b_k$ when $\alpha=+$ and $b_k^\dagger$ when $\alpha=-$.
The Lamb shift rates are given by \cref{eq:phi_alpha}, and \emph{vanish}:
\begin{align}
\label{eq:LS=0}
    \phi_\alpha = \sum_{(\alpha' k')}g^{\alpha'}_{k'} \expval{B^{\alpha'}_{k'}}_B \Gamma^{\alpha,(\alpha' k')}_{\alpha,(\alpha' k')}(\tau) = 0\ ,
\end{align}
since the expectation values of creation and annihilation operators between vacuum states vanish, as indicated by \cref{eq:useful_b}. This result will be seen to undermine the quality of the CG-LE and C-LE when we perform a comparison with the exact results in \cref{Analysis}.

The decoherence rates are: 
\begin{align}
    a_{\alpha\beta}(\tau) &= \tau \sum_{(\alpha' k')(\beta' l')} g^{\alpha'}_{k'}g_{l'}^{-\beta'} \expval{B_{l'}^{-\beta'}B^{\alpha'}_{k'}}_B \nonumber\\
    &\qquad \times \Gamma_{\alpha,(\alpha' k')}^{\alpha,(\alpha' k')}(\tau) \Gamma_{\beta,(\beta' l')}^{\beta, (\beta' l')}(\tau)^* \ .
\end{align}
Using \cref{eq:useful_b,eq:Gamma_CG9}, we obtain:
\begin{align}
    a_{++}(\tau)&=\sum_{k} |g_{k}|^2 \tau 
\sinc^2\left(\frac{(\Omega_0+\omega_{k})\tau}{2}\right)\ ,\\
a_{--}(\tau)&=\sum_{k} |g_{k}|^2 \tau 
\sinc^2\left(\frac{(\Omega_0-\omega_{k})\tau}{2}\right)\ ,
\end{align}
\begin{align}
    \g(\tau) &\equiv a_{--}(\tau)=\sum_{k} |g_{k}|^2 \tau 
\sinc^2\left(\frac{(\Omega_0-\omega_{k})\tau}{2}\right)\ ,
\end{align}
where $\sinc(x) \equiv \sin(x)/x$. Introducing the spectral density
\beq
J(\o) = \sum_k |g_k|^2 \d(\o-\o_k)\ ,
\eeq
we can write this as
\beq
\label{eq:gamma-CG-LE}
    \g(\tau)= \int_0^\infty d\omega J(\omega)\tau \sinc^2\left(\frac{(\Omega_0-\omega)\tau}{2}\right)\ .
\eeq
Let us now define
\beq
\bar{\d}(x,y) \equiv \frac{1}{2\pi}y \sinc^2\left(\frac{x y}{2}\right)\ , \quad y\ge 0  \ .
\eeq
This function behaves similarly to the Dirac-$\delta$ function:
\bes
\begin{align}
& \int_{-\infty}^\infty \bar{\d}(x, y)dx = 1\\
\label{eq:bdelta-delta}
& \lim_{y\to\infty}\bar{\d}(x,y) = \d(x) \ ,
\end{align}
\ees
i.e., it is sharply peaked at $x=0$, and the peak becomes sharper as $y$ grows. The peak width is $\sim 1/y$. 
We can thus also write
\beq
\label{eq:gamma-CG-LE-delta-bar}
    \g(\tau)= 2\pi \int_0^\infty d\omega J(\omega)\bar{\d}\left(\Omega_0-\omega,\tau\right)\ .
\eeq
We will show below that this representation allows us to express the RWA-LE as the $\tau\to\infty$ limit of the CG-LE and C-LE results, as expected on general grounds~\cite{Majenz:2013qw}.

Finally, we obtain the interaction picture Lindblad equation as:
\begin{align}
\label{eq:LE-IP}
    \dot{\tilde{\rho}}(t) &= \g(\tau) \left(\sigma_-\tilde{\rho}(t)\sigma_+-\frac{1}{2}\{\sigma_+\sigma_-,\tilde{\rho}(t)\}\right)\ .
\end{align}
Taking matrix elements, we find that the populations and coherences are decoupled. Solving for the excited state population and coherence, we obtain, respectively:
\bes
\label{eq:r-CG-LE}
\begin{align}
\label{eq:r11-CG-LE}
    \tilde{\rho}_{11}(t) &= \rho_{11}(0)e^{-\g(\tau) t}\\
\label{eq:r01-CG-LE}
    \tilde{\rho}_{01}(t) &= \rho_{01}(0)e^{-\frac{1}{2}\g(\tau) t}\ ,
\end{align}
\ees
which is to be contrasted with the exact solution given in \cref{eq:rho-exact}.
The coarse-graining time $\tau$ can be chosen to optimize the agreement with the exact solution.

\subsubsection{$J_1(\omega)=|g|^2\delta(\omega-\omega_c)$}
For the impulse spectral density we find, using \cref{eq:gamma-CG-LE-delta-bar}:
\bes
\label{eq:gamma-J1-CG-LE}
\begin{align}
    \g(\tau) &=  |g|^2 \tau \sinc^2\left(\frac{(\Omega_0-\omega_c)\tau}{2}\right) \\
\label{eq:gamma-J1-CG-LE-delta-bar}
    &= 2\pi |g|^2 \bar{\d}\left(\Omega_0-\omega,\tau\right)\ .
\end{align}
\ees

\subsubsection{$J_2(\omega)=\eta\omega e^{-\omega/\omega_c}$}
For the Ohmic spectral density we find, using \cref{eq:gamma-CG-LE-delta-bar}:

\bes
\label{eq:gamma-J2-CG-LE}
\begin{align}
    \g(\tau) &=  \eta \int_0^\infty \omega e^{-\omega/\omega_c}\tau \sinc^2\left(\frac{(\Omega_0-\omega)\tau}{2}\right)d\omega  \\
\label{eq:gamma-J2-CG-LE-delta-bar}
    &= 2\pi\eta \int_0^\infty \omega e^{-\omega/\omega_c} \bar{\d}\left(\Omega_0-\omega,\tau\right)d\omega \\
\label{eq:gamma-J2-CG-LE-3}
&=
\frac{\eta}{\tau} e^{-\frac{\Omega_0}{\omega_c}}\left[\left(1-\frac{\Omega_0}{\omega_c}-i\Omega_0 \tau\right)\Ei\left(\frac{\Omega_0}{\omega_c}+i\Omega_0 \tau\right)+\right.\nonumber\\
&\left(1-\frac{\Omega_0}{\omega_c}+i\Omega_0 \tau\right)\Ei\left(\frac{\Omega_0}{\omega_c}-i\Omega_0 \tau\right)\\ 
&\left.+
2\left(\frac{\Omega_0}{\omega_c}-1\right)\Ei\left(\frac{\Omega_0}{\omega_c}\right)\right]+\frac{2\eta}{\tau} (1-\cos \Omega_0 \tau) \ , \nonumber
\end{align}
\ees
where the last equality is derived in \cref{app:CG-LE-integrals-J2}.

\subsubsection{
    $J_3(\omega)=\eta \omega \Theta (\omega_c-\omega)$}
For the triangular spectral density we find, using \cref{eq:gamma-CG-LE-delta-bar}:

\bes
\label{eq:gamma-J3-CG-LE}
\begin{align}
    \g(\tau)&=
    \eta \int_0^{\o_c} \omega \tau \sinc^2\left(\frac{(\Omega_0-\omega)\tau}{2}\right)d\omega\\
\label{eq:gamma-J3-CG-LE-delta-bar}
     &= 2\pi\eta \int_0^{\o_c} \omega  \bar{\d}\left(\Omega_0-\omega,\tau\right)d\omega \\
\label{eq:gamma-J3-CG-LE-3}
&=
\frac{2 \eta}{\tau}  \Bigg(\ln \left(\frac{\omega _c}{\Omega
   _0}-1\right)+\Ci\left(\tau  \Omega _0\right)-\Ci\left(\tau  \left(\omega _c-\Omega _0\right)\right)\notag\\
   &+\frac{\omega _c \left(\cos \left(\tau  \left(\omega _c-\Omega
   _0\right)\right)-1\right)}{\omega _c-\Omega _0}-\cos \left(\tau 
   \left(\omega _c-\Omega _0\right)\right)\notag \\
   &+\cos \left(\tau  \Omega _0\right)\Bigg)+2
   \eta  \Omega _0 \left(\Si\left(\tau  \left(\omega _c-\Omega _0\right)\right)+\Si\left(\tau 
   \Omega _0\right)\right) \ ,
\end{align}
\ees
where the last equality is derived in \cref{app:CG-LE-integrals-J3}.

\subsection{Cumulant Lindblad Equation C-LE}
\label{sec:CGLch10}

This section briefly reviews an alternative derivation of the Lindblad equation, based on a cumulant expansion~\cite{Majenz:2013qw}. Similarly to the CG-LE, the C-LE approach also uses a coarse-graining time scale that can be optimized to approximate the exact result. Despite using a rather different approach to deriving the Lindblad equation, we will show that for the problem we study in this work, the C-LE ultimately results in identical expressions for the master equation and its parameters (and hence also the solution, of course) as the CG-LE.

We start by writing the system-bath interaction Hamiltonian of \cref{eq:general_H} as: 
	\begin{align}
	\label{eq:HSB_AB}
		H_{SB}=\lambda\sum_\alpha S_\alpha\otimes B_\alpha
	\end{align} 
where $S_\alpha$ and $B_\alpha$ are the system and bath operators, respectively (not necessarily Hermitian), and $\lambda$ is a dimensionless parameter to be used below for a series expansion, which we eventually set equal to $1$. Note that unlike \cref{eq:formHSB}, the $S_\alpha$ are now not a basis, and the $B_\alpha$ have dimensions of energy since they include the coupling constants $g_\a$.
Assuming a factorized initial condition $\rho_{SB}(0)=\rho(0)\otimes \rho_B(0)$, 
we associate a CPTP map $\Lambda_{\lambda}$ to the reduced density matrix of \cref{eq:ip_rho_s}:
\begin{align}
\label{eq:CPmap_CG}
\tilde{\rho}(t)=\Lambda_{\lambda}(t)\rho(0)\ .
\end{align}
This CPTP map can be related to \cref{eq:Liouville-exp} by introducing superoperators $K^{(n)}$, which collect terms with matching power of $\lambda$: 
\begin{align}
\Lambda_{\lambda}(t)=\exp \left(\sum_{n=1}^{\infty}\lambda^n K^{(n)}(t)\right)
\end{align}
This is known as the \textit{cumulant expansion}. The first-order term is then:
\begin{align}
K^{(1)}(t)\rho(0)=-i\int_0^tds \Tr_B\left([\tilde{H}(s),\rho_{SB}(0)]\right)\ ,
\end{align}
which can be eliminated by shifting the bath operators $B_{\alpha}$, assuming stationarity, i.e., $[H_B,\rho_B(0)]=0$ (see below and \cref{app:shift}). Moving on to the second order in $\lambda$, we have:
\begin{align}
\label{eq:cumulant_two}
K^{(2)}(t)&\rho(0)=-\int_0^tds \int_0^s ds' \Tr_B[\tilde{H}(s),[\tilde{H}(s'),\rho_{SB}(0)]]\ ,
\end{align}
where the double commutator can be rearranged using:
\begin{align}
\label{eq:second_commutator}
    \Tr_B&[ S_{\alpha}(s)\otimes B_{\alpha}(s),[S_{\beta}(s')\otimes B_{\beta}(s'), \rho(0)\otimes \rho_B(0)]]\nonumber\\
    =&[S_{\alpha}(s),S_{\beta}(s')\rho(0)]\Tr(B_{\alpha}(s)B_{\beta}(s') \rho_B)-\nonumber\\
    &[S_{\alpha}(s),\rho(0)S_{\beta}(s')]\Tr(B_{\beta}(s')B_{\alpha}(s) \rho_B)
\end{align}
Introducing the bath correlation function
\begin{align}\label{eq:Corr_func}
		\mathcal{B}_{\alpha\beta}(s,s')\equiv \Tr[B_{\alpha}(s) B_{\beta}(s') \rho_B ]\ ,
	\end{align}
we have:
\begin{align}
\label{conj_Bab}
    \mathcal{B}_{\alpha\beta}(s,s')=\mathcal{B}_{\alpha\beta}^*(s',s)\ .
\end{align}

To obtain a reduced form of the second-order cumulant in \cref{eq:cumulant_two}, it is useful to define a new variable that includes the double integration of the bath correlation function as follows:
\begin{align}
    \mathcal{B}_{\alpha\beta\omega}(t)\equiv \int_0^t ds \int_0^s ds' e^{i\omega(s-s')}\mathcal{B}_{\alpha\beta}(s,s') 
    \label{eq:93}
\end{align}
We introduce two more variables that will be associated with the Lamb shift and the decoherence rate:
\begin{align}
\label{eq:Q-C-LE}
   &Q_{\alpha\beta\omega}(t)\equiv\frac{1}{2i}\left(\mathcal{B}_{\alpha\beta\omega}(t)-\mathcal{B}^*_{\alpha\beta\omega}(t)\right)=\Im (\mathcal{B}_{\alpha\beta\omega}(t))\ ,
\end{align}
and
\bes
\begin{align}
   b_{\alpha\beta\omega}(t)&\equiv\int_0^tds\int_0^tds'e^{i\omega(s-s')}\mathcal{B}_{\alpha\beta}(s,s')\\
\label{eq:b_CG10}
   &=\mathcal{B}_{\alpha\beta\omega}(t)+\mathcal{B}^*_{\alpha\beta\omega}(t)\\
\label{eq:b-C-LE}
   &=2\Re (\mathcal{B}_{\alpha\beta\omega}(t))\ ,
\end{align}
\ees
where the equality in the second line is shown in \cref{app:C}.

Explicitly, in this section, $\alpha \in \{+,-\}$, and as in \cref{eq:HSBt}, in the interaction picture  the system operators are $S_{\pm}(t)=\sigma_{\pm} e^{\pm i\Omega_0 t}$ and the bath operators are 
$B_{+}(t)=\sum_k g_k e^{-i \omega_k t} b_k,~B_{-}(t)=\sum_k g_k^* e^{i \omega_k t} b_k^{\dagger}$.
The initial bath state $\rho_B=\ketb{v}{v}$ and the commutation rules \cref{eq:useful_b} give us just one non-zero bath correlation function:
\bes
 \begin{align}
 \mathcal{B}_{+-}(s,s')&=\Tr[\rho_B B_{+}(s)B_{-}(s')]\\
\label{eq:Bplusminus2}
 &=\sum_{l,l'}g_lg_{l'}^*e^{-i(\omega_l s -\omega_{l'}s')} \Tr[\rho_B(0) b_l b_{l'}^{\dagger}]\\
\label{eq:Bminusplus}
     \mathcal{B}_{-+}(s,s')&=\mathcal{B}_{++}(s,s')=\mathcal{B}_{--}(s,s')=0\ .
 \end{align}
 \ees
The double commutator \cref{eq:second_commutator} can now be simplified as follows:
\begin{align}
    \sum_{\alpha,\beta}\Tr_B&[ S_{\alpha}(s)\otimes B_{\alpha}(s),[S_{\beta}(s')\otimes B_{\beta}(s'), \rho(0)\otimes \rho_B(0)]]\nonumber\\
    &=e^{i\Omega_0(s-s')}[\sigma_+,\sigma_-\rho(0)]\mathcal{B}_{+-}(s,s')\nonumber\\
    &\quad -e^{-i\Omega_0(s-s')}[\sigma_-,\rho(0)\sigma_+]\mathcal{B}_{+-}(s',s)\ .
\end{align}

Using \cref{conj_Bab}, we have:
\bes
\begin{align}
&K^{(2)}(t)\rho(0) \\
&\quad =-\mathcal{B}_{+-,\Omega_0}(t)[\sigma_+,\sigma_-\rho(0)]+\mathcal{B}_{+-,\Omega_0}^*(t)[\sigma_-,\rho(0)\sigma_{+}] \notag\\
&\quad = -i \Im(\mathcal{B}_{+-,\Omega_0}(t)) [\sigma_+\sigma_-,\rho(0)]\\
&\qquad+2\Re(\mathcal{B}_{+-,\Omega_0}(t))\left(\sigma_-\rho(0)\sigma_+-\frac{1}{2}\{\sigma_+\sigma_-,\rho(0)\}\right)\ .\notag
\end{align}
\ees

Hence, the second-order cumulant takes the form:
\begin{align}
&K^{(2)}(t)\rho(0)= -i Q_{+-,\Omega_0}(t) [\sigma_+\sigma_-,\rho(0)]\nonumber\\
&\quad +b_{+-,\Omega_0}(t)\left(\sigma_-\rho(0)\sigma_+-\frac{1}{2}\{\sigma_+\sigma_-,\rho(0)\}\right)\ .
\end{align}

Hence, the state in the interaction picture after the CP map in \cref{eq:CPmap_CG} using a truncation up to the second order in the cumulant expansion, is:
\bes
\begin{align}
    \tilde{\rho}(t)&=\Lambda_{\lambda}(t)\rho(0)\approx\exp \left( \lambda^2 K^{(2)}(t)\right)\rho(0)\\
    \label{eq:100b}
    &= {\rho}(0) -i \lambda^2 Q_{+-,\Omega_0}(t) [\sigma_+\sigma_-,\rho(t)]\\
    &\quad +\lambda^2b_{+-,\Omega_0}(t)\left(\sigma_-\rho(t)\sigma_+-\frac{1}{2}\{\sigma_+\sigma_-,\rho(t)\}\right)\ . \nonumber
\end{align}
\ees

Now we use the coarse-graining method by averaging over the coarse-graining timescale $\tau$ as in \cref{eq:CG9_rate}:
\bes
\begin{align}
\langle\dot{b}_{+-,\Omega_0}(t)\rangle_0 &= \frac{b_{+-,\Omega_0}(\tau)}{\tau} \\
\langle\dot{Q}_{+-,\Omega_0}(t)\rangle_0 &= \frac{Q_{+-,\Omega_0}(\tau)}{\tau} \ ,
\end{align}
\ees
which, when applied to \cref{eq:100b}, yields:
\begin{align}
\label{eq:102}
&\langle\dot{\tilde{\rho}}(t)\rangle_0=-i \lambda^2 \langle \dot{Q}_{+-,\Omega_0}(t)\rangle_0 [\sigma_+\sigma_-,\rho(0)]\\
&+\lambda^2\langle\dot{b}_{+-,\Omega_0}(t)\rangle_0\left(\sigma_-\rho(0)\sigma_+-\frac{1}{2}\{\sigma_+\sigma_-,\rho(0)\}\right)\ ,\nonumber
\end{align}
where we also used \cref{eq:65}.
Let us now define  
\bes
\begin{align}
\mathcal{S}(\tau)&\equiv 2 \langle\dot{Q}_{+-,\Omega_0}(t)\rangle_0 = \frac{2\Im (\mathcal{B}_{+-,\Omega_0}(\tau))}{\tau}\\
\gamma(\tau)&\equiv  \langle\dot{b}_{+-,\Omega_0}(t)\rangle_0 = \frac{2\Re (\mathcal{B}_{+-,\Omega_0}(\tau))}{\tau}\ ,
\end{align}
\ees
where we used \cref{eq:Q-C-LE,eq:b-C-LE}.

For a general $\rho_B(0)$ obtained by tracing out the system from \cref{eq:state_t} we find that $[\r_B(0),H_B]\ne 0$ (as shown in \cref{app:Dawei}), which means that the bath correlation function $\mathcal{B}_{\alpha \beta}(s,s')$ is not stationary. However, it is for the vacuum bath state $\rho_B(0)=\ketb{v}{v}$ which we assume to be the case throughout, so we can write
\beq
 \label{eq:Bplusminus}
\mathcal{B}_{+-}(s,s') = \mathcal{B}_{+-}(s-s') = \int_{0}^{\infty}d\omega J(\omega)e^{-i\omega(s-s')}\ .
\eeq
We then obtain, using \cref{eq:Q-C-LE}:
\bes
\label{eq:C-LE_LS=0}
\begin{align}
 S(\tau) &= \frac{1}{\tau}\Im \int_0^\tau ds \int_0^s ds' e^{i\omega(s-s')} \int_{0}^{\infty}d\o J(\o) e^{-i\omega(s-s')} \\
 &= \frac{\tau}{2}\Im \int_{0}^{\infty}d\o J(\o) = 0\ ,
 \end{align}
 \ees
since the spectral density is real. I.e., just like in the CG-LE case [\cref{eq:LS=0}], \emph{the Lamb shift vanishes}. 

For the decay rate we now have, using \cref{eq:b_CG10}:
\bes
 \begin{align}
\g(\tau)&=\frac{1}{\tau}\int_{0}^{\infty}d\o J(\o)\int_0^{\tau}ds\int_0^{\tau}ds'e^{i(\Omega_0-\o)(s-s')}\\
&=\int_0^{\infty} d\o J(\o) \tau \sinc^2\left(\frac{(\Omega_0-\omega)\tau}{2}\right)\ ,
\label{eq:gamma-C-LE}
 \end{align}
 \ees
 which is identical to the CG-LE result, \cref{eq:gamma-CG-LE}.
 
Moreover, similar to how we arrived at \cref{eq:60}, assuming Markovianity in the sense that \cref{eq:102} can be extended to any interval $[t,t+\tau]$ we again arrive at the Lindblad equation in the interaction picture, after replacing $\langle\dot{\tilde{\rho}}(t)\rangle_0 \mapsto \dot{\tilde{\rho}}(t)$, and setting $\lambda=1$. The form of this Lindblad equation is identical to \cref{eq:LE-IP}. In particular, both the excited state population and the coherence are the same as in \cref{eq:r-CG-LE}. Thus, the end results of the C-LE and CG-LE are identical for the model considered in this work.

\subsection{Rotating Wave Approximation Lindblad Equation (RWA-LE)}
\label{sec:RWA-LE}

The rotating wave approximation (RWA) drops the non-secular (off-diagonal) frequency terms which appear in the C-LE [see, e.g., \cref{eq:Bplusminus2}]. This approximation is based on the idea that the terms with $\o\neq \o'$ are rapidly oscillating if $t\gg |\o-\o'|^{-1}$, which thus (roughly) average to zero. Since we will assume that $t\gg \tau_B$, where $\tau_B$ is the bath correlation time (the time over which the bath correlation function decays), the former assumption is consistent provided we also assume that the Bohr frequency differences satisfy $\min_{\o\neq\o'} |\o-\o'| > 1/\tau_B$.
Combining this with the weak coupling assumption, we obtain:
\begin{align}
\label{eq:RWA_assum}
g\ll \frac{1}{\tau_B}<\min_{\omega\neq \omega'}|\omega-\omega'|\ .
\end{align}

By considering the weak coupling limit, taking the bath correlation timescale as the inverse of the cutoff frequency, and considering the Bohr frequencies $\{0,\pm \Omega_0\}$, \cref{eq:RWA_assum} becomes:
\begin{align}
\label{eq:RWA_ass_Ohmic}
\eta \ll 1<\Omega_0/\omega_c\ .
\end{align}

Furthermore, the Born approximation states that for a sufficiently large bath, the composite state factorizes: 
\begin{align}
\tilde{\rho}_{SB}(t)\approx \tilde{\rho}(t)\otimes \rho_B.
\end{align}
Thus, up to second order in the Dyson series, the system state evolves according to:
\bes
\label{eq:2ordcumm}
\begin{align}
\dot{\tilde{\rho}}&=-\Tr_B[\tilde{H}(t),\int_0^t d\tau [\tilde{H}(t-\tau),\tilde{\rho}_{SB}(t-\tau)]]\\
&=-\sum_{\alpha,\beta}\Tr_B[A_{\alpha}(t)\otimes B_{\alpha}(t),\int_0^td\tau[A_{\beta}(t-\tau)\nonumber\\
&\qquad \otimes B_{\beta}(t-\tau),\tilde{\rho}(t-\tau)\otimes \rho_B]  ,
\end{align}
\ees
where in our case $\alpha,\beta \in \{+,-\}$. It is useful to define the stationary (single-variable) bath correlation, a special case of \cref{eq:Corr_func}: 
\begin{align}\label{eq:B_tau}
     \mathcal{B}_{+-}(t,t-\tau)=\int_{0}^{\infty}d\omega J(\omega)e^{-i\omega\tau}\equiv\mathcal{B}_{+-}(\tau)\ .
 \end{align}
Now if we assume that $t\gg\tau_B$,  then $\tilde{\rho} (t-\tau)\approx \tilde{\rho}(t)$. We discuss the limitations of this approximation in \cref{app:RWA_useless}, where we show that it can lead to an unbounded error. 

Expanding the double commutator in terms of the bath correlation function, we obtain:
\begin{align}
\label{eq:Born_comm}
&\sum_{\alpha,\beta}\Tr_B [S_{\alpha}(t)\otimes B_{\alpha},[S_{\beta}(t-\tau)\otimes B_{\beta}(t-\tau),\tilde{\rho}(t)\otimes \rho_B]]\nonumber\\
&\quad =[S_+(t),S_-(t-\tau)\tilde{\rho}(t)]\mathcal{B}_{+-}(\tau)\nonumber\\
&\quad \quad -[S_-(t),\tilde{\rho}(t)S_+(t-\tau)]\mathcal{B}_{+-}(-\tau)\ .
\end{align}
Consequently, substituting \cref{eq:Born_comm} back into \cref{eq:2ordcumm}, we can write:
\begin{align}
\label{eq:redfield}
&\dot{\tilde{\rho}}(t)=-\int_0^t d\tau \mathcal{B}_{+-}(\tau) e^{i\Omega_0 \tau}[\sigma_+,\sigma_-\tilde{\rho}(t)]
\nonumber\\
&\quad +\int_0^t d\tau \mathcal{B}_{+-}(-\tau) e^{-i\Omega_0 \tau}[\sigma_-,\tilde{\rho}(t)\sigma_+]\ .
\end{align}
To arrive at a Lindblad form, we complete the Markovian approximation by setting the upper limit of the integral to be $\infty$. This is justified since the bath correlation decays rapidly to zero for $t\gg 1/\omega_c$. 
Now, let
\bes
\label{eq:Gamma}
\begin{align}
\label{eq:Gamma-1}
\Gamma_{\alpha \beta}(\omega)&\equiv\int_0^{\infty}d\tau \mathcal{B}_{\alpha \beta}(\tau)e^{i\omega \tau} \\
\label{eq:Gamma-2}
&= \int_0^{\infty}d\o' J(\o') \int_0^{\infty}d\tau e^{i(\o-\o')\tau} \\
\label{eq:Gamma-3}
&= \pi J(\o) + i\int_0^\infty d\o'\ J(\o') \mathcal{P}\left(\frac{1}{\o-\o'}\right)\ ,
\end{align}
\ees
where we used the identity
\beq
\int_{0}^\infty d\tau e^{ix\tau} = \pi \delta(x) + i \mathcal{P}\left(\frac{1}{x}\right)\ ,
\label{eq:Cauchy-P}
\eeq
and where the Cauchy principal value is defined as
\beq
\mathcal{P}\left(\frac{1}{x}\right)[f] = \lim_{\epsilon\to 0} \int_{-\epsilon}^\epsilon \frac{f(x)}{x}dx \ ,
\eeq
for smooth functions $f$ with compact support on the real line $\mathbb{R}$.

We show in \cref{app:Gamma-cc} that taking the complex conjugate of $\Gamma_{\alpha\beta}$ yields:
\begin{align}
\Gamma^*_{{\pm\mp}}(\omega) = \int_0^{\infty}d\tau \mathcal{B}_{{\pm\mp}}(-\tau)e^{-i\omega \tau}\ .
\label{eq:Gamma-cc}
\end{align}

This simplifies \cref{eq:redfield} into Lindblad form: 
\bes
\label{RWA_LE_final}
\begin{align}
\dot{\tilde{\rho}}(t) &=-\Gamma_{+-}(\Omega_0)[\sigma_+,\sigma_-\tilde{\rho}]+\Gamma_{+-}^*(\Omega_0)[\sigma_-,\tilde{\rho}\sigma_+] \\
&= -i \Im[\Gamma_{+-}(\Omega_0)] [\sigma_+\sigma_-,\rho(t)]\\
&\quad +2\Re[\Gamma_{+-}(\Omega_0)]\left(\sigma_-\rho(t)\sigma_+-\frac{1}{2}\{\sigma_+\sigma_-,\rho(t)\}\right)\ . \nonumber
\end{align}
\ees
This result has the same form as the exact \cref{eq:ansatz}, but with a time-independent Lamb shift and decay rate:
\begin{subequations}
\label{RWA_S_gamma}
\begin{align}
\mathcal{S} &=2\Im[\Gamma_{+-}(\Omega_0)] = 2\int_0^\infty d\o'\ J(\o') \mathcal{P}\left(\frac{1}{\Omega_0-\o'}\right)\\
\gamma &=2\Re[\Gamma_{+-}(\Omega_0)] = 2\pi J(\Omega_0) \ .
\end{align}
\end{subequations}
This last result is consistent with the finding that the RWA-LE is the $\tau\to\infty$ limit of the C-LE~\cite{Majenz:2013qw}, since it follows from \cref{eq:bdelta-delta,eq:gamma-CG-LE-delta-bar} that
\beq
\lim_{\tau\to\infty} \g(\tau)= 2\pi \int_0^\infty d\omega J(\omega)\d(\Omega_0-\omega) = 2\pi J(\Omega_0)\ .
\eeq

The population and coherence are given by the $\tau\to\infty$ limit of \cref{eq:r-CG-LE}, i.e.,
\bes
\label{eq:r-RWA-LE}
\begin{align}
\label{eq:r11-RWA-LE}
    \tilde{\rho}_{11}(t) &= \rho_{11}(0)e^{-\g t}\\
\label{eq:r01-RWA-LE}
    \tilde{\rho}_{01}(t) &= \rho_{01}(0)e^{-\frac{1}{2}\g t}\ ,
\end{align}
\ees

We are now ready to present the results for the three spectral densities.

\subsubsection{$J_1(\omega)= |g|^2\delta(\omega-\omega_c)$}

For the impulse spectral density we find, using \cref{RWA_S_gamma}:
\bes
\label{eq:gamma-J1-RWA-LE}
\begin{align}
\label{eq:gamma-J1-RWA-LE-LS}
\mathcal{S} &= 2|g|^2 \mathcal{P}\left(\frac{1}{\Omega_0-\omega_c}\right)\\
\label{eq:gamma-J1-RWA-LE-gamma}
\g &= 2\pi |g|^2\delta(\Omega_0-\omega_c)\ .
\end{align}
\ees
This means that the decay rate either vanishes or is singular at $\Omega_0=\omega_c$.
Hence, the RWA-LE is unsuitable for describing the model with this spectral density. 

\subsubsection{$J_2(\omega)=\eta \omega e^{-\omega/\omega_c}$}
We can immediately write down the decay rate as $\gamma = 2\pi J_2(\Omega_0)$. However, the Cauchy principal value complicates the calculation of the Lamb shift, so we use a direct method instead. 

For the Ohmic spectral density, the bath correlation in \cref{eq:B_tau} takes the form:
 \begin{align}
 \label{eq:Corr_Ohmic}
\mathcal{B}_{+-}(\tau)=\eta\int_0^{\infty} \omega e^{-\omega/\omega_c} e^{-i\omega \tau}d\omega=\frac{\eta \omega_c^2}{(1+i\omega_c \tau)^2}\ .
 \end{align}
The one-sided Fourier integral in \cref{eq:Gamma-1} becomes:
\bes
\begin{align}
\label{eq:Gamma-RWA-J2-1}
&\Gamma_{+-}(\Omega_0)=\eta\omega_c \int_0^{\infty}d(\omega_c\tau) \frac{e^{i\Omega_0 \tau}}{(1+i\omega_c\tau)^2}\\
\label{eq:Gamma-RWA-J2-2}
&\quad = -i\eta\omega_c+\eta\Omega_0e^{-\Omega_0/\omega_c}(\pi+i \Ei\left(\Omega_0/\omega_c\right))\ ,
\end{align}
\ees
where we derive the second equality in \cref{app:RWA-LE-integrals-J2}.

Thus, the Lamb shift and the decay rate are:
\bes
\label{eq:gamma-J2-RWA-LE}
\begin{align}
\label{eq:gamma-J2-RWA-LE-LS}
\mathcal{S} &=2J_2(\Omega_0)\Ei\left(\frac{\Omega_0}{\omega_c}\right)-2\eta \omega_c\\
\label{eq:gamma-J2-RWA-LE-gamma}
\gamma &= 2\pi J_2(\Omega_0) \ .
\end{align}
\ees

\subsubsection{$J_3(\omega)=\eta \omega \Theta(\omega_c-\omega)$}
We can once more immediately write down the decay rate as $\gamma = 2\pi J_3(\Omega_0)$, but a direct calculation is again advantageous for arriving at the form of the Lamb shift.

For the triangular spectral density, the bath correlation in \cref{eq:B_tau} takes the form:
 \begin{align}
 \label{eq:B+-J3-RWA-LE}
 \mathcal{B}_{+-}(\tau)=\eta\int_0^{\omega_c} \omega e^{-i\omega \tau}d\omega=\eta \frac{e^{-i \omega_c \tau}(1+i \omega_c\tau)-1}{\tau^2}
 \end{align}
The one-sided Fourier integral in \cref{eq:Gamma-1} becomes:

\begin{align}
\label{eq:Gamma-RWA-J3}
&\Gamma_{+-}(\Omega_0) =\eta\omega_c \int_0^{\infty}d(\omega_c\tau) \frac{e^{-i \omega_c \tau}(1+i \omega_c\tau)-1}{(\o_c\tau)^2}e^{i\Omega_0 \tau}
\end{align}

Thus, as we derive in \cref{app:RWA-LE-integrals-J3}, the Lamb shift and the decay rate are:
\bes
\label{eq:gamma-J3-RWA}
\begin{align}
\label{eq:gamma-J3-RWA-LE-LS}
\mathcal{S} &=-2\Omega_0\ln\left|\frac{\omega_c}{\Omega_0}-1\right|-2\eta \omega_c \\
\label{eq:gamma-J3-RWA-LE-gamma}
\gamma &=2\pi J_3(\Omega_0)\ .
\end{align}
\ees
Note that \cref{eq:RWA_ass_Ohmic} requires $\Omega_0>\omega_c$, but in this case it follows from \cref{eq:gamma-J3-RWA-LE-gamma} that $\gamma$ vanishes. 
This breakdown of the validity conditions of the Markov approximation, along with the issue of the potentially unbounded approximation error discussed in \cref{app:RWA_useless}, highlights that the RWA-LE has limited validity for the model we study here. Our simulation results reinforce these conclusions, as shown in \cref{Analysis}.

\subsection{TCL}
\label{sec:TCL}
In this section we briefly review the time-convolutionless formalism, closely following the presentation of \cite{Breuer:book} while adding a few pertinent details.

We can rewrite the Liouville equation in \cref{eq:Liouville} as:
\begin{align}
\frac{d\tilde{\rho}_{SB}}{dt}\equiv\lambda \mathcal{L}\tilde{\rho}_{SB}\ \ , \qquad  \mathcal{L}\equiv [\tilde{H}_{SB},\cdot]\ ,
\label{eq:133}
\end{align}
where we introduce the Feshbach projection superoperator $\mathcal{P}$ via
\begin{align}
\label{eq:NZ_Liouville}
\mathcal{P}\rho_{SB}\equiv \Tr_{B} (\rho)_{SB}\otimes \rho_B \ ,
\end{align}
and its orthogonal complement $\mathcal{Q}=\mathcal{I}-\mathcal{P}$. For an arbitrary operator $A$, we define:
\begin{align}
\hat{A}\equiv \mathcal{P} A\ , \qquad \bar{A}\equiv \mathcal{Q} A \ ,
\end{align}
which leads via \cref{eq:133} to:
\bes
\label{eq:projeqs}
\begin{align}
\partial_t \hat{\rho}_{SB}&=\lambda \hat{\mathcal{L}}\hat{\rho}_{SB}+\lambda \hat{\mathcal{L}}\bar{\rho}_{SB}\\
\partial_t \bar{\rho}_{SB}&=\lambda \bar{\mathcal{L}}\hat{\rho}_{SB}+\lambda \bar{\mathcal{L}}\bar{\rho}_{SB}\ .
\end{align}
\ees
The second equation has a solution given by:
\begin{align}
\label{eq:1322}
\bar{\rho}_{SB}(t)=\mathcal{G}(t,0)\bar{\rho}_{SB}(0)+\lambda\int_0^t \mathcal{G}(t,t')\bar{\mathcal{L}}(t')\bar{\rho}_{SB}(t')dt'\ ,
\end{align}
where:
\begin{align}
\mathcal{G}(t,0)\equiv T_+e^{ \lambda\int_0^t \bar{\mathcal{L}}(t')dt'}\ .
\end{align}
From \cref{eq:133}, we obtain:
\begin{align}
    \tilde{\rho}_{SB}(t)=\mathcal{U}_{+}(t,t')\tilde{\rho}_{SB}(t')\ ,
\end{align}
where
\begin{align}
    \mathcal{U}_{+}(t,t')=T_+e^{\lambda \int_{t'}^t \bar{\mathcal{L}}(s)ds}\ .
\end{align}
We can back-propagate the system state as:
\begin{align}
\label{eq:back_propagate}
    \tilde{\rho}_{SB}(t')=\mathcal{U}_{-}(t',t)\tilde{\rho}_{SB}(t')\ ,
\end{align}
where $\mathcal{U}_{-}(t',t)=[\mathcal{U}_{+}(t,t')]^{-1}$.
When we substitute \cref{eq:back_propagate} back into \cref{eq:1322}, it yields:
\bes
\begin{align}
\bar{\rho}_{SB}(t)&=\mathcal{G}(t,0)\bar{\rho}_{SB}(0)+\Sigma(t)\rho_{SB}(t)\\
\Sigma(t)&\equiv \lambda \int_0^t \mathcal{G}(t,t')\bar{\mathcal{L}}(t')\hat{\mathcal{U}}_{-}(t',t)dt'\ .
\end{align}
\ees
Assuming that $\mathcal{I}-\Sigma$ is invertible and using \cref{eq:projeqs}, we arrive at:
\bes
\begin{align}
\partial_t \hat{\rho}_{SB}(t)&=\mathcal{J}(t)\bar{\rho}_{SB}(0)+\mathcal{K}(t)\hat{\rho}_{SB}(t)
\\
\mathcal{J}(t)&\equiv \lambda \hat{\mathcal{L}}(t)[\mathcal{I}-\Sigma]^{-1}\mathcal{G}(t,0)\mathcal{Q}\\
\mathcal{K}(t)&\equiv\lambda \hat{\mathcal{L}}[\mathcal{I}-\Sigma]^{-1}\mathcal{P}\ .
\end{align}
\label{eq:invertibility}
\ees
If the inhomogeneity $\mathcal{J}(t)$ vanishes (as is the case for a factorized initial system-bath state), then the resulting master equation is time-local and known as the time-convolutionless (TCL) master equation.
By expressing $\mathcal{I}-\Sigma$ as a geometric series, we obtain:
\begin{align}
    [\mathcal{I}-\Sigma]^{-1}=\sum_{n=0}^{\infty}\Sigma^n(t)\ .
\end{align}
This allows us to write the TCL generator as an expansion in powers of $\lambda$:
\begin{align}
\label{eq:Kexpansion}
\mathcal{K}(t)=\lambda \hat{\mathcal{L}}(t)\left(\sum_{n=0}^{\infty}\Sigma^n(t)\right)\mathcal{P}=\sum_{n=1}^{\infty}\lambda^n \mathcal{K}_n(t)\ .
\end{align}
For Gaussian baths, like the one considered here, the odd-order terms vanish, i.e., $\mathcal{K}_{2n+1}=0$, which follows from the vanishing multi-time bath correlation function utilizing Wick's theorem~\cite{Gaudin:1960aa,Fogedby:2022aa}. 
In general, the $n$'th order term is given by~\cite{Kam74-1,Kam74-2}:
\begin{align}
\mathcal{K}_n(t)=\int_0^{t}\!\!\! dt_1\int_0^{t}\!\!\! dt_2\cdots\int_0^{t_{n-2}}\hspace{-0.5cm}dt_{n-1}\langle\mathcal{L}(t)\mathcal{L}(t_1)...\mathcal{L}(t_{n-1})\rangle_{\text{oc}}\ ,
\end{align}
where the ordered cumulants are defined as:
\begin{align}
\langle\mathcal{L}(t)&\mathcal{L}(t_1)...\mathcal{L}(t_{n-1})\rangle_{\text{oc}}=\nonumber\\&\sum (-1)^{q}\mathcal{P}\mathcal{L}(t)\cdots\mathcal{L}(t_i)\mathcal{P}\mathcal{L}(t_j)\cdots\mathcal{L}(t_m)\mathcal{P}\ ,
\end{align}
where the sum is over all possible arrangements of $q$ $\mathcal{P}$'s and $n$ $\mathcal{L}$'s such that there is at least one $\mathcal{L}$ between $\mathcal{P}$'s and there is a time ordering $t\geq ...\geq t_i\geq t_j...\geq t_m$.

Since \cref{eq:Ks_gen} is already time-local, we can relate it to the TCL generator $\mathcal{K}(t)$ via~\cite{Vacchini:2010fv}:
\begin{align}
\mathcal{K}_S(t)\rho(t)=\Tr_B[\mathcal{K}(t)(\rho(t)\otimes \rho_B)]\ .
\end{align}
This leads exactly to the ansatz given in \cref{eq:ansatz}, i.e., the TCL master equation is given by 
\begin{align}
\label{eq:TCL}
\dot{\tilde{\rho}} &= \mathcal{K}_S(t)\tilde{\rho} = -\frac{i}{2} \mathcal{S}(t)[\sigma_+\sigma_-,\tilde{\rho}(t)]\nonumber\\
    &\qquad+\gamma(t)\left(\sigma_-\tilde{\rho}(t)\sigma_+-\frac{1}{2}\{\sigma_+\sigma_-,\tilde{\rho}(t)\}\right)\ .
\end{align}
The solution is given by \cref{eq:rho-exact}, to the desired order in perturbation theory, i.e.:
\bes
\label{eq:rho-TCLn}
\begin{align}
\label{eq:rho11-TCLn}
  \tilde{\rho}_{11}(t)&=\rho_{11}(0) \exp{-\int_0^{t}\gamma^{(n)}(t')dt'}\\
\label{eq:rho01-TCLn}
  \tilde{\rho}_{01}(t)&= \rho_{01}(0) \exp{\frac{1}{2}\int_0^{t}\left(i\mathcal{S}^{(n)}(t')-\gamma^{(n)}(t')\right)dt'} \ ,
\end{align}
\ees
where $n=2$ for TCL2, $n=4$ for TCL4, etc.

To find the perturbative expansion of the Lamb shift and the decay rate, we observe that $\sigma_+$ is an eigenoperator of the generator:
\begin{align}
\mathcal{K}_S(t)\sigma_+=-\frac{1}{2}(\gamma(t)+iS(t))\sigma_+\ .
\label{eq:147}
\end{align}
Using the superoperator $\mathcal{L}(t)=-i[H_{SB}(t),\cdot]$, we can verify \cite{Breuerbook} that $\sigma_+\otimes \rho_B$ is an eigenoperator of $\mathcal{L}(t)\mathcal{L}(t_1)$ with eigenvalue $-f(t-t_1)$, where $f(t)$ is defined in \cref{eq:general_kernel}. Substituting \cref{eq:Kexpansion} into \cref{eq:147}, we obtain:
\begin{align}
&\mathcal{K}_S(t)\sigma_+=\sum_{n=1}^{\infty}(-\lambda^2)^n\int_{t_0}^{t}dt_1\int_{t_0}^{t_1}dt_2\cdots\int_{t_0}^{t_{2n-2}}dt_{2n-1} \nonumber\\
&\quad \langle f(t-t_1)f(t_2-t_3)\cdots f(t_{2n-2}-t_{2n-1}) \rangle_{\text{oc}} \sigma_+\ . 
\end{align}
Here, the terms in the summation follow the rule of considering all possible arrangements of the memory kernel $f(t_i-t_j)$. It is important to ensure that the times are properly ordered as $t_m\leq...\leq t_i\leq t_j\leq...\leq t$. Consequently, we can expand $\gamma(t)$ and $S(t)$ as:
\begin{align}
\label{eq:144}
\gamma(t)=\sum_{n=1}^{\infty}\lambda^{2n}\gamma_{2n}(t)\ , \quad \mathcal{S}(t)=\sum_{n=1}^{\infty}\lambda^{2n}\mathcal{S}_{2n}(t)\ .
\end{align}
In particular, if we define the function
\begin{align}
Z(t,t')\equiv \int_{0}^t dt_1f(t'-t_1) \notag
\end{align}
we obtain, as shown in Ref.~\cite{Vacchini:2010fv}, the following expressions
\bes
\begin{align}
\label{eq:TCL2}
\gamma_{2}(t)+iS_{2}(t)&=2\int_0^{t}dt_1f(t-t_1) = 2\lim_{t'\rightarrow t}Z(t,t')\\
\label{eq:TCL4}
\gamma_{4}(t)+iS_{4}(t)&=2\int_0^{t}dt_1\int_0^{t_1}dt_2\int_0^{t_2}dt_3 \\ &\!\!\!\!\! \left[f(t-t_2)f(t_1-t_3)+f(t-t_3)f(t_1-t_2)\right]\ . \notag
\end{align}
\ees
Thus, 
\begin{subequations}
\begin{align}
\label{eq:gamma2_Ohmic}
&\mathcal{S}_2(t) [\gamma_2(t)] =2\Im [\Re] \lim_{t'\rightarrow t} Z(t,t') \\
\label{eq:gamma4_Ohmic}
&\mathcal{S}_4(t) [\gamma_4(t)] =
2\Im [\Re]\int_ 0^{t} dt_ 1\int_ 0^{t_ 1}dt_2 \left(f(t-t_2)Z(t_1,t_2) \right.\nonumber\\
 &\qquad \left. + 
f(t_1-t_2)Z(t,t_2)\right)\ ,
\end{align}
\end{subequations}
where the imaginary and real parts are taken for $\mathcal{S}$ and $\g$, respectively.
The TCL2 and TCL4 cases correspond, respectively, to the following substitutions into \cref{eq:rho-TCLn}:
\begin{subequations}
\begin{align}
\label{eq:S-g-TCL2}
&\text{TCL2:}
\begin{cases}
\mathcal{S}^{(2)}(t) = \mathcal{S}_2(t)   \\
\g^{(2)}(t) = \g_2(t)
\end{cases} \\
\label{eq:S-g-TCL4}
&\text{TCL4:} 
\begin{cases}
\mathcal{S}^{(4)}(t) = \mathcal{S}_2(t)+\mathcal{S}_4(t)  \\
g^{(4)}(t) = \g_2(t)+\g_4(t)
\end{cases}
\end{align}
\end{subequations}
This follows from \cref{eq:144} with $\lambda=1$ (recall that $\lambda$ is a formal expansion parameter).

\begin{figure*}
\begin{subfigure}{.45\textwidth}
  \centering
\includegraphics[width=1\linewidth]{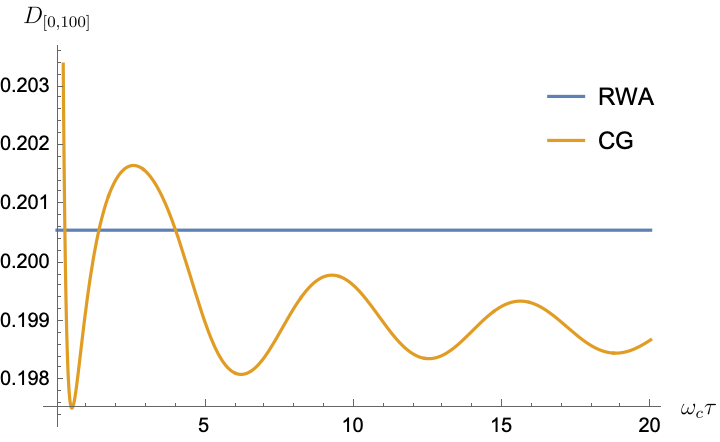}
  \caption{}
  \label{fig:distanceVSTau}
\end{subfigure}%
\begin{subfigure}{.45\textwidth}
  \centering 
\includegraphics[width=1\linewidth]{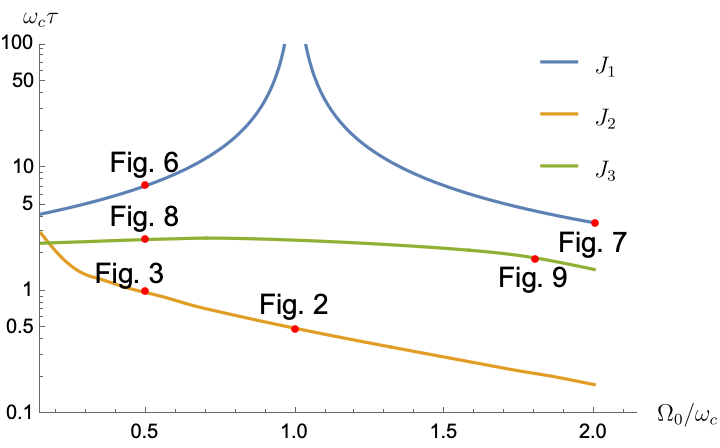}
  \caption{}
  \label{fig:CGtauVSa}
\end{subfigure}
\caption{\raggedright \small{(a) Integrated trace-norm distance $D_{[0,100]}$ between the exact solution and the Markovian approximations: CG-LE and RWA-LE, as a function of the dimensionless coarse-graining time $\omega_c \tau$ for the Ohmic bath spectral density $J_2$ with $\eta=1$ and $\Omega_0/\omega_c=1$. The minimum is obtained for a coarse-graining time $\tau=0.501939/\omega_c$. (b) The coarse-graining at which $D_{[0,100]}$ is minimized as a function of the qubit frequency $\Omega_0/\omega_c$ for $\eta=1$ for the three different spectral densities. The labeled red dots indicate the values of $\Omega_0/\omega_c$ shown in each of the corresponding figures.}
}
\end{figure*}

Next, we consider our three spectral densities.

\subsubsection{$J_1= |g|^2\delta(\omega-\omega_c)$}
We use the integral of the shifted memory kernel from \cref{eq:kernel_J1}:
\begin{align}
Z(t,t_1) = \frac{2|g|^2\sin\frac{(\Omega_0-\omega_c)t_1}{2}}{\Omega_0-\omega_c}e^{i\frac{(\Omega_0-\omega_c)(2t-t_1)}{2}}\ .
\end{align}
The second-order Lamb shift and decay rate are, using \cref{eq:gamma2_Ohmic}:
\bes
\label{eq:TCL2-J1}
\begin{align}
\label{eq:TCL2-J1-LS}
\mathcal{S}_2(t)&= 2|g|^2 \frac{1-\cos((\Omega_0-\omega_c)t)}{\Omega_0-\omega_c}\\
\label{eq:TCL2-J1-gamma}
\gamma_2(t) &=2|g|^2 t \sinc[(\Omega_0-\omega_c)t)]\ .
\end{align}
\ees 
The fourth-order Lamb shift and decay rate are, using \cref{eq:gamma4_Ohmic}:
\bes
\label{eq:TCL4-J1}
\begin{align}
\label{eq:TCL4-J1-LS}
\mathcal{S}_4(t)&= 4|g|^4\frac{-\sin^2(\Omega_0-\omega_c)t+(\Omega_0-\omega_c)t\sin(\Omega_0-\omega_c)t}{(\Omega_0-\omega_c)^3}\\
\label{eq:TCL4-J1-gamma}
\gamma_4(t) &=2|g|^4\frac{2(\Omega_0-\omega_c)t\cos(\Omega_0-\omega_c)t-\sin 2(\Omega_0-\omega_c)t}{(\Omega_0-\omega_c)^3}
\end{align}
\ees

\subsubsection{$J_2(\omega) = \eta \omega e^{-\omega/\omega_c}$}
Again we use the integral of the shifted memory kernel from \cref{eq:kernel_Ohmic}:
\begin{align}
\label{eq:Zfunction_J2}
Z(t,t_1) &= \frac{i\eta\omega_c e^{i\Omega_0 t}}{1+i\omega_c t} -\frac{i\eta\omega_c e^{i\Omega_0 (t-t_1)}}{1+i\omega_c (t-t_1)}\nonumber \\
&\quad +iJ(\Omega_0)\Big(\Ei\left(\frac{\Omega_0}{\omega_c}(1+i\omega_c(t-t_1))\right) \notag \\
&\qquad -\Ei\left(\frac{\Omega_0}{\omega_c}(1+i\omega_ct)\right)\Big)\ .
\end{align}
The second-order Lamb shift and decay rate are, using \cref{eq:gamma2_Ohmic}:
\bes
\label{eq:TCL2-J2}
\begin{align}
\label{eq:TCL2-J2-LS}
\mathcal{S}_2(t)&= -2\eta \omega_c+2\eta \omega_c\frac{\cos(\Omega_0 t)+\omega_c t \sin(\Omega_0 t)}{1+(\omega_ct)^2}\nonumber\\
&-2J_2(\Omega_0)\left(\Re\left[\Ei\left(\frac{\Omega_0}{\omega_c}(1+i\omega_c t)\right)\right]-\Ei \left(\frac{\Omega_0}{\omega_c}\right)\right)\\
\label{eq:TCL2-J2-gamma}
\gamma_2(t)&=2\eta\omega_c\frac{\omega_c t \cos(\Omega_0t)-\sin(\Omega_0 t)}{1+(\omega_c t)^2} \nonumber \\ 
&\quad +2J_2(\Omega_0)\Im\left[\Ei\left(\frac{\Omega_0}{\omega_c}(1+i\omega_ct)\right)\right]\ .
\end{align}
\ees
The Markovian decay rate is obtained in the long time limit. Using $\lim_{x\rightarrow \infty}\Ei(i x+ 1)=i\pi$, we obtain:
\begin{align}
\gamma_M=\lim_{t\rightarrow \infty}\gamma_2(t)=2\pi\eta \Omega_0e^{-\frac{\Omega_0}{\omega_c}}=2\pi J(\Omega_0)\ ,
\end{align}
in agreement with \cref{eq:gamma-J2-RWA-LE-gamma}.

For the fourth-order decay rate, \cref{eq:gamma4_Ohmic} does not admit a closed-form solution and needs to be evaluated numerically.

\subsubsection{$J_3(\omega)=\eta \omega \Theta(\omega_c-\omega)$}
We use the shifted integral of the memory kernel in \cref{eq:kernel_J3}:
\begin{align}
\label{eq:Zfunction_J3}
Z(t,t_1)&=\frac{\eta}{t(t-t_1)}\left(e^{i(\Omega_0-\omega_c)(t-t_1)}t-e^{i\Omega_0(t-t_1)}t\right.\nonumber\\
&\left.+(e^{i\Omega_0 t}-e^{i(\Omega_0-\omega_c)t})(t-t_1)\right) \\
&+i\eta \Omega_0\left[\Ei(i(\Omega_0-\omega_c)t)-\Ei(i\Omega_0 t)\right.\nonumber\\
&\left.-\Ei(i(\Omega_0-\omega_c)(t-t_1))+\Ei(i\Omega_0 (t-t_1))\right] \ . \notag
\end{align}
Thus:
\bes
\label{eq:TCL2-J3}
\begin{align}
\label{eq:TCL2-J3-LS}
\mathcal{S}_2(t)&= -2i\eta\Big(\frac{\sin(\Omega_0 t)-\sin((\Omega_0-\omega_c) t)}{t}\notag\\
&\quad-\Omega_0\ln |1-\frac{\omega_c}{\Omega_0}|-\omega_c\notag\\
&\quad+\Re(\Omega_0\left(\Ei(i(\Omega_0-\omega_c)t)-\Ei(i\Omega_0 t))\right)\Big)\\
\label{eq:TCL2-J3-gamma}
\gamma_2(t)&=2\eta\Big(\frac{\cos(\Omega_0 t)-\cos((\Omega_0-\omega_c) t)}{t}\notag\\
&\quad-\pi\Omega_0 \Theta(\omega_c-\Omega_0)\notag \\
&\quad +\Omega_0\Im \left[\Ei(i\Omega_0 t)-\Ei(i(\Omega_0-\omega_c)t)\right]\Big)\ .
\end{align}
\ees
We recover the Markov approximation in the large $t$ limit:
\begin{align}
\lim_{t\rightarrow \infty}\gamma_2(t)=2\pi \eta J_3(\Omega_0)\ .
\end{align}
While the TCL result \cref{eq:TCL2-J3-gamma} does not entirely match the exact result, it does describe an oscillatory behavior similar to the exact solution, which is entirely absent in the Markovian limit. A similar phenomenon has been described in the context of non-equilibrium dynamics in Ref.~\cite{Ban:2010aa}.

Moreover, recall that the Markovian rate vanishes when $\Omega_0>\omega_c$ [\cref{eq:gamma-J3-RWA-LE-gamma}], resulting in the absence of decay. In contrast, when $\Omega_0>\omega_c$ we find:
\begin{align}
\label{eq:asymptoteTCL2}
\int_0^{\infty}\gamma_2(t')dt'=2\eta \left(\frac{\omega_c}{\Omega_0-\omega_c}+\ln \left(\frac{\Omega_0-\omega_c}{\Omega_0}\right)\right)\ ,
\end{align}
so that the asymptotic limit for the population in the TCL2 approximation (which coincides with the Redfield equation), $e^{-\int_0^{\infty}\gamma_2(t')dt'}$, is non-zero. This qualitatively recovers the non-zero decay property of the exact solution.

For the fourth-order decay rate, \cref{eq:gamma4_Ohmic} again does not admit a closed-form solution and needs to be evaluated numerically.

\begin{figure*}
\vspace{-0.5cm}
$$  \boxed{J_2(\omega)=\eta \omega e^{-\omega/\omega_c} \qquad \eta=1 \qquad \Omega_0/\omega_c=1}$$
\begin{subfigure}{\textwidth}
\vspace{-0.3cm}
  \centering 
  \includegraphics[width=0.6\linewidth]{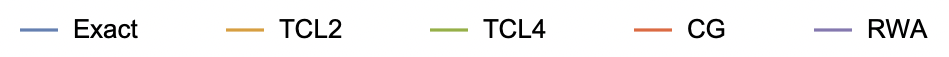}
\end{subfigure}%
\vspace{-0.5cm}
   \vskip\baselineskip
\begin{subfigure}{.33\textwidth}
  \centering \hspace{-1cm}
  \includegraphics[width=1\linewidth]{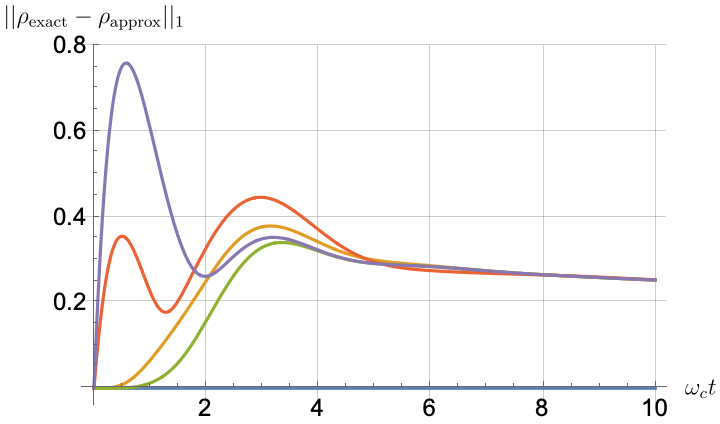}
  \caption{Trace-norm distance}
  \label{fig:traced_a1}
\end{subfigure}%
\begin{subfigure}{.33\textwidth}
  \centering 
  \includegraphics[width=0.92\linewidth]{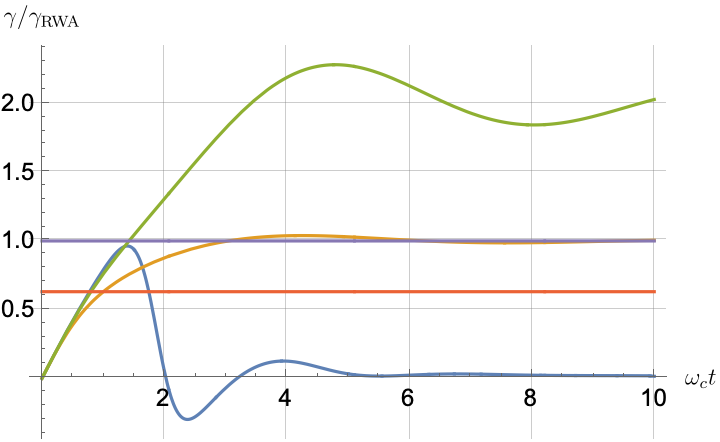}
  \caption{Decay rate}
  \label{fig:gam_a1}
\end{subfigure}%
\begin{subfigure}{.33\textwidth}
  \centering 
\includegraphics[width=0.92\linewidth]{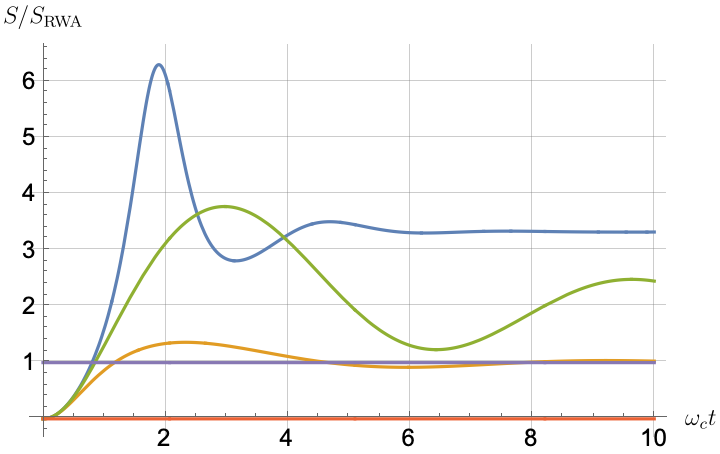}
  \caption{Lamb shift }
  \label{fig:S_a1}
\end{subfigure}
   \vskip\baselineskip
   \vspace{-0.3cm}
\begin{subfigure}{.33\textwidth}
  \centering   \hspace{-1cm}
  \includegraphics[width=0.92\linewidth]{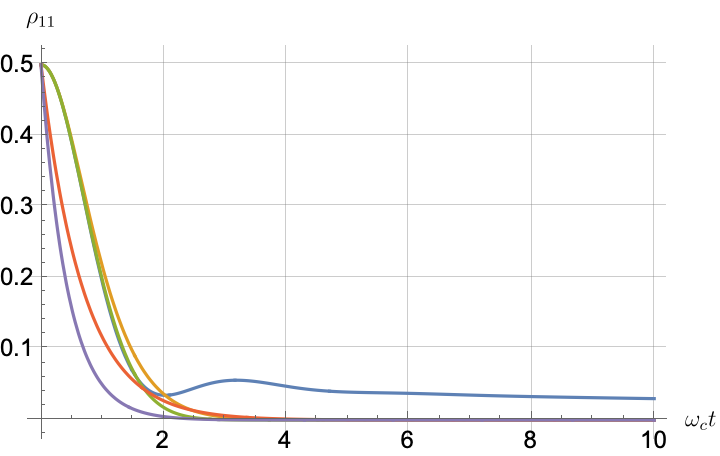}
  \caption{Population }
  \label{fig:popul_a1}
\end{subfigure}%
\begin{subfigure}{.33\textwidth}
  \centering
  \includegraphics[width=0.92\linewidth]{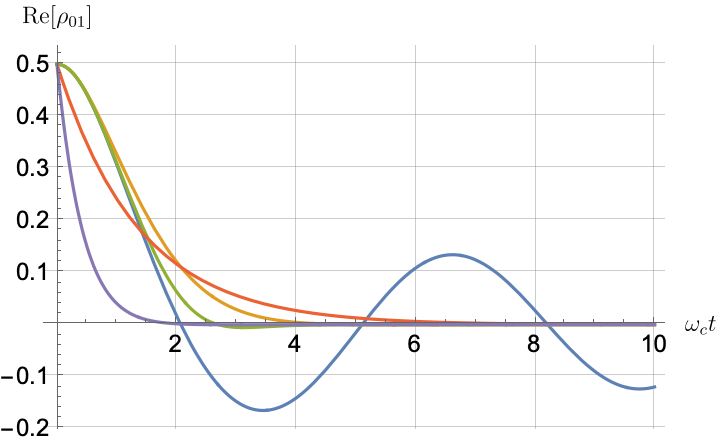}
  \caption{Coherence -- real part}
  \label{fig:Re_coher_a1}
\end{subfigure}
\begin{subfigure}{.33\textwidth}
  \centering
  \includegraphics[width=0.92\linewidth]{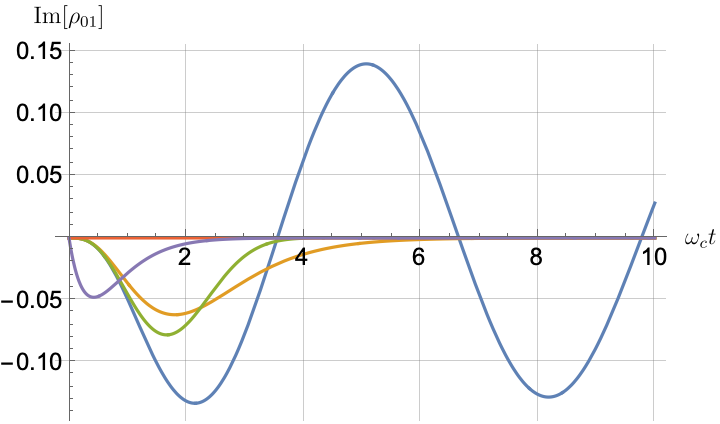}
  \caption{Coherence -- imaginary part}
  \label{fig:Im_coher_a1}
\end{subfigure}
\caption{\raggedright
\small{The trace-norm distance $\|\rho_{\text{exact}}-\rho_{\text{approx}}\|_1$ (a), decay rate ratio $\gamma/\gamma_{\text{RWA}}$ (b), Lamb shift ratio $S/S_{\text{RWA}}$ (c), population $\rho_{11}$ (d) and coherence in its real part $\Re(\rho_{01})$ (e) and imaginary part $\Im(\rho_{01})$ (f) with an initial state $\rho(0)=\ketb{+}{+}$ for Ohmic spectral density $J_2(\omega)=\eta \omega e^{-\omega/\omega_c}$ as a function of dimensionless time $\omega_c t$ for coupling $\eta=1$ and qubit frequency ratio $\Omega_0/\omega_c=1$. Five different approaches are depicted in the plots: the exact solution (Exact), TCL2 (Redfield), TCL4, CG-LE and RWA-LE (Markov). The CG-LE coarse graining time is $\tau=0.501939/\omega_c$.}}
  \label{fig:dynamics_a1}
\end{figure*}

\begin{figure*}
$$\boxed{J_2(\omega)=\eta \omega e^{-\omega/\omega_c} \qquad  \eta=1/2 \qquad \Omega_0/\omega_c=1}$$
\begin{subfigure}{\textwidth}
\vspace{-0.3cm}
  \centering 
  \includegraphics[width=0.6\linewidth]{figure3.png}
\end{subfigure}%
\vspace{-0.5cm}
   \vskip\baselineskip
\begin{subfigure}{.33\textwidth}
  \centering \hspace{-1.6cm}
  \includegraphics[width=1.05\linewidth]{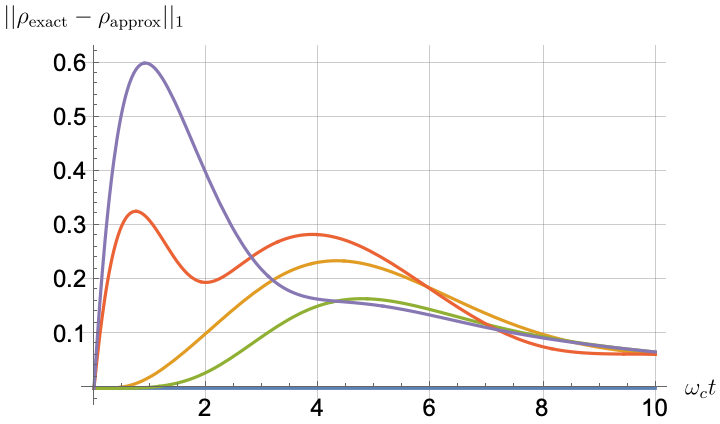}
  \caption{Trace-norm distance}
  \label{fig:traced_eta05}
\end{subfigure}%
\begin{subfigure}{.33\textwidth}
  \centering 
  \includegraphics[width=0.92\linewidth]{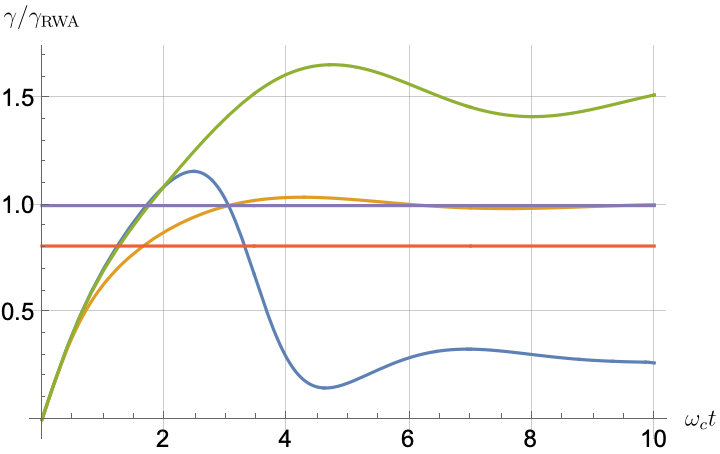}
  \caption{Decay rate}
  \label{fig:gam_eta05}
\end{subfigure}%
\begin{subfigure}{.33\textwidth}
  \centering 
\includegraphics[width=0.92\linewidth]{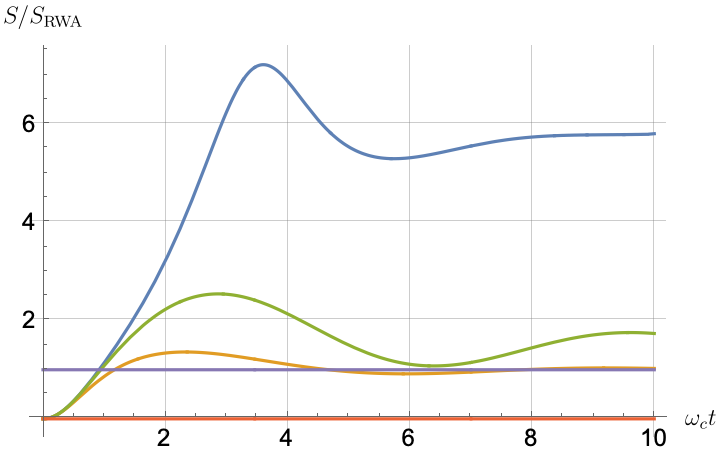}
  \caption{Lamb shift }
  \label{fig:S_eta05}
\end{subfigure}
   \vskip\baselineskip
   \vspace{-0.3cm}
\begin{subfigure}{.33\textwidth}
  \centering   \hspace{-1.6cm}
  \includegraphics[width=0.92\linewidth]{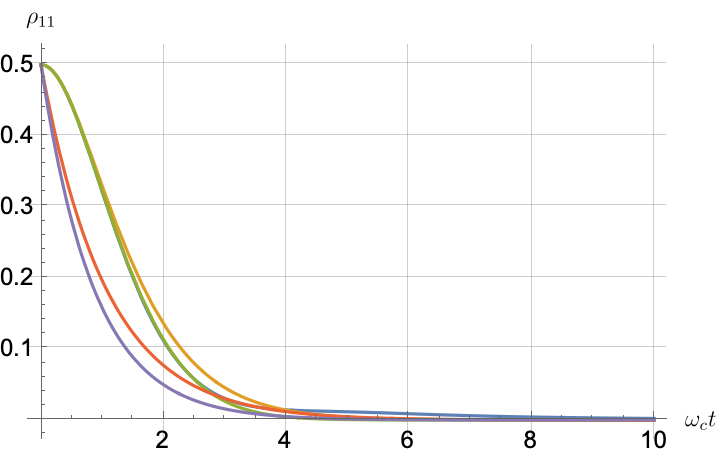}
  \caption{Population }
  \label{fig:popul_eta05}
\end{subfigure}%
\begin{subfigure}{.33\textwidth}
  \centering
  \includegraphics[width=0.92\linewidth]{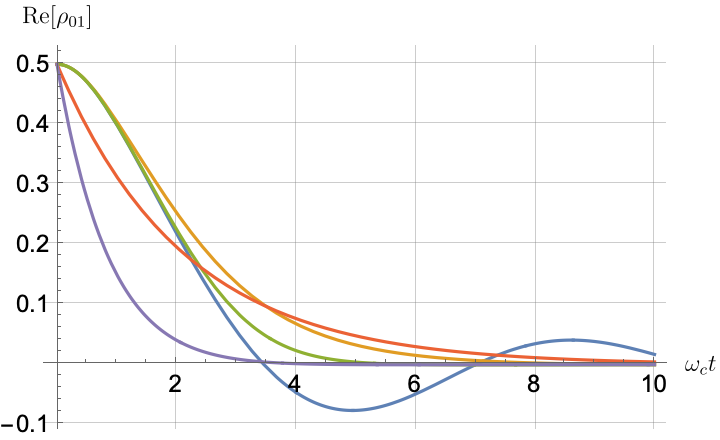}
  \caption{Coherence -- real part}
  \label{fig:Re_coher_eta05}
\end{subfigure}
\begin{subfigure}{.33\textwidth}
  \centering
  \includegraphics[width=0.92\linewidth]{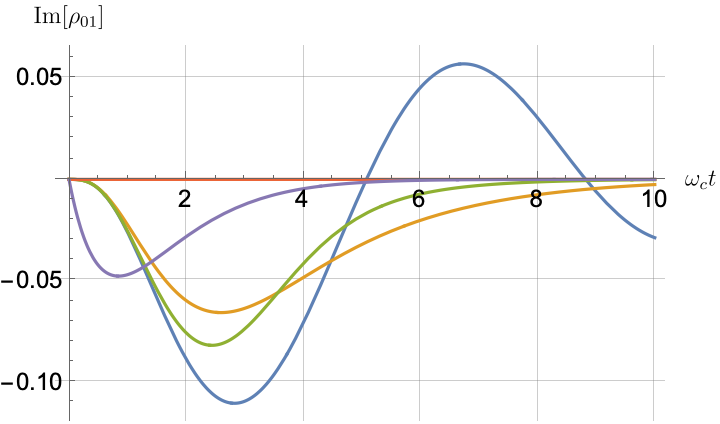}
  \caption{Coherence -- imaginary part}
  \label{fig:Im_coher_eta05}
\end{subfigure}
\caption{\raggedright
\small{Same as \cref{fig:dynamics_a1} with weak coupling $\eta=1/2$ and a qubit frequency equal to the cutoff: $\Omega_0/\omega_c=1$. The CG-LE coarse graining time is $\tau=0.659732/\omega_c$.}
}
 \label{fig:dynamics_eta05}
\end{figure*}

\begin{figure*}
$$\boxed{J_2(\omega)=\eta \omega e^{-\omega/\omega_c} \qquad  \eta=1 \qquad \Omega_0/\omega_c=1/2}$$
\begin{subfigure}{\textwidth}
\vspace{-0.3cm}
  \centering 
  \includegraphics[width=0.6\linewidth]{figure3.png}
\end{subfigure}%
\vspace{-0.5cm}
   \vskip\baselineskip
\begin{subfigure}{.33\textwidth}
  \centering \hspace{-1cm}
  \includegraphics[width=1\linewidth]{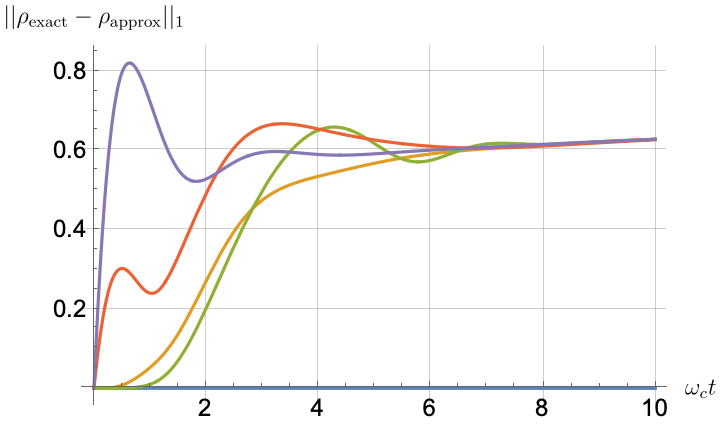}
  \caption{Trace-norm distance}
  \label{fig:traced_a05}
\end{subfigure}%
\begin{subfigure}{.33\textwidth}
  \centering 
  \includegraphics[width=0.92\linewidth]{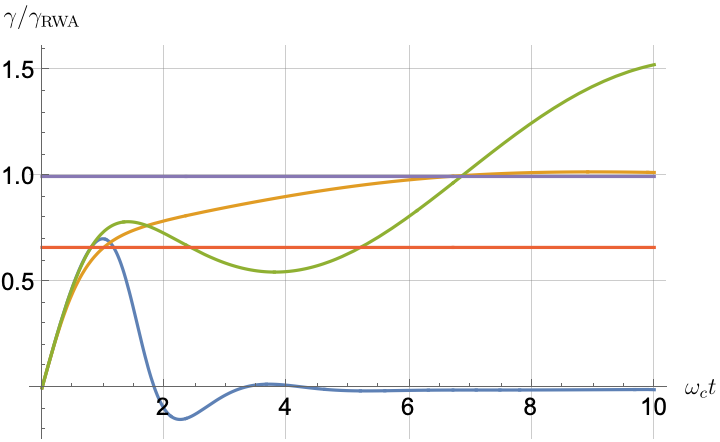}
  \caption{Decay rate}
  \label{fig:gam_a05}
\end{subfigure}%
\begin{subfigure}{.33\textwidth}
  \centering 
\includegraphics[width=0.92\linewidth]{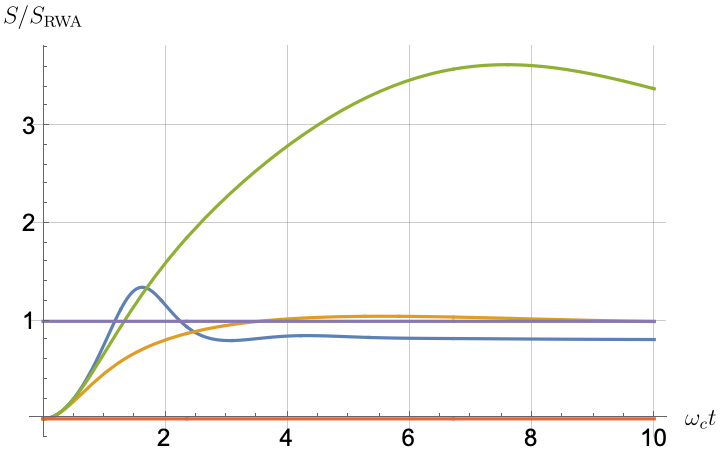}
  \caption{Lamb shift }
  \label{fig:S_a05}
\end{subfigure}
   \vskip\baselineskip
   \vspace{-0.3cm}
\begin{subfigure}{.33\textwidth}
  \centering   \hspace{-1.6cm}
  \includegraphics[width=0.92\linewidth]{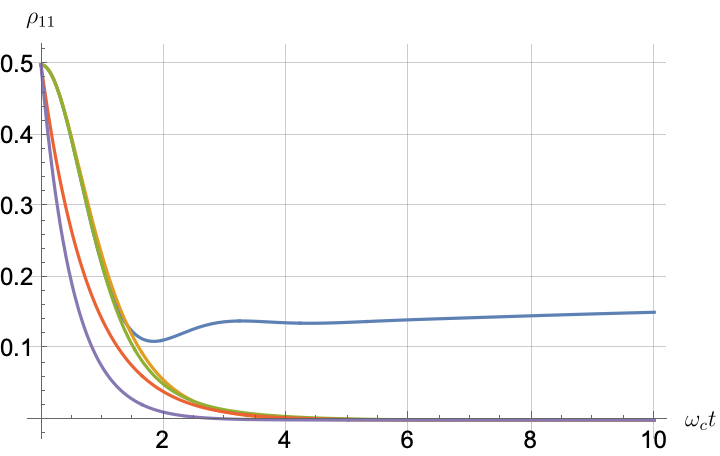}
  \caption{Population }
  \label{fig:popul_a05}
\end{subfigure}%
\begin{subfigure}{.33\textwidth}
  \centering
  \includegraphics[width=0.92\linewidth]{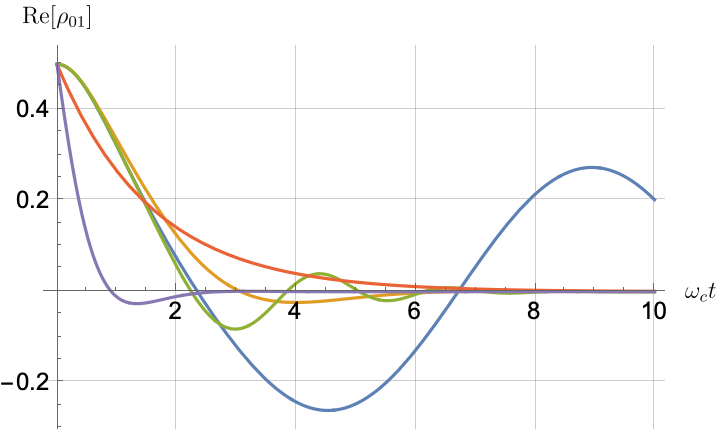}
  \caption{Coherence -- real part}
  \label{fig:Re_coher_a05}
\end{subfigure}
\begin{subfigure}{.33\textwidth}
  \centering
  \includegraphics[width=0.92\linewidth]{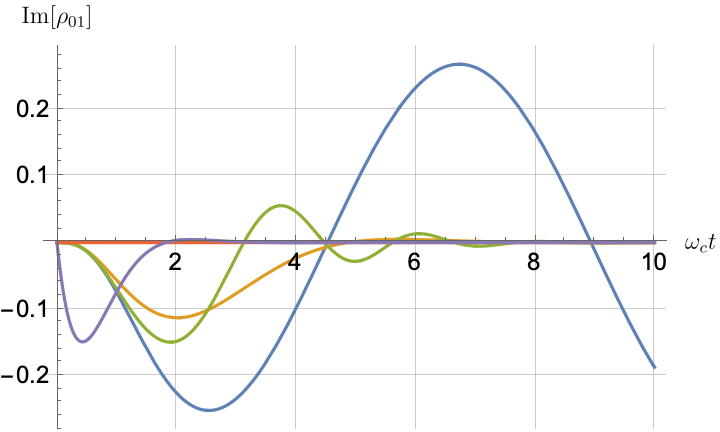}
  \caption{Coherence -- imaginary part}
  \label{fig:Im_coher_a05}
  
\end{subfigure}
\caption{\raggedright
\small{Same as \cref{fig:dynamics_a1} with an Ohmic spectral density $J_2(\omega)$ and the same coupling $\eta=1$, but with a lower qubit frequency: $\Omega_0/\omega_c=1/2$. The CG-LE coarse graining time is $\tau=0.999876/\omega_c$.}
}
 \label{fig:dynamics_a05}
\end{figure*}

\begin{figure*}
$$\boxed{J_2(\omega)=\eta \omega e^{-\omega/\omega_c} \qquad \eta=1 \qquad \Omega_0/\omega_c=4}$$
\begin{subfigure}{\textwidth}
\vspace{-0.3cm}
  \centering 
  \includegraphics[width=0.6\linewidth]{figure3.png}
\end{subfigure}%
\vspace{-0.5cm}
   \vskip\baselineskip
\begin{subfigure}{.33\textwidth}
  \centering \hspace{-1cm}
  \includegraphics[width=1\linewidth]{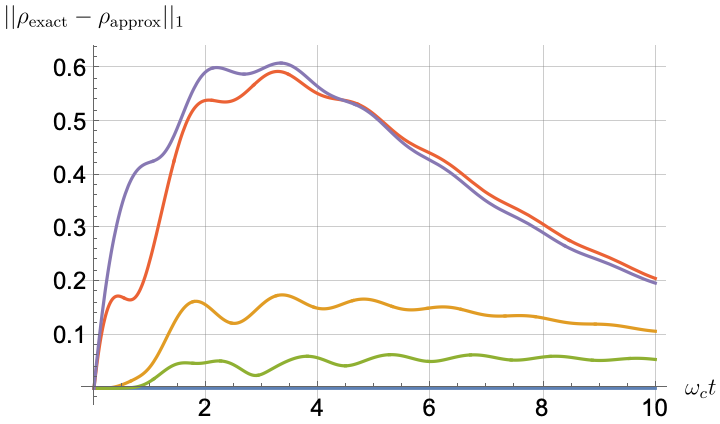}
  \caption{Trace-norm distance}
  \label{fig:traced_a4}
\end{subfigure}%
\begin{subfigure}{.33\textwidth}
  \centering 
  \includegraphics[width=0.92\linewidth]{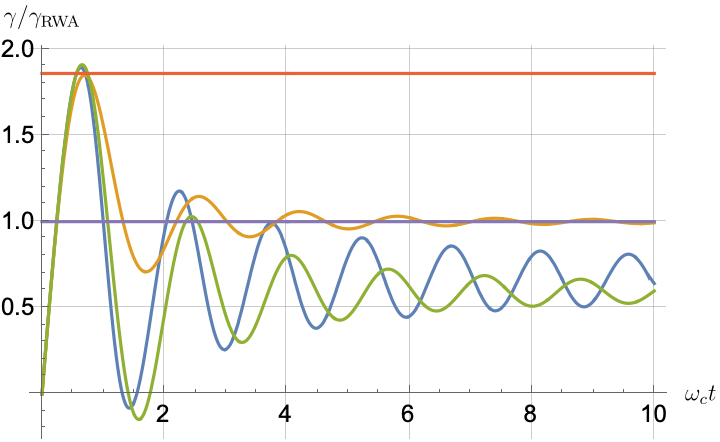}
  \caption{Decay rate}
  \label{fig:gam_a4}
\end{subfigure}%
\begin{subfigure}{.33\textwidth}
  \centering 
\includegraphics[width=0.92\linewidth]{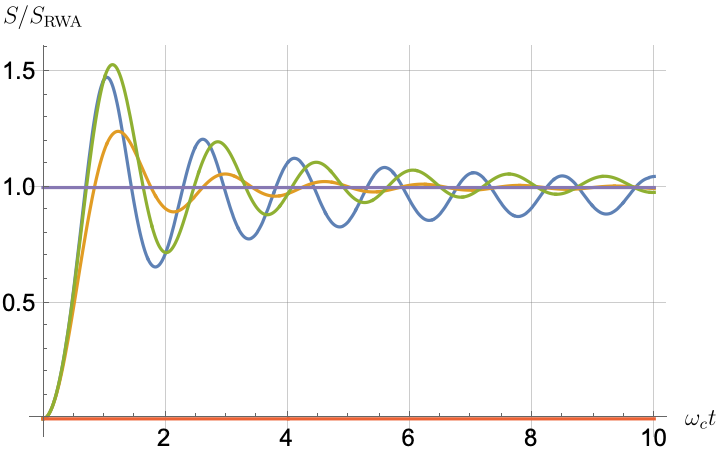}
  \caption{Lamb shift }
  \label{fig:S_a4}
\end{subfigure}
   \vskip\baselineskip
   \vspace{-0.3cm}
\begin{subfigure}{.33\textwidth}
  \centering   \hspace{-1.6cm}
  \includegraphics[width=0.92\linewidth]{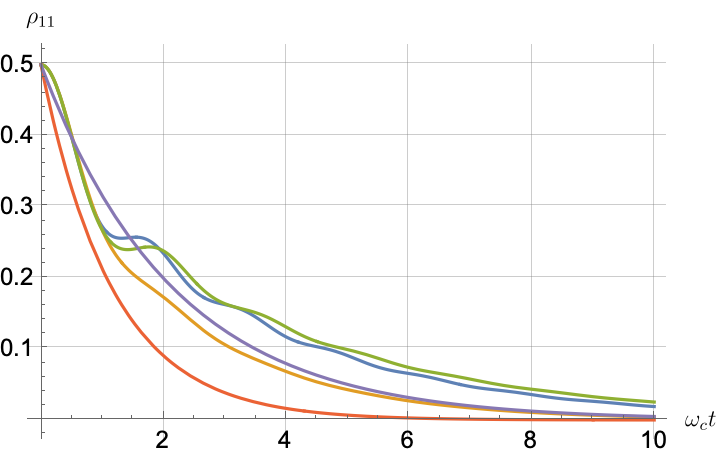}
  \caption{Population }
  \label{fig:popul_a4}
\end{subfigure}%
\begin{subfigure}{.33\textwidth}
  \centering
  \includegraphics[width=0.92\linewidth]{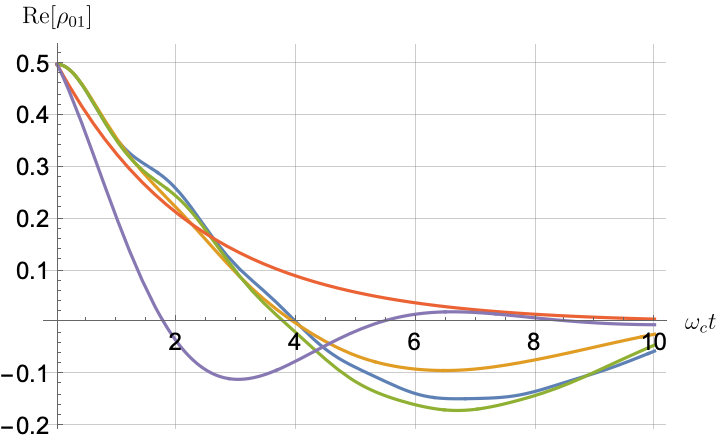}
  \caption{Coherence -- real part}
  \label{fig:Re_coher_a4}
\end{subfigure}
\begin{subfigure}{.33\textwidth}
  \centering
  \includegraphics[width=0.92\linewidth]{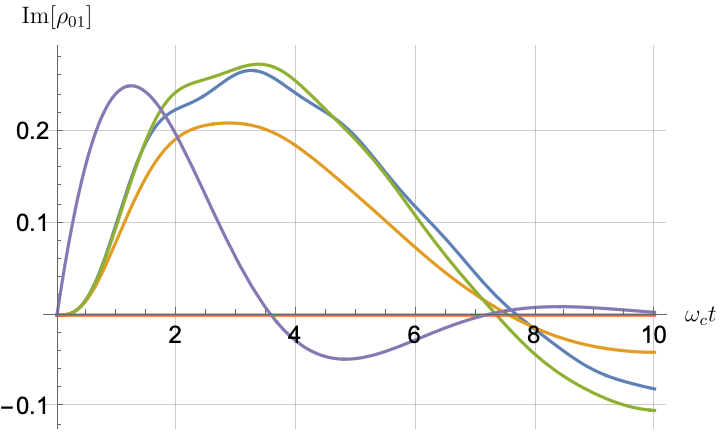}
  \caption{Coherence -- imaginary part}
  \label{fig:Im_coher_a4}
\end{subfigure}
\caption{\raggedright
\small{Same as \cref{fig:dynamics_a1} with an Ohmic spectral density $J_2(\omega)$ and the same coupling $\eta=1$, but with a larger qubit frequency: $\Omega_0/\omega_c=4$. The CG-LE coarse graining time is $\tau=0.025953/\omega_c$.}
}
 \label{fig:dynamics_a4}
\end{figure*}

\begin{table*}
\centering
\begin{tabular}{|cc|c|c|c|c|c|}
\hline
& & exact & C-LE/CG-LE & RWA-LE & TCL2 & TCL4 \\
\hline
{population} \(\r_{11}\)&  & \cref{eq:rho11} & \cref{eq:r11-CG-LE} & \cref{eq:r11-RWA-LE} & \cref{eq:rho11-TCLn,eq:S-g-TCL2} & \cref{eq:rho11-TCLn,eq:S-g-TCL4} \\
\hline
{coherence} \(\r_{01}\)& & \cref{eq:rho01} & \cref{eq:r01-CG-LE} & \cref{eq:r01-RWA-LE} & \cref{eq:rho01-TCLn,eq:S-g-TCL2} & \cref{eq:rho01-TCLn,eq:S-g-TCL4} \\
\hline
\multirow{3}{*}{{Lamb shift \(\mathcal{S}\)}} 
& {\(J_1\)} & \cref{eq:exact-S-J1} & $0$ & \cref{eq:gamma-J1-RWA-LE-LS} & \cref{eq:TCL2-J1-LS} &  \cref{eq:TCL4-J1-LS}\\
& {\(J_2\)} & & $0$ & \cref{eq:gamma-J2-RWA-LE-LS} & \cref{eq:TCL2-J2-LS} &   \\
& {\(J_3\)} & & $0$ & \cref{eq:gamma-J3-RWA-LE-LS} & \cref{eq:TCL2-J3-LS} &  \\
\hline
\multirow{3}{*}{{decay rate \(\g\)}} 
& {\(J_1\)} & \cref{eq:exact-gamma-J1} & \cref{eq:gamma-J1-CG-LE} & \cref{eq:gamma-J1-RWA-LE-gamma} & \cref{eq:TCL2-J1-gamma} & \cref{eq:TCL4-J1-gamma}\\
& {\(J_2\)} & & \cref{eq:gamma-J2-CG-LE} & \cref{eq:gamma-J2-RWA-LE-gamma} & \cref{eq:TCL2-J2-gamma} & \\
& {\(J_3\)} & & \cref{eq:gamma-J3-CG-LE} & \cref{eq:gamma-J3-RWA-LE-gamma} & \cref{eq:TCL2-J3-gamma} &  \\
\hline
\end{tabular}
\caption{Analytical results. The equations listed in the table are explicit analytical results. The results for empty cells are obtained numerically.}
\label{tab:summary}
\end{table*}

\section{Analysis}
\label{Analysis}

In this section, we compare the exact solution for the excited state population $\r_{11}(t)$ and the coherence to the results of the various approximation schemes we described above. We also compare the exact Lamb shift and decay rate $\g(t)$ to the corresponding quantities predicted by these approximation schemes. We do this for all three spectral densities where possible, for the CG-LE/C-LE, 
RWA-LE, TCL2 (Redfield), and TCL4. These results are summarized with corresponding equation references in \cref{tab:summary}.

\subsection{Computation of the optimal coarse graining time in CG-LE}\label{sec:compare_2}

Before presenting the comparison to the exact results, we first explain our methodology for optimizing the coarse-graining time $\tau$ within the framework of the CG-LE, since in the ensuing comparison we use the optimal $\tau$ values. 
Recall that the coarse-graining time needs to satisfy the condition $\o_c\tau \ll 1$ [\cref{eq:CG_cond}].

To determine the appropriate coarse-graining timescale $\tau$, we minimize a metric that quantifies the deviation from the exact solution. We employ the integrated trace-norm distance norm \cite{Majenz:2013qw} as our chosen metric, defined as:
\begin{align} \label{eq:distance_CG}
\mathcal{D}_{[0,T]}\equiv \frac{1}{2T}\int_{0}^T dt \|\rho_{\text{exact}}(t)-\rho_{\text{approx}}(t)\|_1 \ ,
\end{align}
where $\|M\|_1\equiv \Tr \sqrt{M^{\dagger}M}$. Since $M=\rho_{\text{exact}}-\rho_{\text{approx}}$ is Hermitian and traceless, it can be written as $\left(
\begin{array}{cc}
 a & b \\
 b^* & -a \\
\end{array}
\right)$ in the qubit case, so that $\Tr \sqrt{M^2} = 2\sqrt{a^2+|b|^2}$, and we can simplify the integrand as follows:
\begin{align}
   &\|\rho_{\text{exact}}-\rho_{\text{approx}}\|_1\\ \nonumber
   &=2\sqrt{(\rho_{11,\text{exact}}-\rho_{11,\text{approx}})^2 +|\rho_{01,\text{exact}}-\rho_{01,\text{approx}}|^2}\ .
\end{align}

In \cref{fig:distanceVSTau}, we illustrate the distance $\mathcal{D}$ between the exact solution and both the CG-LE and the RWA-LE as a function of the dimensionless quantity $\omega_c\tau$, where we vary the coarse-graining time $\tau$. In this example, we use the parameters $\eta=1$ and $\Omega_0/\omega_c=1$, and the numerical integration is performed up to a total time of $\omega_cT=100$. This upper limit is justified by the observation that the distance is minimized at relatively short times. For example, in \cref{fig:distanceVSTau} the minimum occurs at $\omega_c\tau=0.50139$, thus satisfying \cref{eq:CG_cond}. Notably, the coarse-graining solution outperforms the RWA-LE solution, consistent with the findings reported in \cite{Majenz:2013qw}.

In \cref{fig:CGtauVSa}, we present the minimum coarse-graining time as a function of the qubit frequency $\Omega_0/\omega_c$ for $\eta=1$ across the three different spectral models. The coarse-graining times used in the later figures are indicated by the red dots in the plot. We selected $\Omega_0/\omega_c\in[0.15,2]$ to account for a range of small to large qubit frequencies.

The most notable conclusion from \cref{fig:CGtauVSa} is that condition $\o_c\tau \ll 1$ cannot be satisfied for the impulse ($J_1$) and triangular ($J_3$) spectral densities within the range of $\Omega_0/\omega_c$ values shown. However, it is satisfied for the Ohmic spectral density ($J_2$), the most physically relevant of the three. We thus expect the CG-LE to perform poorly for $J_1$ and $J_3$, but to perform relatively well for $J_2$. These expectations are borne out in our results below.

Note further that for the Ohmic spectral density, the coarse-graining timescale increases as the qubit frequency decreases, tending to the RWA-LE solution in line with \cref{RWA_S_gamma}. Conversely, for the spectral density $J_1 = |g|^2\d(\o-\o_c)$, characterized by strong non-Markovian behavior and oscillations that prevent the fitting of a Markovian exponential decay, the coarse-graining time diverges at $\o-\o_c$ to attempt to fit the exact solution.

We remark that to ensure the validity of RWA-LE, the weak coupling limit \cref{eq:RWA_ass_Ohmic} needs to be satisfied, a point we comment on in more detail below.

\subsection{Exact solution \textit{vs} TCL and Markov approximations for the Ohmic spectral density}
\label{Comparison}

We now present the results of a comparison between the exact solution and the different master equations for the Ohmic spectral density, the most physically interesting of the three densities. Our general expectations are that the TCL approximations will capture some of the aspects of the non-Markovian dynamics and will thus outperform the two Markovian master equations, and that the CG-LE will outperform the RWA-LE due to the ability to optimize the coarse-graining timescale $\tau$ in the former. This expectation depends on the validity condition $\o_c\tau \ll 1$ not being violated, as explained in the previous section. Indeed, we find that this holds for all the examples we consider in this section, at least in the sense that we find $\o_c\tau < 1$ in all cases.

For the initial condition, we consider a simple scenario where $c_1(0)=c_0(0)=\frac{1}{2}$ and $c_k(0)=0$. In this setup, the system is initially in the state $\ket{+}=\frac{1}{\sqrt{2}}(\ket{0}+\ket{1})$, while the bath is initially in the vacuum state $\ket{v}$ of the cavity. To determine the amplitude for the exact solution, from which we obtain the population $\rho_{11}(t)=|c_{11}(t)|^2$ of the excited state $\ket{1}$, we need to apply a numerical inverse Laplace transform to \cref{eq:laplace_c1}, where the Laplace transform of the memory kernel $\hat{f}(s)$ is given by \cref{eq:laplace_Ohmic}. Irrespective of the specific positive values of $\eta$, $\omega_c$, and $\Omega_0$, the roots of \cref{eq:laplace_c1} are found to be complex. This means that the amplitude $c_1(t)$ is oscillatory. This trend persists even in cases of weak coupling, resulting in a damped oscillation of the amplitude.

\subsubsection{General observations}

\cref{fig:dynamics_a1,fig:dynamics_eta05,fig:dynamics_a05,fig:dynamics_a4} offer a comparison of the temporal dynamics against the exact solution across four distinct approximations: TCL2, TCL4, CG-LE ($=$C-LE), and RWA-LE. The comparison metrics comprise the trace-norm distance $\|\rho_{\text{exact}}-\rho_{\text{approx}}\|_1$, decay rate $\gamma$, Lamb shift $S$, population $\rho_{11}$, and coherence $\rho_{01}$, all with the Ohmic spectral density $J_2$, but with different values of the coupling $\eta$ and qubit frequency ratio $\Omega_0/\omega_c$. The equations plotted for each curve are listed in \cref{tab:summary}.

During the initial time intervals ($\o_c t \approx 1$), the TCL2 and TCL4 solutions generally exhibit much closer agreement with the exact solution than the Markovian CG-LE and RWA-LE, as evidenced especially by the trace-norm distance curves. The plots representing population and the real part of the coherence reveal that the TCL approximations aptly capture the Zeno effect observed in the exact solution -- characterized by a gradual concave decay at short times. This stands in contrast to the monotonic exponential decay exhibited by the Markovian approaches. 

Note that, as mentioned above in the discussion of \cref{RWA_S_gamma}, the TCL2 decay rate $\gamma$ gradually approaches the asymptotic behavior of the RWA-LE decay rate. Concerning the CG-LE, we note that it exhibits a smaller trace-norm distance from the exact solution compared to the RWA-LE, as anticipated in Ref.~\cite{Majenz:2013qw}. This is due to the aforementioned ability to optimize the coarse-graining time. Both Markovian approximations exhibit a constant decay rate and Lamb shift, but in the case of the CG-LE, the Lamb shift $S$ is zero, resulting in the coherence being purely real.

\subsubsection{Weak and strong coupling}

Recall that the RWA-LE needs to satisfy \cref{eq:RWA_ass_Ohmic}, in particular $\eta \ll 1$. In practice, we will use both weak ($\eta=0.5$) and strong ($\eta=1$) coupling. We thus expect the RWA-LE to be a relatively poor approximation in the later case, at least for relatively short evolutions.

\cref{fig:dynamics_a1,fig:dynamics_eta05} compare the results for weak and strong coupling. Using a qubit frequency equal to the cutoff frequency $\Omega_0/\omega_c=1$, \cref{fig:dynamics_a1} shows the strong coupling results, while \cref{fig:dynamics_eta05} shows the weak coupling results.

After rising first, the exact decay rate [\cref{eq:sandgamma-g}] for strong coupling ($\eta=1$) in \cref{fig:gam_a1} exhibits a negative trend and continues to exhibit non-monotonic behavior for even longer evolution times. Intervals during which $\gamma$ turns negative correspond to non-Markovian dynamics, leading to an increase in population, as evidenced in \cref{fig:popul_a1}. Consequently, during intermediate times, the approximation methods struggle to accurately match the exact solution, which displays a remarkably slow population decay. 

In contrast, for weak coupling ($\eta=1/2$), the decay rate in \cref{fig:gam_eta05} remains positive, causing the population and coherence to approach zero more rapidly. This aligns closely with the behavior exhibited by the approximation methods, as demonstrated in \cref{fig:popul_eta05,fig:Re_coher_eta05,fig:Im_coher_eta05}. This trend is also affirmed by the trace-norm distance plots in \cref{fig:traced_a1,fig:traced_eta05}, where weaker coupling results in a smaller trace-norm distance difference between the approximation methods and the exact solution.

\subsubsection{Weak and strong qubit frequency}

Fixing the coupling at $\eta=1$ [technically at the upper limit of \cref{eq:RWA_ass_Ohmic}], we compare the cases where the qubit frequency $\Omega_0$ is either less than or greater than the cutoff frequency $\omega_c$, as shown in \cref{fig:dynamics_a05,fig:dynamics_a4}, respectively. Recall that \cref{eq:RWA_ass_Ohmic} also imposes the validity condition $\Omega_0/\omega_c > 1$ on the RWA-LE, so we expect better agreement in \cref{fig:dynamics_a4}, as in indeed the case.

In more detail, from the trace-norm distance plots in \cref{fig:traced_a05,fig:traced_a4}, we observe that all the approximations exhibit closer agreement with the exact result when the qubit frequency is relatively high, especially at longer evolution times. This trend is also noticeable in the population (\cref{fig:popul_a05,fig:popul_a4}) and coherence decay (\cref{fig:Re_coher_a05,fig:Re_coher_a4,fig:Im_coher_a05,fig:Im_coher_a4}). 
For low qubit frequencies, as in \cref{fig:gam_a05}, the exact decay rate $\gamma$ exhibits non-Markovian negative phases. Correspondingly, the population in \cref{fig:popul_a05} increases initially and then stabilizes at a non-zero value. The approximations fail to match the exact solution's behavior for extended periods, with the TCL approximation being effective primarily during short times ($\o_c t < 1.5$). This behavior is characteristic of strong non-Markovian behavior, which is expected in models with non-flat spectral densities such as the ones considered here.

Conversely, for higher qubit frequencies in \cref{fig:gam_a4}, the behavior of the TCL decay rate resembles that of the exact solution. Even the exact Lamb shift in \cref{fig:S_a4} closely aligns with the approximation methods, with the notable exception of CG-LE (where the Lamb shift is zero). Consequently, in this scenario, the approximations reasonably reproduce the exact behavior, as demonstrated in \cref{fig:popul_a4,fig:Re_coher_a4,fig:Im_coher_a4}. The TCL4 approximation is particularly good according to all six of our metrics.

\begin{figure*}
$$\boxed{J_1(\omega)=|g|^2\delta(\omega-\omega_c) \qquad |g|=1 \qquad \Omega_0/\omega_c=1/2}$$
\begin{subfigure}{\textwidth}
\vspace{-0.3cm}
  \centering 
\includegraphics[width=0.6\linewidth]{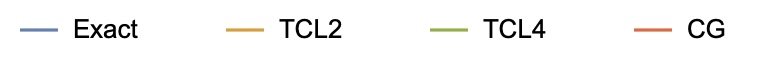}
\end{subfigure}%
\vspace{-0.5cm}
   \vskip\baselineskip
\begin{subfigure}{.33\textwidth}
  \centering \hspace{-1cm}
  \includegraphics[width=1\linewidth]{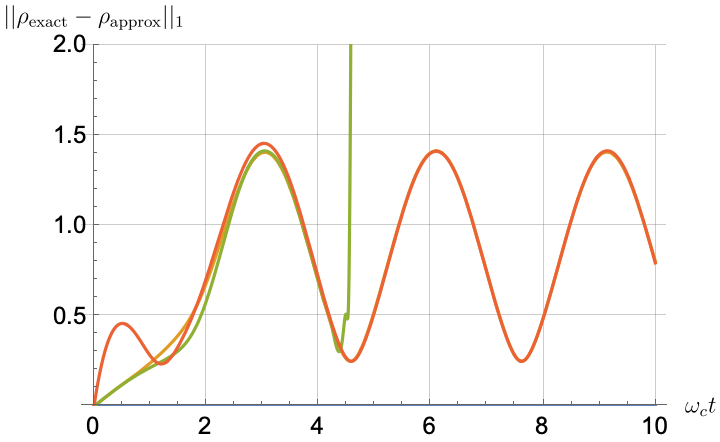}
  \caption{Trace-norm distance}
  \label{fig:traced_J1_a05}
\end{subfigure}%
\begin{subfigure}{.33\textwidth}
  \centering 
  \includegraphics[width=0.92\linewidth]{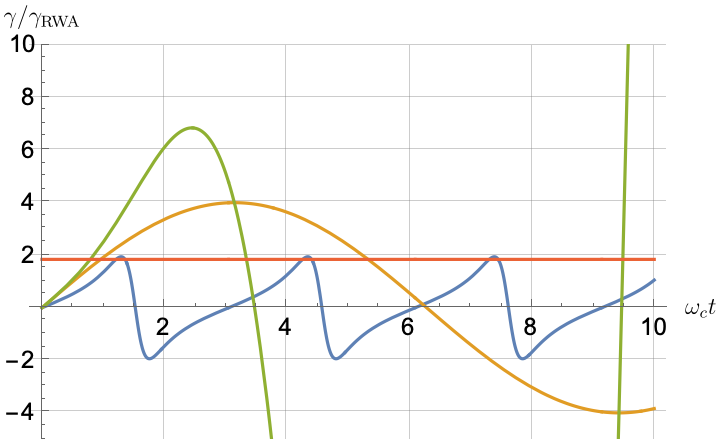}
  \caption{Decay rate}
  \label{fig:gam_J1_a05}
\end{subfigure}%
\begin{subfigure}{.33\textwidth}
  \centering 
\includegraphics[width=0.92\linewidth]{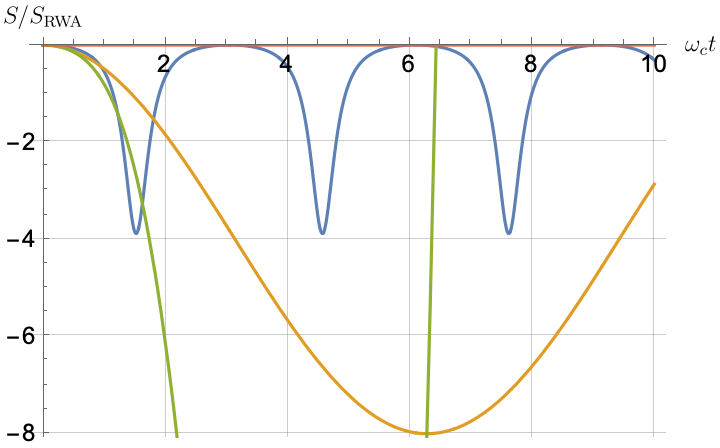}
  \caption{Lamb shift }
  \label{fig:S_J1_a05}
\end{subfigure}
   \vskip\baselineskip
   \vspace{-0.3cm}
\begin{subfigure}{.33\textwidth}
  \centering   \hspace{-1cm}
\includegraphics[width=0.92\linewidth]{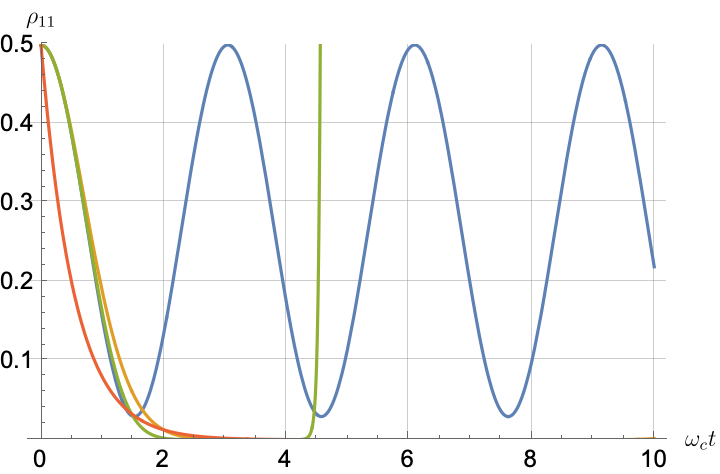}
  \caption{Population }
  \label{fig:popul_J1_a05}
\end{subfigure}%
\begin{subfigure}{.33\textwidth}
  \centering
\includegraphics[width=0.92\linewidth]{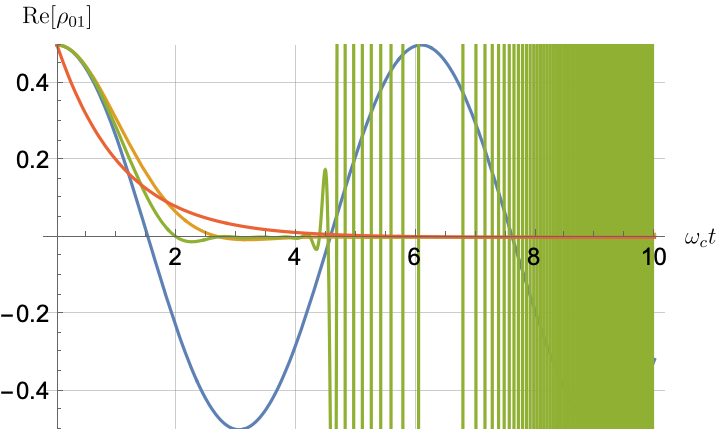}
  \caption{Coherence -- real part}
  \label{fig:Re_coher_J1_a05}
\end{subfigure}
\begin{subfigure}{.33\textwidth}
  \centering
  \includegraphics[width=0.92\linewidth]{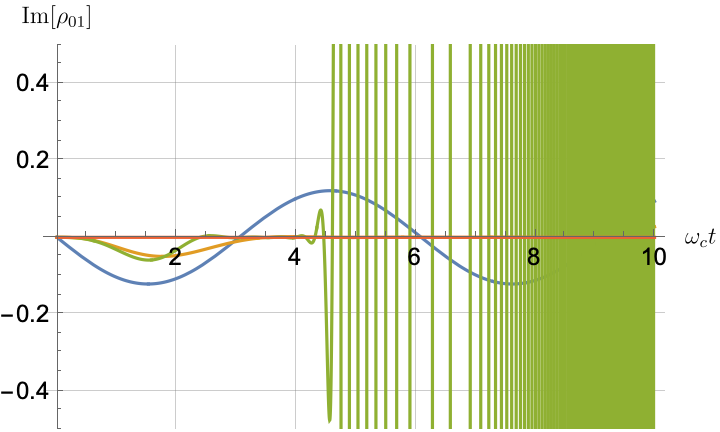}
  \caption{Coherence -- imaginary part}
  \label{fig:Im_coher_J1_a05}
\end{subfigure}
\caption{\raggedright
\small{Same as \cref{fig:dynamics_a1} but with an impulse bath spectral density $J_1(\omega)=|g|^2\delta(\omega-\omega_c)$ with coupling $|g|=1$ and a low qubit frequency: $\Omega_0/\omega_c=1/2$. The CG-LE coarse graining time is $\tau=7.197305/\omega_c$.}
}
 \label{fig:dynamics_J1_a05}
\end{figure*}

\begin{figure*}
$$\boxed{J_1(\omega)=|g|^2\delta(\omega-\omega_c) \qquad |g|=1 \qquad \Omega_0/\omega_c=2}$$
\begin{subfigure}{\textwidth}
\vspace{-0.3cm}
  \centering 
\includegraphics[width=0.6\linewidth]{figure28.jpg}
\end{subfigure}%
\vspace{-0.5cm}
   \vskip\baselineskip
\begin{subfigure}{.33\textwidth}
  \centering \hspace{-1cm}
  \includegraphics[width=1\linewidth]{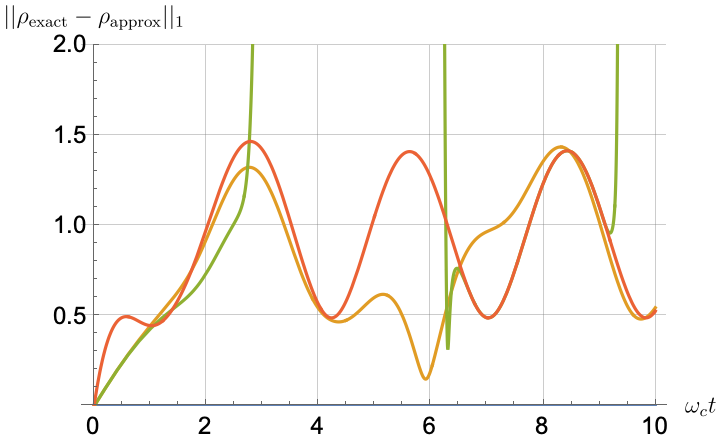}
  \caption{Trace-norm distance}
  \label{fig:traced_J1_a2}
\end{subfigure}%
\begin{subfigure}{.33\textwidth}
  \centering 
  \includegraphics[width=0.92\linewidth]{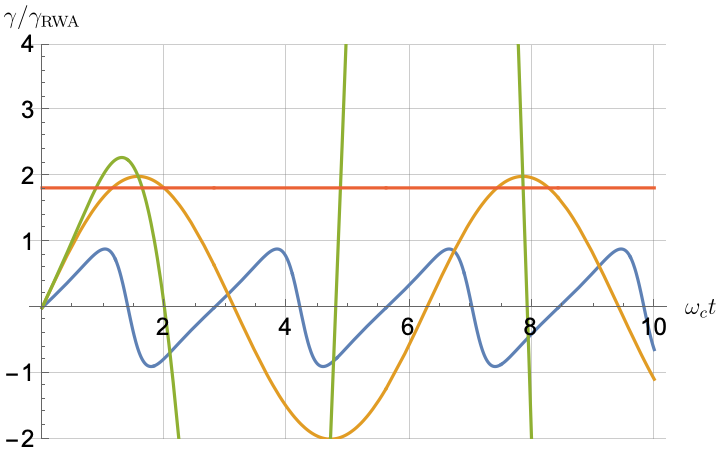}
  \caption{Decay rate}
  \label{fig:gam_J1_a2}
\end{subfigure}%
\begin{subfigure}{.33\textwidth}
  \centering 
\includegraphics[width=0.92\linewidth]{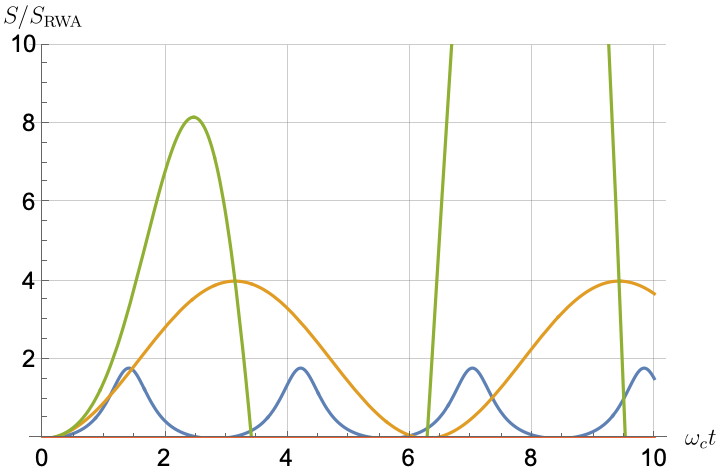}
  \caption{Lamb shift }
  \label{fig:S_J1_a2}
\end{subfigure}
   \vskip\baselineskip
   \vspace{-0.3cm}
\begin{subfigure}{.33\textwidth}
  \centering   \hspace{-1cm}
\includegraphics[width=0.92\linewidth]{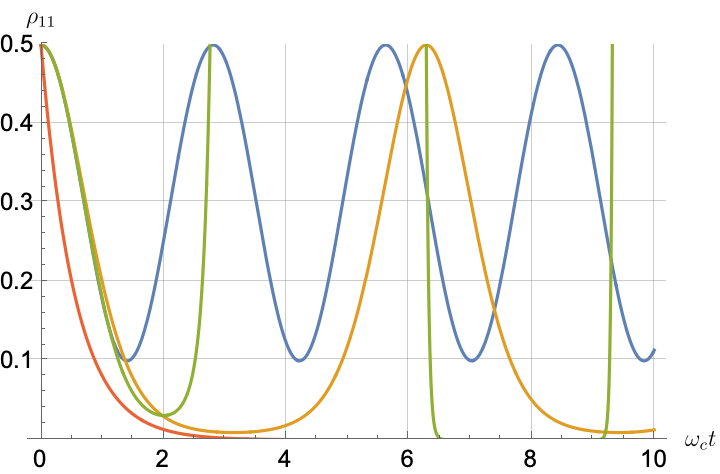}
  \caption{Population }
  \label{fig:popul_J1_a2}
\end{subfigure}%
\begin{subfigure}{.33\textwidth}
  \centering
\includegraphics[width=0.92\linewidth]{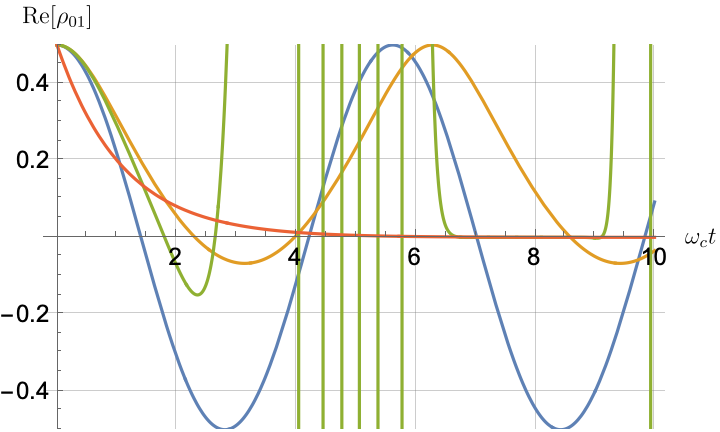}
  \caption{Coherence -- real part}
  \label{fig:Re_coher_J1_a2}
\end{subfigure}
\begin{subfigure}{.33\textwidth}
  \centering
  \includegraphics[width=0.92\linewidth]{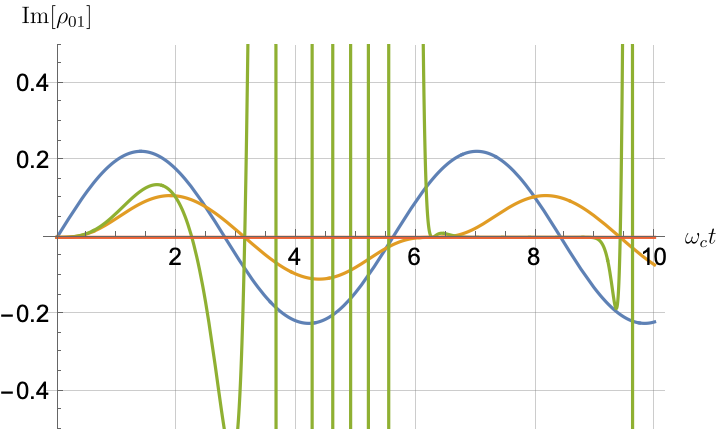}
  \caption{Coherence -- imaginary part}
  \label{fig:Im_coher_J1_a2}
\end{subfigure}
\caption{\raggedright
\small{Same as \cref{fig:dynamics_a1} but with an impulse bath spectral density $J_1(\omega)$ with coupling $|g|=1$ and a high qubit frequency: $\Omega_0/\omega_c=2$. The CG-LE coarse graining time is $\tau=3.609881/\omega_c$.}
}
 \label{fig:dynamics_J1_a2}
\end{figure*}

\begin{figure*}
$$\boxed{J_3(\omega)=\eta\Theta(\omega_c-\omega) \qquad \eta=1 \qquad \Omega_0/\omega_c=1/2}$$
\begin{subfigure}{\textwidth}
\vspace{-0.3cm}
  \centering 
\includegraphics[width=0.6\linewidth]{figure3.png}
\end{subfigure}%
\vspace{-0.5cm}
   \vskip\baselineskip
\begin{subfigure}{.33\textwidth}
  \centering \hspace{-1cm}
  \includegraphics[width=1\linewidth]{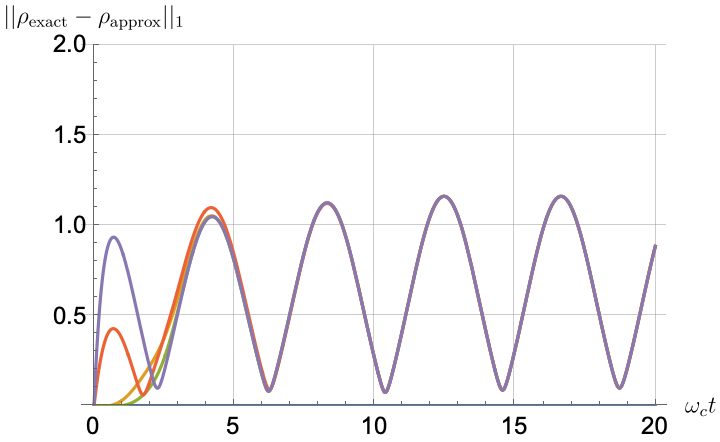}
  \caption{Trace-norm distance}
  \label{fig:traced_J3_a05}
\end{subfigure}%
\begin{subfigure}{.33\textwidth}
  \centering 
  \includegraphics[width=0.92\linewidth]{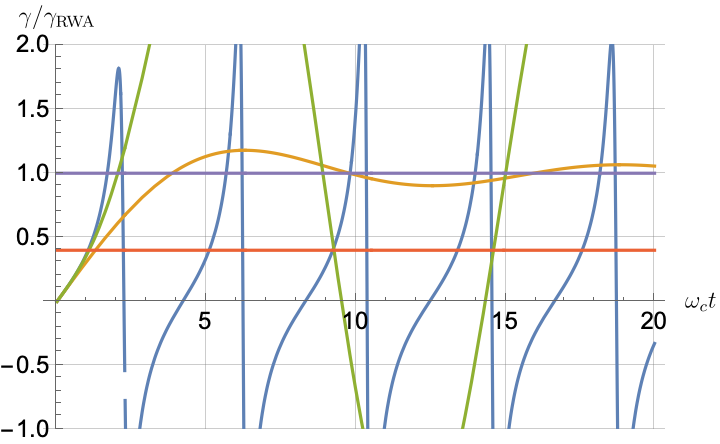}
  \caption{Decay rate}
  \label{fig:gam_J3_a05}
\end{subfigure}%
\begin{subfigure}{.33\textwidth}
  \centering 
\includegraphics[width=0.92\linewidth]{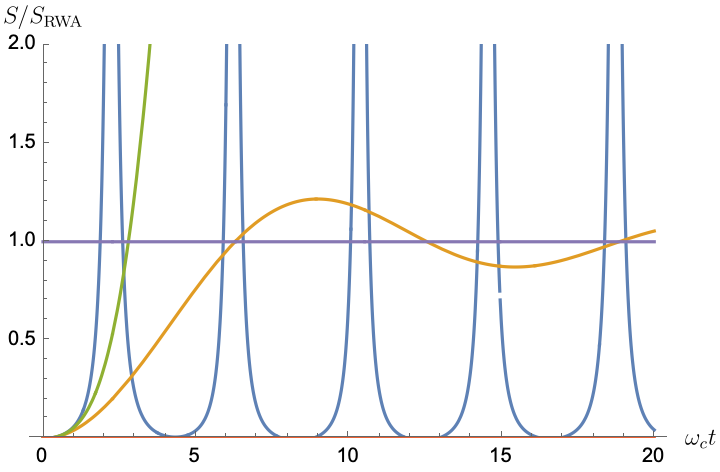}
  \caption{Lamb shift }
  \label{fig:S_J3_a05}
\end{subfigure}
   \vskip\baselineskip
   \vspace{-0.3cm}
\begin{subfigure}{.33\textwidth}
  \centering   \hspace{-1cm}
\includegraphics[width=0.92\linewidth]{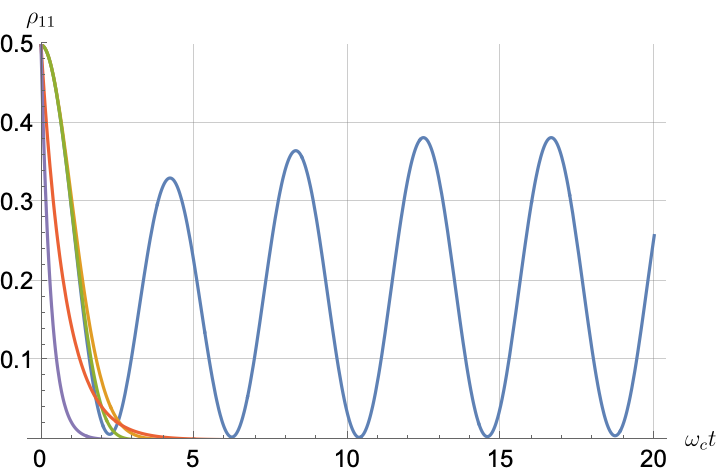}
  \caption{Population }
  \label{fig:popul_J3_a05}
\end{subfigure}%
\begin{subfigure}{.33\textwidth}
  \centering
\includegraphics[width=0.92\linewidth]{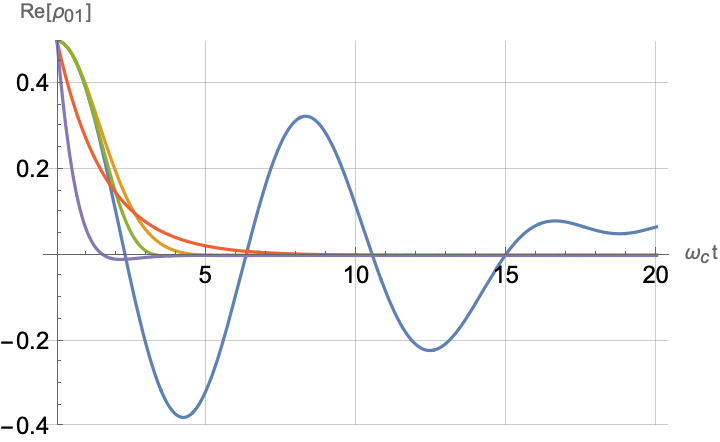}
  \caption{Coherence -- real part}
  \label{fig:Re_coher_J3_a05}
\end{subfigure}
\begin{subfigure}{.33\textwidth}
  \centering
  \includegraphics[width=0.92\linewidth]{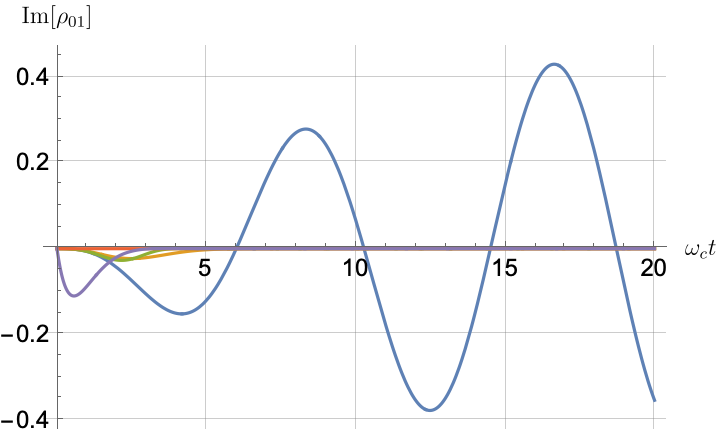}
  \caption{Coherence -- imaginary part}
  \label{fig:Im_coher_J3_a05}
\end{subfigure}
\caption{\raggedright
\small{Same as \cref{fig:dynamics_a1} but with a triangular bath spectral density $J_3(\omega)=\eta\omega\Theta(\omega_c-\omega)$ with coupling $\eta=1$ and a low qubit frequency: $\Omega_0/\omega_c=1/2$. The CG-LE coarse graining time is $\tau=2.631777/\omega_c$.}
}
 \label{fig:dynamics_J3_a05}
\end{figure*}

\begin{figure*}
$$\boxed{J_3(\omega)=\eta\Theta(\omega_c-\omega) \qquad \eta=1 \qquad \Omega_0/\omega_c=1.8}$$
\begin{subfigure}{\textwidth}
\vspace{-0.3cm}
  \centering 
\includegraphics[width=0.6\linewidth]{figure3.png}
\end{subfigure}%
\vspace{-0.5cm}
   \vskip\baselineskip
\begin{subfigure}{.33\textwidth}
  \centering \hspace{-1cm}
  \includegraphics[width=1\linewidth]{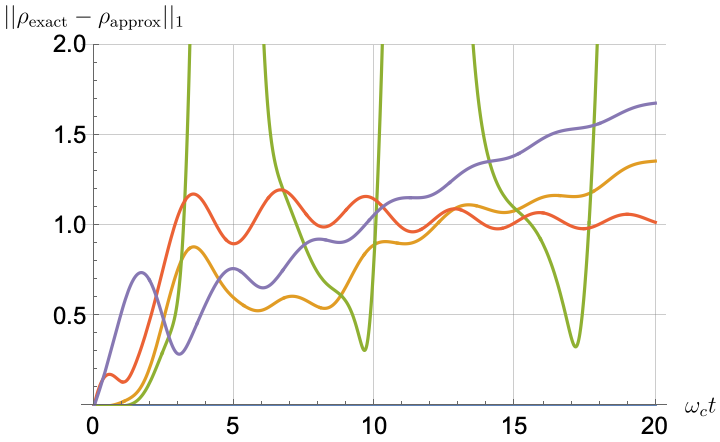}
  \caption{Trace-norm distance}
  \label{fig:traced_J3_a18}
\end{subfigure}%
\begin{subfigure}{.33\textwidth}
  \centering 
  \includegraphics[width=0.92\linewidth]{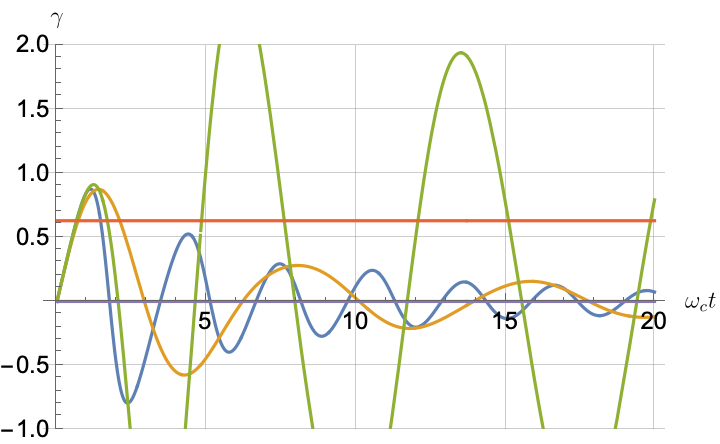}
  \caption{Decay rate}
  \label{fig:gam_J3_a18}
\end{subfigure}%
\begin{subfigure}{.33\textwidth}
  \centering 
\includegraphics[width=0.92\linewidth]{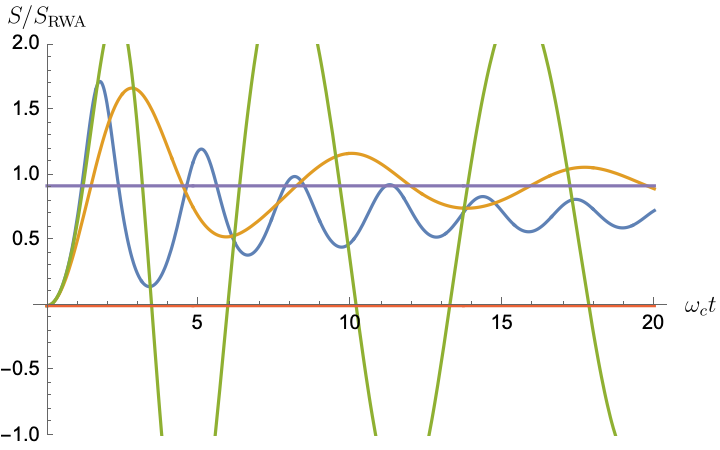}
  \caption{Lamb shift }
  \label{fig:S_J3_a18}
\end{subfigure}
   \vskip\baselineskip
   \vspace{-0.3cm}
\begin{subfigure}{.33\textwidth}
  \centering   \hspace{-1cm}
\includegraphics[width=0.92\linewidth]{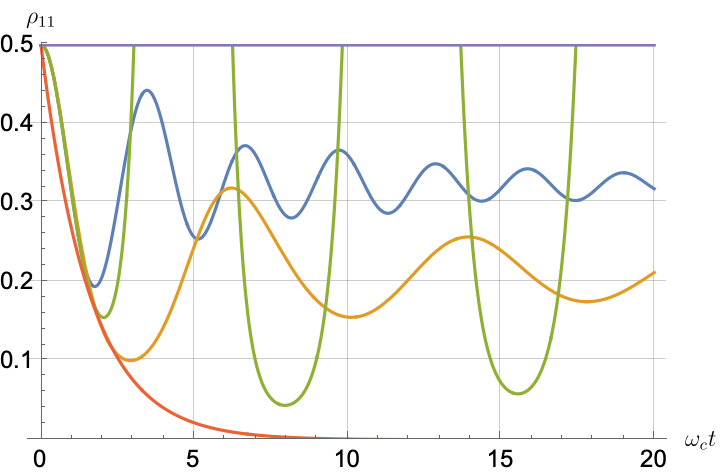}
  \caption{Population }
  \label{fig:popul_J3_a18}
\end{subfigure}%
\begin{subfigure}{.33\textwidth}
  \centering
\includegraphics[width=0.92\linewidth]{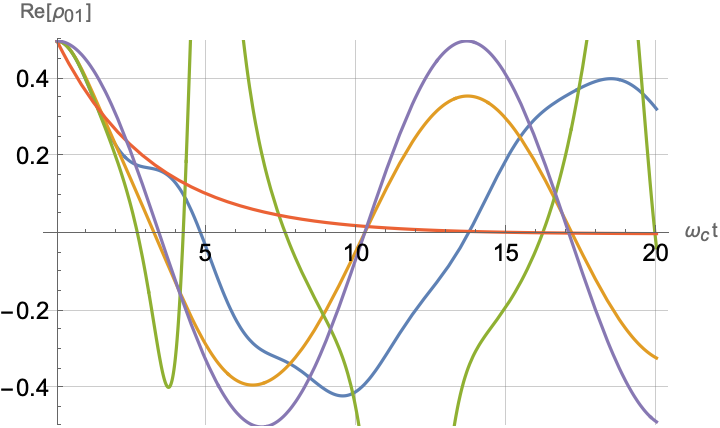}
  \caption{Coherence -- real part}
  \label{fig:Re_coher_J3_a18}
\end{subfigure}
\begin{subfigure}{.33\textwidth}
  \centering
\includegraphics[width=0.92\linewidth]{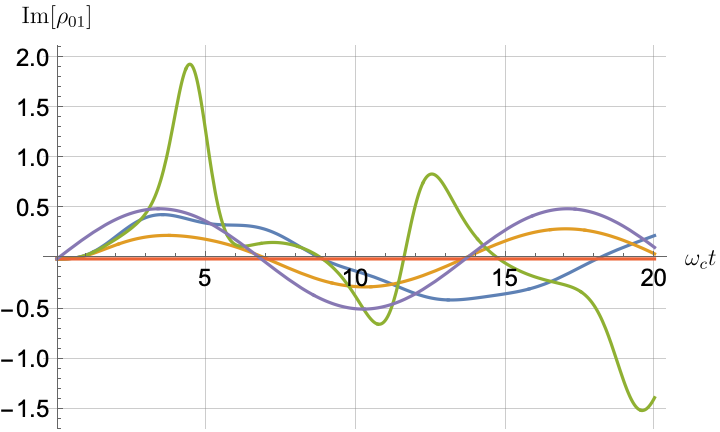}
  \caption{Coherence -- imaginary part}
  \label{fig:Im_coher_J3_a18}
\end{subfigure}
\caption{\raggedright
\small{Same as \cref{fig:dynamics_a1} but with a triangular bath spectral density $J_3(\omega)=\eta\omega\Theta(\omega_c-\omega)$ with coupling $\eta=1$ and a high qubit frequency: $\Omega_0/\omega_c=1.8$. The CG-LE coarse graining time is $\tau=1.8629886/\omega_c$. Note that in panel (b), the RWA-LE decay rate is zero due to \cref{eq:gamma-J3-RWA-LE-gamma}.}
}
 \label{fig:dynamics_J3_a18}
\end{figure*}

\subsection{Exact solution \textit{vs} TCL and Markov approximations for the impulse spectral density}
\label{sec:compare_JI}

For the impulse spectral density $J_1(\omega)=|g|^2\delta(\omega-\omega_c)$, the exact solution for the population $|c_1(t)|^2$ of the excited population [see \cref{eq:exactJ1}] is purely periodic. Consequently, no Markovian approximation with exponential decay can accurately capture the exact solution. This renders the RWA-LE inadequate for the description, as remarked in the discussion of \cref{eq:gamma-J1-RWA-LE-gamma}. However, since the CG-LE retains the coarse-graining time-scale as an optimization parameter, we use it to understand how close a Markovian approximation can still be to the exact solution in this rather extreme non-Markovian case, even though we have already concluded that the validity condition of the CG-LE cannot be satisfied [recall the discussion in \cref{sec:compare_2}].

\cref{fig:dynamics_J1_a05,fig:dynamics_J1_a2} depict the time evolution of the same metrics as in the previous section but for the impulse spectral density $J_1$ with two different qubit frequencies: $\Omega_0/\omega_c=1/2$ and $\Omega_0/\omega_c=2$. For this spectral density, we have closed analytical expressions for the TCL Lamb shift [see \cref{eq:TCL2-J1-LS,eq:TCL4-J1-LS}] and decay rate [see \cref{eq:TCL2-J1-gamma,eq:TCL4-J1-gamma}]. In general, we observe that as expected, the CG-LE is a poor approximation to both the short-time and longer-time oscillatory behavior of the decay rate, population, and coherence. 

In contrast, the TCL2 and TCL4 approximations in \cref{eq:TCL2-J1-gamma,eq:TCL4-J1-gamma} are fairly accurate for short-time dynamics. However, the populations and coherences under TCL4 deviate rapidly from the exact solution for longer times, as demonstrated in \cref{fig:popul_J1_a05,fig:popul_J1_a2,fig:Re_coher_J1_a05,fig:Im_coher_J1_a05,fig:Re_coher_J1_a2,fig:Im_coher_J1_a2}. The deviation is accentuated by the recurrent intervals of negative decay rates in the exact solution and TCL approximations. This is especially noticeable in TCL4, where the oscillations of the decay rates increase with time, taking on higher negative values, as seen in \cref{fig:gam_J1_a05,fig:gam_J1_a2}. Consequently, TCL4 develops an instability reflected in the diverging oscillation frequency of the coherence seen in \cref{fig:Re_coher_J1_a05,fig:Im_coher_J1_a05}. An explanation of the breakdown of the TCL approximation is given in \cref{app:Breakdown_TCL}.

When comparing the low qubit frequency case ($\Omega_0/\omega_c=1/2$) in \cref{fig:popul_J1_a05,fig:Re_coher_J1_a05,fig:Im_coher_J1_a05} with the high qubit frequency case ($\Omega_0/\omega_c=2$) in \cref{fig:popul_J1_a2,fig:Re_coher_J1_a2,fig:Im_coher_J1_a2}, a distinction emerges. In the former, TCL2 exhibits a decay to zero similar to CG-LE, while in the latter, TCL2 mirrors the oscillatory behavior of the exact solution. This observation underscores the higher accuracy of the TCL approximation as the qubit frequency increases. However, while TCL4 is a better approximation at short times $\o_c t < 1.5$, it is worse than TCL2 at long times for the impulse spectral density.

\subsection{Exact solution \textit{vs} TCL and Markov approximations for the triangular spectral density}
\label{sec:compare_J3}

The triangular spectral density $J_3(\omega)=\eta \omega \Theta(\omega_c-\omega)$ is intermediate between the impulse density $J_1(\omega)$ and the Ohmic density $J_2(\omega)$ because it matches $J_2(\omega)$ for low frequencies and drops to zero for frequencies exceeding the cutoff. 
In \cref{fig:dynamics_J3_a05,fig:dynamics_J3_a18}, we perform a comparison for $\Omega_0/\omega_c=1/2$ and $\Omega_0/\omega_c=1.8$. Recall that here too, the validity condition of the CG-LE cannot be satisfied, as discussed in \cref{sec:compare_2}.

Similar to the impulse case $J_1$, for $J_3$, we observe a strong non-Markovian oscillatory behavior in the exact solution's population and coherence, in \cref{fig:popul_J3_a05,fig:Re_coher_J3_a05,fig:Im_coher_J3_a05,fig:popul_J3_a18,fig:Re_coher_J3_a18,fig:Im_coher_J3_a18}, respectively. The decay rate (as seen in \cref{fig:gam_J3_a05,fig:gam_J3_a18}) oscillates between positive and negative values.

As previously discussed, the TCL approximation is capable of capturing the quantum Zeno effect, which leads to a better agreement with the exact solution than the Markovian approximation does for short evolution times. However, a significant difference emerges in the exact decay rate between low and high qubit frequencies, as shown in \cref{fig:gam_J3_a05,fig:gam_J3_a18}. In the latter case, the decay rate oscillates and converges to zero (equivalent to the rate of the Markov approximations). Conversely, in the former case, the decay rate exhibits discontinuous behavior, rendering the approximation methods unsuitable for fitting the exact solution. In general, when $\Omega_0<\omega_c$, the discontinuous behavior of the decay rate leads to oscillatory population dynamics, where the population first decays to zero, then revives, and subsequently decays again. However, for $\Omega_0>\omega_c$, the decay rate converges to zero, resulting in non-decaying population dynamics, similar to what was observed for $J_1$ in \cref{fig:popul_J1_a2}.

Furthermore, as depicted in \cref{fig:S_J3_a05}, the exact Lamb shift $S$ diverges when $\Omega_0<\omega_c$ in contrast to the zero Lamb shift in the CG-LE, the constant non-zero $S$ in the RWA-LE, and TCL2 (which converges to RWA). However, TCL4 diverges in an attempt to match the exact $S$. In contrast, for the Lamb shift $S$ when $\Omega_0>\omega_c$ (as shown in \cref{fig:S_J3_a18}), the TCL approximations align with the oscillatory behavior of the exact solution. Even the constant RWA-LE approximation closely resembles the asymptotic behavior. However, the CG solution maintains a zero Lamb shift.

Moving on to the population and coherence in $\Omega_0<\omega_c$, as illustrated in \cref{fig:popul_J3_a05,fig:Re_coher_J3_a05,fig:Im_coher_J3_a05}, we observe that the approximations rapidly decay to zero, matching the exact solution at short times. However, the oscillatory behavior at later times remains beyond the reach of the approximations, similar to what was observed for $J_1$ in \cref{fig:popul_J1_a05}.

In contrast, for the population and coherence at $\Omega_0>\omega_c$, as depicted in \cref{fig:popul_J3_a18,fig:Re_coher_J3_a18,fig:Im_coher_J3_a18}, the TCL approximation exhibits an oscillatory and non-decaying behavior, akin to the exact solution, matching the Zeno effect at short times. TCL4, in particular, matches the exact solution well up to $\o_c t \approx 2$. The CG-LE approximation performs much better than RWA-LE for short times in describing the excited state population, which remains constant in the latter due to its zero decay rate [as can be seen from \cref{eq:gamma-J3-RWA-LE-gamma}]. The two are comparable in describing the coherence. For long times, the Markov approximations, which display monotonic decay, fail to capture the exact behavior. CG-LE, which aims to minimize the trace-norm distance with the exact solution, decays slowly to achieve its objective.

In summary, for the triangular spectral density, when $\Omega_0>\omega_c$, the non-Markovian oscillatory behavior is captured qualitatively by the TCL approximation, as opposed to CG or RWA. In contrast, when $\Omega_0<\omega_c$, all approximations exhibit rapid decay and fail to accurately capture the exact solution.

\section{Summary and conclusions}
\label{sec:summary}

In this work, we studied the dynamics of a qubit in a cavity interacting with bosonic baths described by three different spectral densities: impulse, Ohmic, and triangular. The model is exactly solvable within the single-excitation subspace, and this allowed us to perform a comprehensive comparison to a number of different master equations, both Markovian (CG-LE, C-LE, and RWA-LE) and non-Markovian (TCL2 and TCL4). Of the three spectral densities, the Ohmic model is the most physically relevant, and the other two were introduced primarily to allow us to reach closed-form analytical results, as well as to study extreme non-Markovian dynamics. 

For weak coupling and a large qubit frequency, the Ohmic case leads to a quasi-exponential decay, characteristic of the Markovian limit. In this regime, therefore, we found that the purely Markovian master equations are able to approximate the exact solution rather well. Outside of this regime, in particular also for the impulse and triangular spectral densities, the Markovian master equations perform poorly, but we found that TCL is still relatively accurate, in particular for short evolution times. The TCL is also able to capture distinctly non-Markovian features such as bath-induced population and coherence oscillations.

Within the regime of validity of the Markovian master equations, we found the CG-LE and C-LE to be better approximations than the standard RWA-LE. This was achieved by optimizing the coarse-graining time to minimize the difference between the CG-LE and the exact solution.

Overall, this work shows that low-order quantum master equations can be accurate when operated in their guaranteed regime of validity (short evolution times, in particular), but significant caution must be exercised in trusting their predictions outside of these regimes, as they can dramatically deviate from the exact dynamics. The TCL approach stands out as significantly more accurate than the Markovian master equations, even when the latter are fine-tuned via the optimization of a free parameter such as the coarse-graining time. The standard Lindblad equation based on the rotating wave approximation is particularly suspect.

Future research may wish to address---within the context of the same analytically solvable model as studied here, or similar analytically solvable models---the accuracy of TCL at higher orders, as well as a variety of other master equations, such as the phenomenological post-Markovian master equation~\cite{ShabaniLidar:05,Haimeng:PMME}, Floquet-based master equations for periodic driving~\cite{PhysRevA.73.052311,Hartmann:2017aa}, time-dependent Markovian quantum master equations~\cite{Haddadfarshi:2015aa,Yamaguchi:2017vu,Dann:2018aa,Mozgunov:2019aa,dimeglio2023time}, the universal Lindblad equation~\cite{nathan2020universal}, the geometric-arithmetic master equation~\cite{Davidovic2020completelypositive,Davidovic:2022aa}, regularized cumulant-based master equations~\cite{winczewski2021bypassing}, as well as various adiabatic master equations~\cite{ABLZ:12-SI,Yip:2018aa,Venuti:2018aa}. Another interesting direction worth considering is to incorporate more general initial states, such as a hermal Gibbs state, a coherent state~\cite{Ahmadi:2024} or a squeezed initial bath state.

\acknowledgments
This research was supported by the ARO MURI grant W911NF-22-S-0007, by the National Science Foundation the Quantum Leap Big Idea under Grant No. OMA-1936388, and by the Defense Advanced Research Projects Agency (DARPA) under Contract No. HR001122C0063.
We thank Prof. Todd Brun for useful discussions and Dawei Zhong for providing the analysis given in \cref{app:Dawei}.

\appendix

\section{Detailed exact solution}
\label{app:deriv_exact_sol}

Considering the Hamiltonian in \cref{eq:general_H}, the general solution for the $1$-excitation subspace is given by \cref{eq:state_t}. This solution can be expressed as a linear combination of the basis eigenvectors $\{\ket{\psi}_0,\ket{\psi}_1,\ket{\varphi_k}\}$, which are tensor products of the qubit system basis $\{\ket{0},\ket{1}\}$ and the $1$-excitation bath basis $\{\ket{v},\ket{k}\}_{k\in 1,2,..}$, as shown in \cref{eq:substates}. Here, $\ket{v}$ denotes the vacuum state with no photons, and $\ket{k}=b_k^{\dagger}\ket{v}$ represents the state with one photon in mode $k$. The coefficients of the linear combination satisfy the normalization condition given in \cref{eq:norm_cond}. For the evolution of the closed system within the $1$-excitation subspace, we utilize the Schr\"odinger equation with the interaction Hamiltonian in \cref{eq:HSBt}. Using
\bes
\begin{align}
\sigma_{-}(t)\ket{0}&=\sigma_{+}(t)\ket{1}=B(t)\ket{v}=0\\
\sigma_{+}(t)\ket{0}&=e^{i\Omega_0 t}\ket{1}\ ,\quad 
\sigma_{-}(t)\ket{1}=e^{-i\Omega_0 t}\ket{0}  \\
B(t)\ket{k}&=g_ke^{-i\omega_k t}\ket{v}\ ,\quad B^{\dagger}(t)\ket{v}=\sum_k g_k^*e^{i\omega_k t}\ket{k}\ ,
\end{align}
\ees
this leads to \cref{eq:closed_evol}:
\bes
\begin{align}
    i \dot{\ket{\phi}}
    &=\dot{c}_0(t)\ket{\psi_0}+\dot{c}_1(t)\ket{\psi_1}+\sum_k \dot{\mathfrak{c}}_k(t)\ket{\varphi_k}\\
    &=\lambda \tilde{H}_{SB}\ket{\phi(t)}\\
    &=\lambda[\sigma_+(t)\otimes B(t)+\sigma_-(t)\otimes B^{\dagger}(t)]\times\notag\\
    &\Big(c_0(t)\ket{\psi_0}+c_1(t)\ket{\psi_1}+\sum_k \dot{\mathfrak{c}}_k(t)\ket{\varphi_k}\Big)\\
    &=\lambda \Big(\sigma_+(t)\ket{0}\otimes B(t)\sum_k \mathfrak{c}_k(t) \ket{k}+\notag\\
&  \qquad\qquad c_1(t)\sigma_-(t)\ket{1}\otimes B^{\dagger}(t)\ket{v} \Big)\\
&=\lambda \Big(\sum_k g_k \mathfrak{c}_k(t)e^{i(\Omega_0-\omega_k) t}\ket{\psi_1}+\notag\\
&\qquad\qquad \sum_k g_k^*c_1(t)e^{-i(\Omega_0-\omega_k)t}\ket{\varphi_k}\Big)
\end{align}
\ees
We can obtain the joint system-bath density matrix using \cref{eq:state_t}:
\begin{align}\label{eq:rhoSBtotal}
    &\rho_{SB}(t)=\left(|c_0|^2\ketb{0}{0}+c_0c_1^*\ketb{0}{1}+c_1c_0^*\ketb{1}{0}\right.\nonumber\\
&\left.+|c_1|^2\ketb{1}{1}\right)\otimes \ketb{v}{v}+ \ketb{0}{0}\otimes \left(\sum_{j,k}\mathfrak{c}_j\mathfrak{c}_k^*\ketb{j}{k}\right.\nonumber\\&\left.+\sum_k (c_0\mathfrak{c}_k^*\ketb{v}{k}+c_0^*\mathfrak{c}_k\ketb{k}{v})\right)+\sum_{k}c_1\mathfrak{c}_k^*\ketb{1}{0}\nonumber\\
&\otimes \ketb{v}{k}+\sum_k \mathfrak{c}_k c_1^*\ketb{0}{1}\otimes \ketb{k}{v}\ .
\end{align}
Taking the partial trace over the bath, we obtain the system state in the matrix form shown in \cref{eq:state_system}. Its time derivative is given in \cref{eq:der_rho}. If we write the system density matrix in terms of the populations $\rho_{00}, \rho_{11}$ and its coherences $\rho_{01}, \rho_{10}$, we have:
\bes
\begin{align}
    \rho_{11}(t)&=|c_1(t)|^2=1-\rho_{00}(t)\\
    \rho_{01}(t)&=c_0\dot{c}^*_1(t)=\rho_{10}^*(t)\ .
\end{align}
\ees
Explicitly substituting the system density matrix into the ansatz \cref{eq:ansatz} along with \cref{eq:Ks_gen}, we obtain:
\begin{align}
\mathcal{K}_S(t)\rho =\left(\begin{array}{cc}
        \gamma(t)\rho_{11}& \frac{1}{2}(i\mathcal{S}(t)-\gamma(t))\rho_{01} \\
      \frac{1}{2}(-i\mathcal{S}(t)-\gamma(t))\rho_{10} &-\gamma(t)\rho_{11}
    \end{array}\right)
\end{align}
Comparing this with \cref{eq:der_rho} gives us the following differential equations:
\bes
\begin{align}
\partial_t |c_1(t)|^2 &= -\gamma(t)|c_1(t)|^2\\
c_0^* \dot{c}_1(t)&=-\frac{1}{2}\left(i\mathcal{S}(t)+\gamma(t)\right)c_0^* c_1(t)\ .
\end{align}
\ees

From the first equation, we obtain the population evolution \cref{eq:rho11}, and from the second one, we can take its real and imaginary parts to obtain \cref{eq:sandgamma}.

\section{Simplification of the decay rate expressions for the CG-LE}
\label{app:CG-LE-integrals}

\subsection{Ohmic spectral density}
\label{app:CG-LE-integrals-J2}

Starting from \cref{eq:gamma-J2-CG-LE} we have:
\bes
\begin{align}
\label{eq:gamma-J2-CG-LE-2}
    \g(\tau) &=  \int_0^\infty \eta\omega e^{-\omega/\omega_c}\tau \sinc^2\left(\frac{(\Omega_0-\omega)\tau}{2}\right)d\omega  \\
&=\frac{2}{\tau} \eta \int_{-\Omega_0}^\infty d\nu \frac{(\nu+\Omega_0)e^{-(\nu+\Omega_0)/\omega_c}}{\nu^2}(1-\cos\nu \tau)\ ,
\end{align}
\ees
where we used the change of variables $\nu=\omega-\Omega_0$.
This integral can be split into two parts. By using  the exponential integral function \cref{eq:Ei}
we have: 
\begin{align}
&\int_{-\Omega_0}^\infty d\nu \frac{e^{-\nu/\omega_c}}{\nu}(1-\cos\nu \tau) = -\Ei\left(\frac{\Omega_0}{\omega_c}\right) \notag\\
&\quad +\frac{1}{2}\left[\Ei\left(\frac{\Omega_0}{\omega_c}+i\Omega_0 \tau\right)+
\Ei\left(\frac{\Omega_0}{\omega_c}-i\Omega_0 \tau\right)\right]\ .
\end{align}
For the second integral, we can utilize integration by parts. Setting $u= e^{-\nu/\omega_c}(1-\cos\nu \tau)$ and $dv =  \nu^{-2} d\nu$, we have:
\begin{align}
&\int_{-\Omega_0}^\infty u dv=
-\frac{e^{\Omega_0/\omega_c}}{\Omega_0}(1-\cos \Omega_0 \tau)\nonumber\\
&+\int_{-\Omega_0}^{\infty}d\nu \frac{e^{-\nu/\omega_c}}{\nu}\left(\tau\sin \nu \tau-\frac{1}{\omega_c}(1-\cos\nu \tau)\right)\ .
\end{align}
Since 
\begin{align}
&\int_{-\Omega_0}^{\infty} d\nu \frac{e^{-\nu/\omega_c}}{\nu}\tau\sin \nu \tau 
=\\ \notag
&
-\frac{i}{2}\tau\left[\Ei\left(\frac{\Omega_0}{\omega_c}+i\Omega_0 \tau\right)
-\Ei\left(\frac{\Omega_0}{\omega_c}-i\Omega_0 \tau\right)\right]\ ,
\end{align}
we can combine all the terms, and arrive at the final expression for $\g(\tau)$ given in \cref{eq:gamma-J2-CG-LE-3}.

\subsection{Triangular spectral density}
\label{app:CG-LE-integrals-J3}

Starting from \cref{eq:gamma-J3-CG-LE} we have:
\bes
\begin{align}
\label{eq:gamma-J3-CG-LE-2}
    \g(\tau)&=
     \int_0^{\o_c} \eta\omega \tau \sinc^2\left(\frac{(\Omega_0-\omega)\tau}{2}\right)d\omega  \\
    &=\frac{2 }{\tau} \int_{-\Omega_0}^{\omega_c-\Omega_0} d\nu \frac{\eta(\nu+\Omega_0)}{\nu^2}(1-\cos\nu \tau)\ . 
\end{align}
\ees
where we again used the change of variables $\nu=\omega-\Omega_0$. Now for $\Omega_0<\omega_c$, we have:
\begin{align}
&\int_{-\Omega_0}^{\omega_c-\Omega_0} d\nu \frac{1-\cos\nu \tau}{\nu}=\notag\\
&\ \ \ \ln \left(\frac{\omega_c-\Omega_0}{\Omega_0}\right)-\Ci((\omega_c-\Omega_0)\tau)+\Ci(\Omega_0\tau)\ ,
\end{align}
where the sine and cosine integral functions are given in \cref{eq:Si,eq:Ci}, respectively. 
The second term can be obtained by using integration by parts:
\begin{align}
\int_{-\Omega_0}^{\omega_c-\Omega_0}d\nu \frac{1-\cos\nu \tau}{\nu^2}&=-\frac{1-\cos\nu \tau}{\nu}\bigg|_{-\Omega_0}^{\omega_c-\Omega_0}\notag \\
&+\tau\int_{-\Omega_0}^{\omega_c-\Omega_0} \frac{\sin (\nu \tau)}{\nu}\ .
\end{align}
The last term is the sine integral function. Combining, we arrive at the total rate as given by \cref{eq:gamma-J3-CG-LE-3}.

\section{Why the first order cumulant $K^{(1)}$ in the C-LE can be made to vanish}
\label{app:shift}

Let $\ave{B} = \Tr(\r_B B)$ and define a new, shifted bath operator:
\beq
B' \equiv B - \ave{B} I_B\ .
\label{eq:XIV.25}
\eeq
Its expectation value vanishes:
\beq
\ave{B'} = \ave{B} - \ave{B}\ave{I_B} =  0\ .
\label{eq:429}
\eeq
The corresponding bath interaction picture operator is:
\begin{align}
B'(t) &= U_B^\dag(t) B' U_B(t) = U_B^\dag(t) (B - \ave{B} I_B) U_B(t) \notag \\
&= B(t)-\ave{B}I_B\ ,
\label{eq:B'-shifted}
\end{align}
where, as before, $U_B(t) = e^{-iH_B t}$.
Assuming stationarity $[H_B,\r_B(0)]=0$ immediately implies $[U_B(t),\r_B]=0$. In this case also $\ave{B'(t)}=0$, since then:
\bes
\label{eq:aveB't=0}
\begin{align}
\ave{B'(t)} &= \Tr[U_B^\dag(t) B' U_B(t)\r_B] \\
&= \Tr[B' U_B(t)\r_B U_B^\dag(t)] = \ave{B'} = 0\ .
\end{align}
\ees
Let 
\beq
H'_{SB} \equiv A\ox B'\ , \quad \Delta H_S' \equiv \ave{B} A \ , \quad H'_S \equiv H_S + \D H'_S \ .
\eeq
Correspondingly, the system-bath interaction can be written as 
\beq
H_{SB} = H_{SB}'+A\ox(B-B') = H_{SB}' + \D H_S'\ox I_B \ .
\eeq
Thus, we can write the full Hamiltonian as:
\begin{subequations}
\begin{align}
H &= H_S\ox I_B + H_{SB} +I_S\ox H_B \\
&= H'_0 + H'_{SB}\ , \quad H'_0 \equiv H'_S\ox I_B +I_S \ox H_B\ ,
\end{align}
\end{subequations}
where $H'_0$ defines a new, shifted interaction picture Hamiltonian.

Correspondingly, in this new interaction picture $\tilde{H}'(t) = U_0^{\prime\dag} (t) H'_{SB} U'_0(t) = A(t)\ox B'(t)$, and we find, using \cref{eq:aveB't=0}:
\begin{align}
\Tr_B [ \tilde{H}'(t) , \rho_{SB}(0) ]&= \Tr_B [ A(t)\ox B'(t) , \r_S(0)\ox \rho_{B}(0) ]\notag \\
&= \ave{B'(t)}[A(t),\r_S(0)] = 0\ .
\end{align}
Therefore, $K^{\prime(1)} (t) \rho(0) = 0$, with $K^{\prime(1)}$ defined within the shifted interaction picture and with the modified system-bath interaction $H'_{SB}$. 

The extension to the case when $H_{SB}$ has the general form  $H_{SB} = \sum_\a A_\a\ox B_\a$ is immediate; in this case $B'_\a = B_\a - \ave{B_\a}I_B$ and:
\beq
H_S' = \sum_{\a} \ave{B_\a} A_\a \ , \quad H'_{SB} =  \sum_{\a}A_\a\ox B'_\a \ .
\label{eq:H_S'}
\eeq

Now, for our bath operators $B_+(t)=\sum_{k}g_k e^{i\omega_k t}b_k$ and $B_{-}=B_{+}^*$ and the bath state $\rho_B=\ketb{v}{v}$ we have that
\beq
 \ave{B_+(t)}=\Tr[\rho_B B_+(t)]=\sum_{k}g_k e^{i\omega_k t} \ave{b_k}=0\ ,
\eeq
and analogously $\ave{B_-(t)}=0$, both arising from the fact that the annihilation and creation operators average to zero [\cref{eq:useful_b}]. As a result, in our case, in fact $B'(t)=B(t)$.

\section{Proof of \cref{eq:b_CG10}}
\label{app:C}

It is useful to relate $\mathcal{B}_{\alpha\beta\omega }(t)$ with $b_{\alpha\beta\omega}(t)$:
\begin{align}&\mathcal{B}_{\alpha\beta\omega}(t)= \int_0^{t}ds\int_0^{s}ds'e^{i(\omega(s-s')}\mathcal{B}_{\alpha\beta}(s,s')\\&=\left[\int_0^t ds \int_0^t ds'-\int_0^t ds \int_s^t ds'\right]  e^{i(\omega's-\omega s')}\mathcal{B}_{\alpha\beta}(s,s')\nonumber\\&=\left[\int_0^t ds \int_0^t ds'-\int_0^t ds' \int_0^{s'} ds\right]  e^{i\omega(s- s')}\mathcal{B}_{\alpha\beta}(s,s')\nonumber\\
&=b_{\alpha \beta \omega }(t)- \int_0^t ds \int_0^s ds'e^{-i\omega(s'-s)}\mathcal{B}_{\alpha\beta}(s',s)\nonumber
\\&=b_{\alpha \beta \omega}(t)-\mathcal{B}^*_{\alpha \beta \omega}(t)\ ,
\end{align} 
where in the last line we used \cref{conj_Bab}.

\section{$[H_B,\rho_B]\neq 0$} 
\label{app:Dawei}

The bath state can be obtained via a partial trace of the system from the state \cref{eq:state_t}, whose explicit pure density matrix is given in \cref{eq:rhoSBtotal}:
 \begin{align} 
 &\rho_B(t)=\Tr_S\ketb{\phi(t)}{\phi(t)}=(|c_0|^2+|c_1|^2)\ketb{v}{v}\\&+\sum_k \mathfrak{c}_k(t)c_0^*\ketb{k}{v}+\sum_k \mathfrak{c}_k^*(t)c_0\ketb{v}{k}+\sum_{k,k'} \mathfrak{c}_k\mathfrak{c}^*_{k'}\ketb{k}{k'}\ . \nonumber 
 \end{align}
Now using the bath Hamiltonian in \cref{eq:HB} and using the identities $n_j\ket{v}=0$ and $n_j\ket{k}=\delta_{jk}\ket{j}$,  we have:
\bes
\begin{align} H_B \rho_B&=\sum_{j,k}\omega_j \mathfrak{c}_j(t)\mathfrak{c}_k^*(t) \ketb{j}{k}+\sum_j \omega_j c_0^*\mathfrak{c}_j(t)\ketb{j}{v} \\
&= (\rho_B H_B)^\dag \neq \rho_B H_B\\
&=\sum_{j,k}\omega_j \mathfrak{c}_k(t) \mathfrak{c}_j^*(t) \ketb{k}{j}+\sum_j \omega_j c_0\mathfrak{c}_j^*(t)\ketb{v}{j}.
\end{align}
\ees
In the particular case where $\rho_B(0)=\ketb{v}{v}$, or $c_0=0$ and $\mathfrak{c}_k(0)=0$, it trivially follows that $[H_B,\rho_B(0)]=0$.

\section{Inadequacy of the Markov approximation}
\label{app:RWA_useless}

In the main text, we showed that we can reduce \cref{eq:2ordcumm} to \cref{RWA_LE_final}. However, this may lead to an unbounded approximation error, as we now show in detail. 

Before the Markov approximation, \cref{eq:2ordcumm} contains terms of the following form:
\begin{align}
    \int_0^t d\tau \mathcal{B}_{\alpha \beta} (\pm \tau) e^{ \pm i \Omega_0 \tau} \sigma_{\alpha}\sigma_{\beta}\tilde{\rho}(t-\tau)\ ,
\end{align}
where $\alpha,\beta\in\{+,-\}$. The Markov approximation replaces the latter with
\begin{align}
    \int_0^{\infty} d\tau \mathcal{B}_{\alpha \beta} (\pm \tau) e^{ \pm i \Omega_0 \tau} \sigma_{\alpha}\sigma_{\beta}\tilde {\rho}(t)\ .
\end{align}
Therefore the approximation error is the difference between these two quantities, which we write as 
\beq
\delta=\|\Delta_1+\Delta_2\|\leq \|\Delta_1\|+\|\Delta_2\|\ ,
\eeq 
where $\| \cdot \|$ represents the operator norm and 
\bes
\begin{align}
    \Delta_1&\equiv\int_0^{\infty} d\tau \mathcal{B}_{\alpha \beta}(\pm \tau) e^{\pm i \Omega_0 \tau} \sigma_{\alpha}\sigma_{\beta}(\tilde{\rho}(t)-\tilde{\rho}(t-\tau))\\
    \Delta_2&\equiv\int_t^{\infty} d\tau \mathcal{B}_{\alpha \beta}(\pm \tau) e^{\pm i\Omega_0 \tau} \sigma_{\alpha}\sigma_{\beta}\tilde{\rho}(t-\tau)\ .
\end{align}
\ees
Let $\|\cdot \|_1$ denote the trace norm and observe that $\|AB\| \le \|A\|\|B\|_1$ for any pair of operators $A$ and $B$. 

For the Ohmic spectral density $J_2(\o)$ we have, using \cref{eq:Corr_Ohmic}:
\bes
\begin{align}
\|\Delta_2\|&\leq\int_t^{\infty} d\tau |\mathcal{B}_{\alpha \beta}(\pm \tau)| \,\|\sigma_{\alpha}\|\, \|\sigma_{\beta}\|\, \|\tilde{\rho}(t-\tau)\|_1\\
    &\leq \int_t^{\infty} d\tau |\mathcal{B}_{+-}(\pm \tau)|=\eta \omega_c \int_{t}^\infty \frac{d(\omega_c \tau)}{1+ (\omega_c \tau)^2}\\
    &=\left(\frac{\pi}{2}-\arctan(\omega_c \tau)\right)\eta \omega_c\ .
    \label{eq:D4c}
\end{align}
\ees
This quantity goes to zero for $\omega_c \tau\gg 1$ as required. On the other hand, the error term $\Delta_1$ is unbounded. First, by the mean value theorem, there is a point $t'\in [t-\tau,t]$ such that:
\begin{align}
    \|\tilde{\rho}(t)-\tilde{\rho}(t-\tau)\|\leq \tau \sup_{t'\in [t-\tau,t]}\|\dot{\tilde{\rho}}(t')\|\ ,
\end{align}
so that:
\begin{align}
    \|\Delta_1\|\leq \int_0^{\infty}d\tau \, \tau |B_{\alpha \beta}|\sup_{t'\in [t-\tau,t]}\|\dot{\tilde{\rho}}(t')\|\ .
\end{align}

We can bound $\|\dot{\tilde{\rho}}(t')\|$ using the state evolution in \cref{eq:redfield}, where we undo the Markovian approximation by replacing $\tilde{\rho}(t)$ with $\tilde{\rho}(t-\tau)$ [i.e., returning to \cref{eq:2ordcumm}]: 
\bes
\begin{align}
\|\dot{\tilde{\rho}}(t)&\|\leq \int_0^t d\tau |B_{+-}(\tau)|\, \|[\sigma_+,\sigma_-\tilde{\rho}(t-\tau)]\| \notag    \\
    &\quad +\int_0^t d\tau |B_{-+}(-\tau)| \, \|[\sigma_+,\tilde{\rho}(t-\tau)\sigma_-]\|\\
    &\leq 4 \int_0^t  d\tau |B_{+-}(\tau)|\, \|\sigma_+\| \, \|\sigma_-\|\|\tilde{\rho}(t-\tau)\|_1\\
    &\leq 4 \int_0^t  d\tau |B_{+-}(\tau)| \leq 4 \int_0^{\infty}  d\tau |B_{+-}(\tau)|\\
    &= 2\pi \eta \omega_c\ ,
\end{align}
\ees
where in the last line we used \cref{eq:D4c} evaluated at $t=0$.

While the integral
\begin{align}
\int_0^{\infty} \tau^{n}|\mathcal{B}_{+-}(\tau)|d\tau=\frac{\pi}{2}\eta \omega_c^{1-n} \sec(n\pi/2)
\end{align}
converges for $|n|<1$, it does not for $|n|\ge 1$. Indeed, for the Ohmic spectral density we have:
\begin{align}
\int_0^{\infty} \tau|\mathcal{B}_{+-}(\tau)|d\tau=\lim_{\tau \rightarrow \infty}\frac{\eta }{2}\ln [1+(\omega_c \tau)^2]\ ,
\end{align}
which diverges. While this diverging upper bound does not prove that the error $\delta$ itself diverges, it does \emph{suggest} that this is indeed the case and that hence 
the approximation of replacing $\rho(t-\tau)$ by $\rho(t)$ is inaccurate. 
Indeed, the exact solution exhibits excited state population oscillations instead of purely Markovian exponential decay for all values of the parameters of the Ohmic density.

A similar situation arises for the triangular spectral density $J_3(\o)$. We find, numerically, that the integral $
\int_0^{\infty} \tau^{n}|\mathcal{B}_{+-}(\tau)|d\tau$ [recall \cref{eq:B+-J3-RWA-LE}] diverges for $n>0$, and therefore the Markov approximation is inadequate. 

For a rigorous error bound, see Ref.~\cite{Mozgunov:2019aa}.

\section{Proof of \cref{eq:Gamma-cc}}
\label{app:Gamma-cc}

Recall that 
$\Gamma_{\alpha \beta}(\omega)\equiv\int_0^{\infty}d\tau \mathcal{B}_{\alpha \beta}(\tau)e^{i\omega \tau}$ and $H_{SB}=\sigma_+\otimes B_+ + \sigma_-\otimes B_-$ and $B_+^\dagger = B_-$. Thus, if $\alpha\neq\beta$, $B_\alpha = B^\dagger_\beta$. It follows that
\bes
\begin{align}
\mathcal{B}_{\beta \alpha}^*(\tau) & =\Tr\left[\left(\rho_B B_\beta(\tau) B_\alpha\right)^{\dagger}\right]\\
& =\Tr\left[\rho_B U_B(\tau) B^\dagger_\alpha U_B^{\dagger}(\tau) B^\dagger_\beta\right]\\
&=\Tr\left[\rho_B U_B(\tau) B_\beta U_B^{\dagger}(\tau) B_\alpha\right]=\mathcal{B}_{\beta\alpha}(-\tau)\ . 
\end{align}
\ees

Hence, $\mathcal{B}_{\pm\mp}^*(\tau)=\mathcal{B}_{\pm\mp}(-\tau)$ [as well as $\mathcal{B}^*_{\pm\pm}(\tau)=\mathcal{B}_{\mp\mp}(-\tau)$, though we don't use this result], which yields $\Gamma^*_{{\pm\mp}}(\omega) = \int_0^{\infty}d\tau \mathcal{B}_{{\pm\mp}}(-\tau)e^{-i\omega \tau}$.

\section{Simplification of the Lamb shift expressions for the RWA-LE}
\label{app:RWA-LE-integrals}

\subsection{$J_2(\omega)=\eta \omega e^{-\omega/\omega_c}$}
\label{app:RWA-LE-integrals-J2}

Here we derive \cref{eq:Gamma-RWA-J2-2}. Our starting point is \cref{eq:Gamma-RWA-J2-1}, which we write as
\bes
\begin{align}
\label{eq:Gamma_RWA_J2}
&\Gamma_{+-}(\Omega_0)/(\eta\omega_c)= \int_0^{\infty}d(\omega_c\tau) \frac{e^{i\Omega_0 \tau}}{(1+i\omega_c\tau)^2}\\
\label{eq:integ_interest1}
&\quad = \int_0^{\infty} 
    \frac{e^{i\alpha x} dx}{(1+ix)^2}\\
\label{eq:integ_interest}
    &\quad =\int_{-\infty}^{\infty} 
    \frac{e^{i\alpha x} dx}{(1+ix)^2}-\int_{-\infty}^0 
    \frac{e^{i\alpha x} dx}{(1+ix)^2}\ ,
\end{align}
\ees
where $\alpha=\Omega_0/\omega_c$ and $x=\omega_c \tau$.

The complex exponential integral is defined as follows \cite{Corrington:61}:
\begin{align}
\label{eq:ExpIntcomp}
\text{E}_1(z)\equiv\int_z^{\infty}\frac{e^{-u}}{u}du=-\int_{-\infty}^{-z}\frac{e^{u}}{u}du\ ,
\end{align}
where $z=x+iy$ and $|\arg(z)|\leq \pi/2$. The following property holds for $y>0$:
\begin{align}
\label{eq:ExpInt_rel}
-\text{E}_1(-y)=\Ei(y)+i\pi\ ,
\end{align}
where the real exponential integral $\Ei$ was defined in \cref{eq:Ei}.

The first integral on the right hand side of \cref{eq:integ_interest} can be computed via the residue theorem:
\begin{align}
\label{eq:integral_first}
\int_{-\infty}^{\infty} 
    \frac{e^{i\alpha x} dx}{(1+ix)^2}=-\int_{-\infty}^{\infty} \frac{ e^{i\alpha x} dx}{(x-i)^2}=-2\pi i (i\alpha e^{-\alpha})\ ,
\end{align}
since the second order pole of the analytic function is at $x=i$, so that the residue is:
\begin{align}
    \text{Res}=\frac{d}{dx}e^{i\alpha x}\bigg|_{x=i}=i\alpha e^{-\alpha}\ .
\end{align}

For the second integral on the right side of \cref{eq:integ_interest} we use a change of variable $z=x-i$ and integrate by parts:
\bes
\begin{align}
    &\int_{-\infty}^0 
    \frac{e^{i\alpha x} dx}{(1+ix)^2}=-\int_{-\infty}^0 
    \frac{e^{i\alpha x} dx}{(x-i)^2}=-e^{-\alpha}\int_{-\infty}^{-i}\frac{e^{i\alpha z}dz}{z^2}\\
    &=-e^{-\alpha}\left[-\frac{e^{i\alpha z}}{z}\bigg|_{-\infty}^{-i}+i\alpha \int_{-\infty}^{-i}\frac{e^{i\alpha z}}{z}dz\right]\\
    &=i-i\alpha e^{-\alpha}\int_{-\infty}^{-i}\frac{e^{i\alpha z}}{z}dz=i-i\alpha e^{-\alpha}\int_{-\infty}^{\alpha}\frac{e^u}{u}du\ .
\end{align}
\ees
Using the complex exponential integral \cref{eq:ExpIntcomp} alongside \cref{eq:ExpInt_rel}, we have:
\bes
\begin{align}
    \int_{-\infty}^0 
    \frac{e^{i\alpha x} dx}{(1+ix)^2}&=i-i\alpha e^{-\alpha}(-\text{E}_1(-\alpha))\\
    &=\pi\alpha e^{-\alpha}+i-i\alpha e^{-\alpha}\Ei(\alpha)\ .
\end{align}
\ees
Hence, the integral of interest, \cref{eq:integ_interest1}, becomes 
\begin{align}
\int_0^{\infty} 
    \frac{e^{i\alpha x} dx}{(1+ix)^2}&=\pi\alpha e^{-\alpha}-i+i\alpha e^{-\alpha}\Ei(\alpha)\nonumber\\
    &=-i+\alpha e^{-\alpha}(\pi+i\Ei(\alpha))\ .
\end{align}
Collecting these results we obtain \cref{eq:Gamma-RWA-J2-2}.

\subsection{$J_3(\omega)=\eta \omega \Theta(\omega_c-\omega)$}
\label{app:RWA-LE-integrals-J3}

Here we derive \cref{eq:gamma-J3-RWA}. Our starting point is \cref{eq:Gamma-RWA-J3}, which we write as
\begin{align}
\Gamma_{+-}(\Omega_0)/(\eta\omega_c) = \int_0^{\infty}dx \frac{e^{-i x}(1+i x)-1}{x^2} e^{i\a x}\ ,
\label{eq:F9}
\end{align}
where, again, $\alpha=\Omega_0/\omega_c$ and $x=\omega_c \tau$.

We start from the following indefinite integral, solved via integration by parts:
\begin{align}
    \int  \frac{e^{i u x}}{x^2}dx=-\frac{e^{iux}}{x}+iu\int \frac{e^{i u x}}{x}dx\ .
\end{align}
Hence the integral in \cref{eq:F9} is:
\bes
\begin{align}
    &=\int_0^\infty\frac{e^{i(\a-1)x}}{x^2}dx-\int_0^\infty\frac{e^{i\a x}}{x^2}dx+i\int_0^\infty\frac{e^{i(\a-1)x}}{x}dx \\
    &=\left.\frac{e^{i\a x}(1-e^{-ix})}{x}\right|_0^\infty-i\a\int_0^\infty \frac{e^{i\a x}(1-e^{-ix})}{x}dx\\
    &=-i + 2\a \int_{0}^{\infty}e^{i(2\a-1)u}\sinc(u)du
\end{align}
\ees
To compute the last integral, we consider the real and imaginary parts separately. The real part is:
\bes
\begin{align}
&\int_0^{\infty}\cos (2\alpha-1)u \frac{\sin u}{u}du\\
&=\begin{cases} \int_0^{\infty}\frac{\sin 2\alpha u+ \sin (2-2\alpha)u }{2u}du=\frac{\pi}{2} & 0<\alpha<1\\ \int_0^{\infty}\frac{\sin 2\alpha u - \sin (2\alpha-2)u }{2u}du=0&\alpha>1 \end{cases}
\end{align}
\ees
The imaginary part is:
\bes
\begin{align}
&\int_0^{\infty}\sin (2\alpha-1)u \frac{\sin u}{u}du\\
&\quad= \int_0^{\infty}\frac{\cos (2\alpha-2)u-\cos 2\alpha u}{2u}du\\
&\quad=\begin{cases}-\frac{1}{2}\ln\left(\frac{1}{\alpha}-1\right) & 0<\alpha<1\\ -\frac{1}{2}\ln\left(1-\frac{1}{\alpha}\right)&\alpha>1 \end{cases}\ .
\end{align}
\ees

Therefore the integral in \cref{eq:F9} is 
\begin{align}
\frac{\Gamma_{+-}(\Omega_0)}{\eta \omega_c}=\begin{cases}\pi \alpha-i\left(1+\alpha\ln\left(\frac{1}{\alpha}-1\right)\right) & 0<\alpha<1\\ -i\left(1+\alpha\ln\left(1-\frac{1}{\alpha}\right)\right)&\alpha>1 \end{cases}\ .
\end{align}

Collecting these results and using \cref{RWA_S_gamma} we obtain \cref{eq:gamma-J3-RWA}.

\section{Breakdown of the TCL approximation}
\label{app:Breakdown_TCL}

For strong coupling or a small gap $\Omega_0<\omega_c$, as in \cref{fig:dynamics_J1_a05,fig:dynamics_J3_a05}, we observe that the TCL approximation is accurate for short times but fails to capture the subsequent oscillatory behavior. For these cases the operator $I-\Sigma$ in \cref{eq:invertibility} is not invertible. For example, in the case of the impulse spectral density $J_1$, there is a common time $t_0$ where the excited state population $\rho_{11}(t_0)=|c_1(t_0)|^2=0$ independently from the initial condition. This can be seen directly from \cref{eq:exactJ1} or from \cref{eq:exact-gamma-J1} when $\gamma$ diverges. The time $t_0$ is the minimum time $t_n$ such that
\begin{equation}
   t_n=\frac{2}{|\delta|} \left( \arctan \frac{|\delta|}{|\Omega_0-\omega_c|}+n\pi\right) ,
\end{equation}
where $n$ is and integer and $\delta$ is given in \cref{eq:delta_J1}. Since TCL is a time-local master equation, it is impossible to invert the evolution for $t\geq t_0$ back to its initial condition. This implies that the TCL gives inconsistent result when the exact solution for the population vanishes.

A similar argument can be given for the triangular spectral density $J_3$ as shown in \cref{fig:dynamics_J3_a05}. In contrast, this argument is not valid for the infinite range Ohmic spectral density $J_2$, where the population decreases but only vanishes as $t\to\infty$.


\providecommand{\noopsort}[1]{}\providecommand{\singleletter}[1]{#1}%

\end{document}